\documentclass[journal]{IEEEtran}
\ifCLASSINFOpdf
\else
\fi
\usepackage[utf8]{inputenc}
\usepackage{graphicx}
\usepackage{tabularx,booktabs}
\usepackage[T1]{fontenc}
\usepackage[export]{adjustbox}
\usepackage{amsmath,cite}
\usepackage{amssymb}
\usepackage{gensymb}
\usepackage{bbm}
\usepackage{subfig}
\usepackage{listings}
\usepackage{mathtools}
\newtheorem{proposition}{\textbf{Proposition}}
\newtheorem{lemma}{\textbf{Lemma}}

\newtheorem{theorem}{\textbf{Theorem}}
\newtheorem{approximation}{\textbf{Approximation}}
\newtheorem{remark}{\textbf{Remark}}
\usepackage{pgfplots}
\usepackage{array, makecell}
  \pgfplotsset{compat=newest}
\usepackage{wasysym} 
\usepackage{graphicx}

\usepackage{tikz}
\usetikzlibrary{patterns}

\usepackage{tikz}

\usepackage{color,soul}
\begin{document}
\title{Ultra-Reliable Device-Centric Uplink Communications in Airborne Networks: A Spatiotemporal Analysis}
\author{Yasser~Nabil, Hesham ElSawy, Suhail Al-Dharrab, Hussein Attia,  and Hassan Mostafa
\thanks{Y. Nabil is with the Department of Electronics and Communication Engineering, Cairo University, Giza 12613, Egypt,  e-mail: \texttt{Yasser.202010367@eng-st.cu.edu.eg}. \\
H.\ ElSawy is with the School of Computing, Queen’s University, Kingston, ON, Canada, e-mail: \texttt{hesham.elsawy@queensu.ca}. \\
S. Al-Dharrab and H. Attia are with the Electrical Engineering Department, and the Center for Communication Systems and Sensing at King Fahd University of Petroleum \& Minerals (KFUPM), Dhahran, 31261, Saudi Arabia, e-mails: \texttt{\{suhaild, hattia\}@kfupm.edu.sa}.\\
H. Mostafa is with the Department of Electronics and Communication
Engineering, Cairo University, Giza 12613, Egypt, and also with the
University of Science and Technology, Nanotechnology and Nanoelectronics Program, Zewail City of Science and Technology, Giza 12578, Egypt, e-mail: \texttt{hmostafa@staff.cu.edu.eg}}}
\maketitle
\begin{abstract}
This paper proposes an ultra-reliable device-centric uplink (URDC-UL) communication scheme for airborne networks. In particular, base stations (BSs) are mounted on unmanned aerial vehicles (UAVs) that travel to schedule UL transmissions and collect data from devices. To attain an ultra-reliable unified device-centric performance, the UL connection is established when the UAV-BS is hovering at the nearest possible distance from the scheduled device. The performance of the proposed URDC-UL scheme is benchmarked against a stationary UAV-centric uplink (SUC-UL) scheme where the devices are scheduled to communicate to UAV-BSs that are continuously hovering at static locations. Utilizing stochastic geometry and queueing theory, novel spatiotemporal mathematical models are developed, which account for the UAV-BS spatial densities, mobility, altitude, antenna directivity, ground-to-air channel, and temporal traffic, among other factors. The results demonstrate the sensitivity of the URDC-UL scheme to the ratio between hovering and traveling time. In particular, the hovering to traveling time ratio should be carefully adjusted to maximize the harvested performance gains for the URDC-UL scheme  in terms of link reliability, transmission rate, energy efficiency, and delay. Exploiting the URDC-UL scheme allows IoT devices to minimize transmission power while maintaining unified reliable transmission. This preserves  the device's battery and addresses a critical IoT design challenge.
\end{abstract}


\IEEEpeerreviewmaketitle

\section{Introduction}

\IEEEPARstart{A}{irborne} networks are envisioned as a fundamental pillar for 5G and beyond systems \cite{cao2018airborne, bor20195g,zeng2019accessing,li2018uav}. Airborne networks may comprise multi-altitude aircraft platforms that carry base stations (BSs) to provide comprehensive and ubiquitous wireless services. In particular, airborne networks include low-orbit satellite constellations, high-altitude platforms (HAPs), and low-altitude platforms (LAPs)~\cite{darwish2021leo, kurt2021vision,alam2021high,mozaffari2019tutorial}. 
Such emerging airborne networking paradigm offers several benefits which include extending wireless network connectivity to rural places, providing rapid network deployment in case of disasters, aggregating data from massive Internet of Things (IoT) devices, and improving the overall network resilience \cite{lin2018sky,azizi2019profit}. In addition, aerial BSs enhance wireless communication performance and provide a flexible network architecture that can adapt to real-time traffic variations~\cite{cicek2019uav,zeng2017energy}.

Concurrently, device-centric communication architectures are viewed as a crucial approach for future wireless networks, particularly IoT Networks  \cite{ahmad2020device,jameel2018survey}. In device-centric networks, the device's performance is of critical importance, with  fairness among devices as the most significant characteristic \cite{ahmad2020device,jameel2018survey}. Using the mobility capabilities of unmanned aerial vehicles (UAVs), they can travel between devices in order to reduce performance disparities between them. In the case of UAVs, the device-centric design will not only provide a short connection distance, but also a very high line-of-sight (LOS) probability for the link between devices and UAVs. This can finally lead us to the ultra-reliable transmission \cite{masaracchia2021uav}.
Inspired by the spatiotemporal models in \cite{gharbieh2017spatiotemporal,mankar2021spatial,nabil2022data}, this paper utilizes stochastic geometry and queuing theory to characterize device-centric ultra-reliable uplink transmission in large-scale airborne networks. The developed model account for the impact of aggregate interference on transmission reliability, energy efficiency, and delay.

\subsection{Related Work}
UAVs are among the most common  LAPs, which are considered to be agile, low-cost, and easy to deploy~\cite{zeng2016wireless,fotouhi2019survey,matracia2021coverage,mei2019cellular,arshad2018integrating}. Such foreseen merits have motivated the academic and industrial societies to propose the UAV-BSs utilization and assess their performance for various application  in 5G and beyond systems. For instance, the authors in \cite{mozaffari2017mobile} exploit the UAV agility to optimize their locations in response to the real-time device's activity  to minimize the uplink transmission power. The authors in \cite{mozaffari2016unmanned} utilize UAVs to provide downlink services to a rural region with no terrestrial BSs coverage. The uplink data aggregation scenario from IoT devices in  rural areas is considered in~\cite{bushnaq2019aeronautical}. In both \cite{mozaffari2016unmanned, bushnaq2019aeronautical} the trajectory and hovering locations for the UAV are optimized to minimize the trip time to circulate the service area. Instead of using a single UAV, the authors in \cite{choi2019modeling} proposed a fleet of UAVs to gather data from IoT devices that begin transmission whenever they are inside a UAV’s coverage
area. In \cite{zhang2019stochastic}, UAVs are used as downlink sky-haul relays to connect IoT devices to the internet through satellites. 

Wireless power transfer from UAVs to IoT devices to enhance battery life or enable local computation is proposed in \cite{xiong2020uav,ng2022uav}. Moreover, joint optimization of performance metrics for a hovering UAV  that offers edge services for mobile users is presented in \cite{yang2021joint}. The
Age of Information (AoI) minimization for UAV-aided edge computing is presented in \cite{han2022age,yang2022aoi,wang2022data}.
The coexistence of UAVs with conventional terrestrial BSs is studied in \cite{9470921, hattab2020energy}. In \cite{9470921}, terrestrial backhauled UAV-BSs are utilized to improve the coverage of cellular devices. In \cite{hattab2020energy}, the UAVs are used to aggregate data from IoT devices and the terrestrial BSs are used to provide conventional cellular service. The authors in \cite{UAV_Dis} utilize UAVs to compensate for destroyed terrestrial BSs to maintain coverage during disasters. The negative impact of the limited battery power on the UAV-BS service is quantified in \cite{qin2020performance}. To overcome such power constraints, the authors in \cite{lahmeri2019stochastic} proposed a laser-powered UAV operation. Tethered UAV-BSs are investigated in \cite{Tethered_kisk} as an alternative solution for the limited UAV battery lifetime problem. Such a tether provides a perpetual energy source and reliable high-bandwidth backhaul for UAVs at the cost of limiting their mobility.  In \cite{zhu2022aerial} serving UAVs are charged on the fly to increase the travel time.

In addition to the aforementioned use cases, UAV-BSs are promoted to offer ultra-reliable communication services \cite{masaracchia2021uav}. Adapting  UAV locations and altitudes, ultra-reliable line-of-sight (LOS) communication links can be established as proposed in works listed in Table  \ref{table_ref}. However, to attain the required $99.999\%$ transmission success probability ~\cite{popovski2014ultra,sutton2019enabling}, interference from other UAVs and devices has to be considered. In all the  works that studied ultra-reliable transmission through UAVs, stochastic geometry analysis that can characterize interference  in large-scale networks is missing. In addition, in the same context, the spatiotemporal models are overlooked, whereas, in this paper, we try to close this gap.
\begin{table*}[h]
  \begin{center}
     \caption{Ultra-reliable UAV efforts}
     \begin{tabular}{|c|c|c|c|c|}
      \hline
      \textbf{Reference} & Scope     & Network scale & Interference analysis\\
      \hline
     \cite{wang2020packet}  &  \thead{compute the average probability of packet error and effective \\ throughput for  short control packets to UAV }     & single link & No\\
       \hline
        \cite{she2018uav}& \thead{optimize the height and bandwidth allocation to reduce\\ the total bandwidth requirement}     & multi-cell  & No\\
          \hline
         \cite{chen2020power}& \thead{optimize the locations of UAVs and device associations to reduce\\ the
total transmit power of the IoT devices}    & single-cell & No\\
           \hline
          \cite{li2021aerial} & \thead{optimize locations of UAVs carrying reconfigurable intelligent\\ surface (RIS) to reflect the signal from BS to far users}    &   \thead{macro cell with\\ relay clusters} &  \thead{only \\intra-cluster}\\
             \hline
           \cite{azari2017ultra} & \thead{UAV-BS serve ground users in the downlink, considering altitude-dependent\\ path loss exponent
and fading function}    & single-cell & No\\
                \hline
                 \cite{ranjha2019quasi} &  \thead{optimize distance and the blocklength of multi-hop downlink\\  between two IoT devices using UAVs relays}   & \thead{multi-hop\\ relay links} & No\\
                   \hline
                    \cite{han2019uav} & \thead{beamwidth optimization for downlink UAV-assisted\\ non-orthogonal
multiple access (NOMA)}  &\thead{single-cell with\\ multiple groups} &  \thead{only \\intra-group}\\
                      \hline
                       \cite{ranjha2021facilitating} & \thead{optimize
location, height, beamwidth, and resource allocation for\\ a relay UAV between a controller and mobile robots}   & \thead{single-cell with\\ relay link}  & No\\
                         \hline
                          \cite{she2019ultra} &  \thead{optimize the UAVs' height, uplink and downlink duration, and antenna configuration\\ to maximize the horizontal distance between UAVs and a ground station}
  & \thead{multi-UAV \\network} & No\\
                            \hline
                             \cite{el2021uav} & \thead{UAVs offer edge computing offloading for IoT
devices while optimizing \\  UAVs locations, resource allocation, and offloading decisions}    &\thead{multi-UAV \\network}  & No\\
                               \hline
                                \cite{xi2020network} &  \thead{jointly  optimize resource allocation for both
payload and ultra-reliable\\  control  links in a multi-UAV relay network}   & \thead{single-cell multi-UAV\\ relay network}  & \thead{only for\\ payload links}\\
                                  \hline
    \end{tabular}
      \label{table_ref}
  \end{center}
\end{table*}

\subsection{Contributions}
This paper studies aerial data aggregation in large-scale IoT networks, where dynamic device-centric and stationary UAV-centric network scenarios are considered. The former is denoted as an ultra-reliable device-centric uplink (URDC-UL) scheme, where each UAV-BS travels to schedule uplink transmissions and collect data from IoT devices with better LOS conditions, but an amount of time is wasted on physical movement. In contrast, the latter is denoted as a stationary UAV-centric uplink (SUC-UL) scheme, where each UAV-BS stays at a fixed location to schedule the uplink transmissions. Here all the time is dedicated to transmission but subjected to a higher  transmission failure. It is shown that the URDC-UL scheme outperforms the SUC-UL in terms of reliability, rate, energy efficiency, delay, and fairness. More importantly, by virtue of the URDC-UL scheme, the devices can operate at very low transmit power and rarely experience decoding errors,\footnote{Decoding errors lead to packet retransmissions resulting in higher power consumption.}which improves energy efficiency and extends the battery lifetime of the devices.
In summary, the main contributions of this paper compared to the previously 
  mentioned works are summarized as follows.
\begin{itemize}
   \item  It presents a novel device-centric spatiotemporal mathematical model for data aggregation in large-scale IoT networks  via UAVs, with trajectory planning taken into account.
    \item It employs stochastic geometry to examine the ultra-reliable transmission perspective through UAVs in large-scale networks while considering  interference from other UAVs and devices.
    \item It accounts for the UAVs' mobility for ultra-reliable data aggregation from IoT devices. This can extend the device's battery lifetime, in addition to achieving fairness between devices through the device-centric approach.
\end{itemize}

The remainder of the paper is divided as follows: Section \ref{sys_model} introduces the URDC-UL and SUC-UL  system models. Section \ref{TSP} presents the transmission success probability analysis. Section \ref{temp_an} presents the temporal  analysis, and some performance metrics such as outage capacity and energy efficiency. Section \ref{res} presents the numerical results and simulations. Finally, Section \ref{conc} concludes and summarizes the work.

\section{System Model} \label{sys_model}

This section describes the system model for both the URDC-UL scheme and the SUC-UL benchmark. We first introduce the considered network parameters including the aerial channel and antennas models. Following that, the temporal and transmission models for the URDC-UL and the SUC-UL schemes  are presented.

\subsection{Network Model}\label{spa_mob}

An infinite network is considered where IoT devices are distributed according to a Poisson point process (PPP) defined as $\bold\Phi_d=\{d_i\in\mathbb{R}^{2},\forall i\in\mathbb{N}^{+}\}$ with intensity $\lambda_d$  $\text{device}/\text{km}^2$. This PPP placement of devices will effectively cover the entire area and provide some redundant sensing in case a node fails \cite{bushnaq2019aeronautical}. A homogeneous coverage  is adopted for aerial uplink services, where the network is partitioned via a hexagonal tessellation with hexagons of radii $R$. Each UAV-BS is assigned to serve the devices within each hexagonal cell. While the area served by each UAV is fixed by design and based on the definition of PPP \cite{elsawy2016modeling}, the number of devices per each hexagonal cell is independent and follows a Poisson distribution with mean $N$. Such that $N$ equals to the  cell area  multiplied by $\lambda_d$, i.e. $N=\frac{3\sqrt{3} \lambda_d R^2}{2}$. Devices have negligible heights and  UAV-BSs are flying at a constant altitude of $h$.


Aerial communication  is characterized by complex propagation conditions, where obstacles in  environments may lead to deep shadow areas and  the absence of LOS. In particular, the probability of a LOS link depends on the ratio of built-up land area to the total area, the average number of buildings per $\text{km}^{2}$, and buildings' height distribution. Such LOS probability is well-approximated by a modified Sigmoid function with proper environment parameters \cite{al2014optimal},  given by
 \begin{equation}\label{LOSP}
p_{\rm{LOS}}\left(r\right)=\frac{1}{1+a\; \exp \left(-b \left(\frac{180}{\pi}\arctan \left(\lvert \frac{h}{r}\rvert \right)-a \right)\right)}\;,
\end{equation}
where $a$ and $b$ denote environment parameters, and $r$ is the ground Euclidean distance between the UAV and the device. 

The uplink propagation channel between devices and UAVs has a distance-dependent power-law path loss function with different exponents, $\alpha_L$ and $\alpha_N$ for LOS and none LOS (NLOS) channels, respectively. We assume a quasi-static Nakagami-$m$ multipath fading model where the channel gains remain constant during one transmission duration but randomly change across different transmissions. All power fading gains are modeled as independent and identically distributed (i.i.d.) Gamma random variables (RVs) with shape parameters $m_L$ and  $m_N$ for LOS and NLOS, respectively. Note that $\alpha_L < \alpha_N$ and $m_L>m_N$, reflect the better LOS channel conditions when compared to their NLOS counterpart. 

Universal frequency reuse of $W$ Hz is adopted by all UAV-BSs and  devices transmitted with a constant power of $P$. To improve the intended signal and mitigate interference, both the UAVs and devices are equipped with directional antennas. For simplicity, the antenna patterns of the UAVs and devices are approximated with the discretized sectored gain model adopted in \cite{andrews2016modeling}, \cite{ratnarajah2016performance}. In particular, the main lobe and side lobe gains of the UAV antennas are denoted by $G_{uM}$ and $G_{um}$, respectively. Similarly, the main lobe and side lobe gains of the device's antennas are denoted by $G_{dM}$ and $G_{dm}$, respectively. Note that the case of simple devices with Omni-directional antennas is captured by  a special case of $G_{dM}=G_{dm}=1$. Perfect beam alignment is assumed between a UAV-BS and its intended devices in both the URDC-UL and the SUC-UL schemes. However, the orientation of the interfering devices beams with respect to other UAVs is assumed to be random with uniform distribution in the range $(0, 2\pi)$. Hence, the beamforming effect between a UAV and interfering devices can be modeled via i.i.d. discrete RVs  
$G \in \{ G_{dM} G_{uM},\; G_{dM} G_{um},\; G_{dm} G_{uM},\; G_{dm} G_{um}\}$, which has a probability mass function 
 $\mathbb{P}\{G=G_{d\{\cdot\}} G_{u\{\cdot\}} \} \in \{ c_{d} c_{u},\; c_{d} (1-c_{u}),\; (1-c_{d}) c_{u},\; (1-c_{d}) (1-c_{u})\}$ such that $c_{d} = \frac{\theta_d}{2\pi}$ and $c_{u} = \frac{\theta_u}{2\pi}$, where $\theta_d$ and $\theta_u$ are the main lobe beamwidth for devices and UAVs.


\subsection{Transmission and Temporal Model}

We consider a time-slotted system with time slot duration $T_s$, where each device generates a packet of size $L$ bits with probability $\alpha$ every time slot\footnote{The developed model can be extended to other traffic generation models as in \cite{emara_Iot}.}. In addition, each device has a buffer that stores generated packets to be transmitted to its serving UAV according to the first in first out (FIFO) discipline. The UAVs schedule the devices to transmit their head  of the buffer (HoB) packets according to periodic transmission cycles. For the sake of fairness, each transmission cycle includes all the devices within the service area of each UAV, where each device is scheduled only once\footnote{ A queue-aware scheduling assumption is technically challenging in the considered uplink scenario because the UAV may be
oblivious of the queue states of the devices. Moreover, buffer state updates from devices to UAVs result in additional power consumption and signaling overhead \cite{gharbieh2018spatiotemporal,kouzayha2017joint}. Instead and to void wasting of resources, if the assigned device has an empty buffer, it is assumed that it will transmit additional secondary data that is not included in queueing analysis. Consequently, there is always transmission and interference at each time slot from all cells.}. Therefore, the duration of the transmission cycle for a randomly selected UAV is equal to $T_c= N_d\times T_s$, where $N_d$ is the number of devices within the service area of the selected UAV. Note that $N_d$ has a Poisson distribution, with mean $N$, across different UAVs. The scheduling and transmission policies for both schemes are given in the sequel. 
 
 \subsubsection{\underline{\textbf{SUC-UL scheme}}}  All UAVs remain stationary at the center of their service area and the devices are scheduled  according to  round-robin  scheduling  as depicted in Fig. \ref{Spatial_sta}. Each device exploits the entire slot duration $T_s$ for transmitting the HoB packet. To transmit the packet size of $L$ bits within $T_s$, the required transmission rate is given by\footnote{Hereafter, the subscripts UC and DC are used to denote, respectively, the UAV centric (i.e., SUC-UL) and the device-centric (i.e., URDC-UL) schemes.}
\begin{equation}\label{rate}
\mathcal{R}_{\text{UC}}= \frac{L}{T_s}= \zeta \; W   \; \log_2\left(1+\theta_{\text{UC}}\right),
\end{equation}
where $\zeta$ is the rate penalty of using practical coding schemes as opposed to the theoretical Shannon capacity. $\theta_{\text{UC}}$ is the signal-to-interference-plus-noise-ratio (SINR) threshold required to correctly decode the packet at the UAV. Hence, the transmitted packet is correctly received at the serving UAV if the received SINR satisfies 
\begin{equation}\label{s_th}
\mathrm{SINR}_{\text{UC}} \geq \theta_{\text{UC}} = 2^{\frac{L}{T_s \zeta \; W}} -1.\\
\end{equation}

 \subsubsection{\underline{\textbf{URDC-UL scheme}}} 
\begin{figure*}
   \begin{minipage}{0.42\textwidth}
     \centering
   \includegraphics[width=0.7\textwidth]{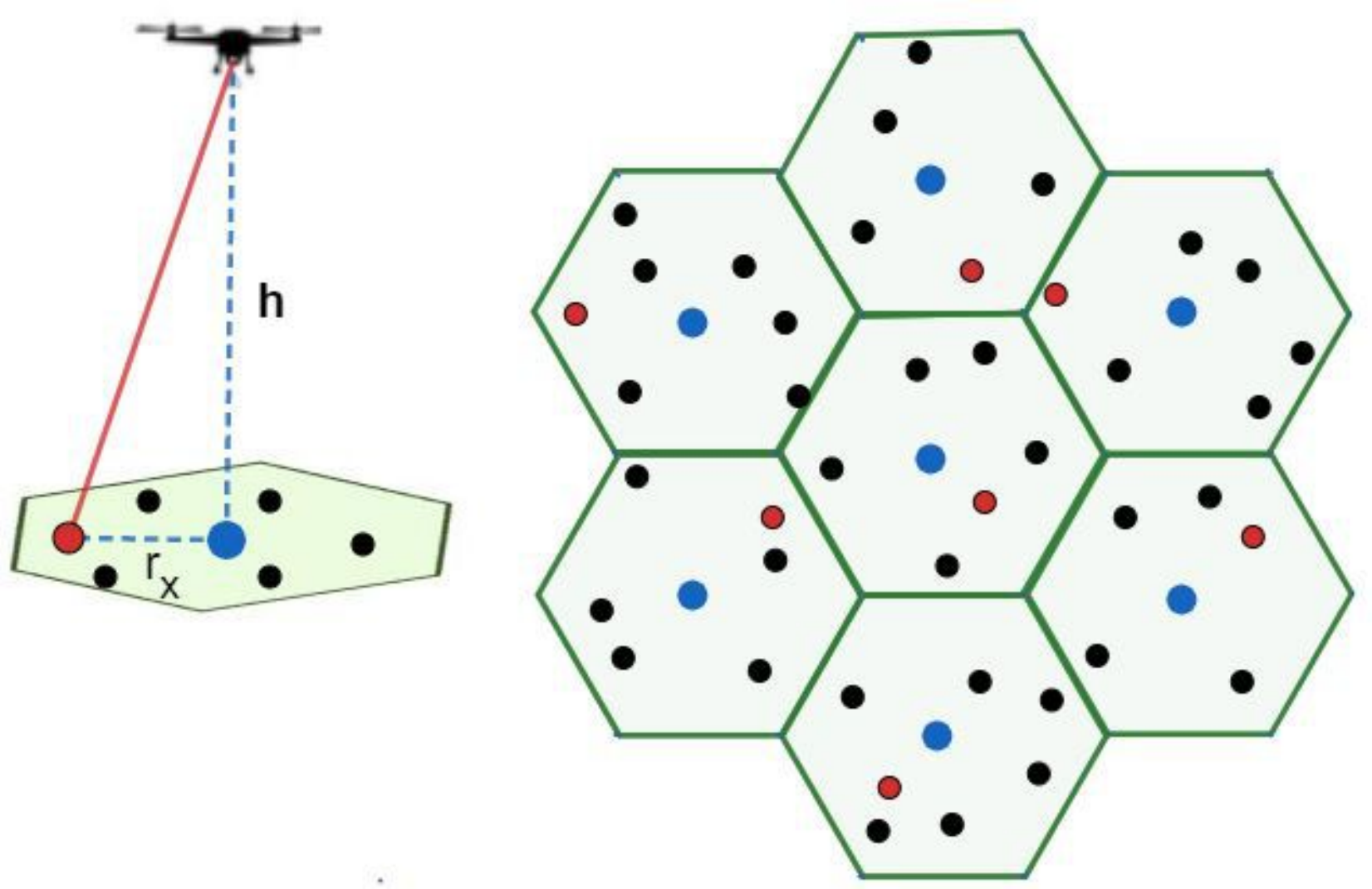}
    \caption{The SUC-UL scheme for $N=6$ (filled circles in blue: UAV fixed locations, red: active devices, black: inactive devices).}
    \label{Spatial_sta}
   \end{minipage}\hfill
   \begin{minipage}{0.42\textwidth}
     \centering
      \includegraphics[width=0.7\textwidth]{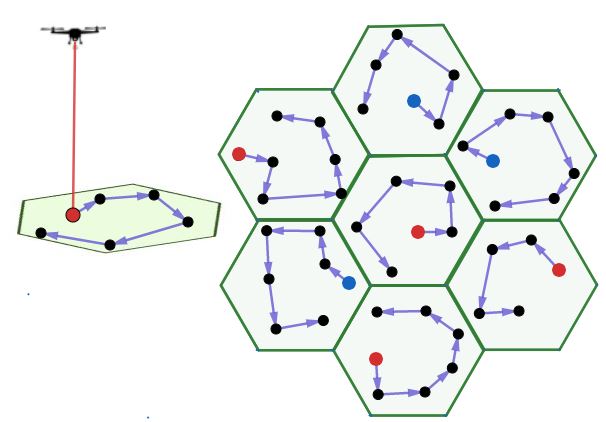}
    \caption{The URDC-UL scheme for $N=6$ (filled circle in blue: traveling UAV, red: UAV hovering close to an active device, and arrows show the trajectory).} 
    \label{Spatial_mob}
   \end{minipage}
\end{figure*}

A trajectory planning is required for the UAV to travel, schedule the devices and collect HoB packets. For simplicity, a greedy scheduling trajectory is employed for the URDC-UL scheme\footnote{Trajectory optimization is a stand-alone problem that is out of the scope of this paper. The main objective of this work is to assess the impact of device-centric uplink scheduling as opposed to the stationary UAV-centric approach. In this context, the gains offered by the URDC-UL scheme will be enhanced  if trajectory optimization or clustered user techniques  are deployed}. In particular, the UAV trajectory starts at a randomly selected device and passes in a hop-by-hop fashion by all devices within the service area of the UAV.  At each hop, the nearest unscheduled device is selected as the next hop in the trajectory, and thus, each device is scheduled once every transmission cycle as shown in Fig. \ref{Spatial_mob}.
Due to practical and environmental constraints and/or possible errors in the localization, sometimes it could be hard for the UAV to hover directly above the device. Thus, the horizontal distance between the UAV and the selected device is assumed not to be zero but is modeled as a Gaussian RV with a zero mean and standard deviation (SD) $\eta$.
 
 The URDC-UL scheme activates a device for uplink transmission once the UAV arrives at its corresponding hop. Hence, there is  the time consumed by the UAV to travel between the devices, where all the devices in the cell remain silent.  For the sake of fair comparison, the transmission cycle duration, $T_c$, is fixed for both the URDC-UL and SUC-UL schemes, where all the devices have to be scheduled. Hence, each device in the URDC-UL is granted $t_{\text{DC}} < T_s$ for packet transmission.  In particular, the total UAV travel time and the total transmission time have to satisfy $\sum_{i=0}^{N_d-1} t_{v, i} + N_d \times t_{\text{DC}} = N_d \times T_{s} = T_c$, where $N_d$ is the number of devices served by a UAV and $t_{v, i}$ is the time it takes the UAV to travel the $i^{\text{th}}$ segment of the trajectory. Note that $t_{\text{DC}}$ is fixed for all devices for the sake of fairness, however, $t_{v, i}$ varies according to the trajectory and the relative locations of devices. Due to the shorter duration allocated to transmit a packet of the same size $L$, the URDC-UL devices have to operate at a higher rate when compared to their SUC-UL counterpart. The required transmission rate for the URDC-UL scheme is
\begin{equation}\label{rate}
\mathcal{R}_{\text{DC}}= \frac{L}{t_{\text{DC}}}= \zeta \; W   \; \log_2\left(1+\theta_{\text{DC}}\right).
\end{equation}
Hence, the URDC-UL has a more strict SINR threshold  which is given by 
\begin{equation}\label{rate_cond_DC}
\mathrm{SINR}_{\text{DC}} \geq \theta_{\text{DC}} = 2^{\frac{L}{t_{\text{DC}} \zeta \; W}} -1.
\end{equation}
In both the URDC-UL and the SUC-UL, packets  correctly received at the UAV are omitted from the  buffer. Otherwise, the HoB packet is retransmitted until it is successfully delivered.

\begin{remark}
While the URDC-UL scheme devices are assigned shorter transmission slots and are required to operate at higher rates, they are privileged with better channel conditions  including high LOS probability and shorter transmission distance when compared to the SUC-UL scheme. Since the LOS links experience less attenuation and/or multipath fading, the URDC-UL scheme offers higher reliability and enables the devices to operate with lower power, which prolongs their battery lifetime.
\end{remark}



\section{Transmission Success Probability Analysis}\label{TSP}

In this section, the transmission success probability for both the URDC-UL and SUC-UL schemes is formulated.

\subsection{URDC-UL }\label{mob_model_ana}

Due to the fact that each UAV serves a single device during each time slot, there could only be one interfering device per hexagonal cell. Note that the travel and transmission time slots of UAVs in different cells are not synchronized due to the different lengths of the trajectory segments in different cells. Hence, the transmission activities of the interfering devices are not synchronized. As such, we adopt the following approximation for the interference devices. 

\begin{approximation}
 Without loss of generality, consider the serving UAV-BS that schedules a typical device to be located at the origin. To ensure mathematical tractability, the interfering devices for a typical device in the URDC-UL  are approximated with a PPP\footnote{The PPP approximation of interferers can cover other scenarios even if devices' spatial deployment is not PPP as in \cite{nabil2022data}.} $\bold\Phi=\{x_i\in\mathbb{R}^{2},\forall i\in\mathbb{N}^{+}\}$ of intensity 

\begin{equation}\label{int_mob}
\lambda_{\text{DC}}= \frac{t_{\text{DC}}} {T_s} \times \frac{2}{3\sqrt{3}  R^2}.
\end{equation}
The intensity in \eqref{int_mob} is equivalent to that of the hovering UAV-BSs, where the term $\frac{t_{\text{DC}}} {T_s}$ captures the asynchronous transmission. Furthermore, due to  the absence of intra-cell interference and the fact that two active devices may simultaneously exist at the margins of two adjacent cells,  but at the same time  other neighboring interference devices will be located at distances  far greater than the cell radius, R. Therefore, the interfering devices are assumed to exist outside an interference-free region, which is approximated via a circle of radius $\frac{R}{2}$ around  the origin. This strikes a compromise between the two extreme cases of no interference protection region at all and the case of an unrealistically large protection region with a radius of R. The accuracy of this approximation will be evaluated in  section \ref{res} of numerical results. 
\label{approx}
\end{approximation}

The SINR of the typical device  is 

\begin{equation}\label{sinr_mob}
 \mathrm{SINR_{\text{DC}}}=\frac{PH_{k,o}G_{dM}G_{uM}(r_x^2+h^2)^{\frac{-\alpha_k}{2}}}{I_L+I_N+\sigma^2},
\end{equation}
where $k\in \{L, N\}$ is an indicator that depends on whether the intended link is LOS or NLOS, $r_x$ is the typical device horizontal distance from the origin which is a RV, $H_{k,o}$ is the intended device channel fading gain, $P$ is the device transmission power, $\sigma^2$ is the noise power, and $I_L$ (resp. $I_N$) is the LOS (resp. NLOS) interference. The LOS and NLOS interference are given by \[I_L=\sum\limits_{\substack{x_i}}\mathbbm{1}_{ \{x_i\in \bold\Phi_L\}} PH_{L,i}G_{\epsilon,i}(r_i^2+h^2)^{\frac{-\alpha_L}{2}},\] 
\[I_N=\sum\limits_{\substack{x_i}} \mathbbm{1}_{ \{x_i\in \bold\Phi_N\}}PH_{N,i}G_{\epsilon,i}(r_i^2+h^2)^{\frac{-\alpha_N}{2}},\]
 where $H_{L,i}$ (resp. $H_{N,i}$) is the channel gain of the $i^{\text{th}}$ LOS (resp. NLOS) interfering device, and
$r_i$ is the horizontal distance between device $x_i$ and the origin.
$\bold\Phi_L$ is the PPP of LOS interfering devices with intensity $ p_{\rm{LOS}}\left(r\right) \times\lambda_{\text{DC}}$, while 
$\bold\Phi_N$ is the PPP of NLOS interfering devices with intensity $ \left(1-p_{\rm{LOS}}\left(r\right)\right)\times \lambda_{\text{DC}}$, and $p_{\rm{LOS}}\left(r\right)$ is given by (\ref{LOSP}).
$G_{\epsilon,i}$ is the device $x_i$ interference antennas gain.
$G_{\epsilon,i} \in \{ G_{dM} G_{uM},\; G_{dM} G_{um}, \;G_{dm} G_{uM},\; G_{dm} G_{um}\}$
with probability $\mathbb{P}_{\epsilon} \in \{ c_{d} c_{u}, \;c_{d} (1-c_{u}),\; (1-c_{d}) c_{u},\; (1-c_{d}) (1-c_{u})\}$, and $\epsilon\in\{1, \;2, \;3, \;4\}$. The indicator function $\mathbbm{1}_{\{\cdot\}}$ takes the value one if the statement $\{\cdot\}$ is true and zero otherwise.
 
To find the transmission success probability of the URDC-UL scheme, we start by formulating the Laplace transform (LT) of the aggregate interference seen at the typical UAV.

\begin{lemma} \label{lemm1}
For the URDC-UL model, the LT of the aggregate LOS and NLOS interference seen by the typical UAV located at the origin is given by (\ref{LOS}) and (\ref{NLOS}) respectively.
\small
\begin{equation}\label{LOS}
\begin{aligned}
\mathcal{L}_{I_L} (s)&=\exp\left\{-2\:\pi\: \lambda_{\text{DC}} \sum_{\epsilon=1}^{4} \mathbb{P}_\epsilon \int_{\frac{R}{2}}^{\infty}p_{\rm{LOS}}\left(r\right) \right.\\
& \quad \quad \left.  \left(1-\left( 1+\frac{sPG_\epsilon(r^2+h^2)^{\frac{-\alpha_L}{2}}}{m_L}\right)^{-m_L}    \right) rdr \right\}.
\end{aligned}
\end{equation}
\begin{equation}\label{NLOS}
\begin{aligned}
\mathcal{L}_{I_N} (s)&=\exp\left\{-2\:\pi\: \lambda_{\text{DC}} \sum_{\epsilon=1}^{4} \mathbb{P}_\epsilon \int_{\frac{R}{2}}^{\infty}\left(1-p_{\rm{LOS}}\left(r\right)\right)\right.\\
& \quad \quad \left.  \left(1-\left( 1+\frac{sPG_\epsilon(r^2+h^2)^{\frac{-\alpha_N}{2}}}{m_N}\right)^{-m_N}    \right) rdr \right\}.
\end{aligned}
\end{equation}
\normalsize
\begin{IEEEproof}
See Appendix A.
\end{IEEEproof}
\end{lemma}

Using the results of Lemma 1, the packet success probability can be formulated as in the following theorem.
\begin{theorem}\label{theorem1}
For the URDC-UL model, the packet success probability is given by
\begin{equation}\label{pcov}
\begin{aligned}
S_p&=\int_{-\infty}^{\infty}\left[\left(S_p|k=L\right) p_{\rm{LOS}}(r_x) + \left(S_p|k=N\right)\right.\\ &\quad \quad \left.\left(1-p_{\rm{LOS}}\left(r_x\right)\right)\right]
\frac{1}{\sqrt{2\pi \eta^2}}\exp\left(\frac{- r_x^2}{2 \eta^2}\right) dr_x,
\end{aligned}
\end{equation}
where $p_{\rm{LOS}}\left(r_x\right)$ is the probability that the intended link is LOS given by (\ref{LOSP}), and
\small
\begin{equation} \label{succ_given_L}
\begin{aligned}
&\!\!(S_p|k)=\sum_{n=1}^{m_k} \left(-1\right)^{n+1} {m_k \choose n}  \exp\left(-\;\frac{g_k\;n \;\theta_{\text{DC}} \; (r_x^2+h^2)^{\frac{\alpha_k}{2}}\; \sigma^2}{PG_{dM}G_{uM}}\right)\\
\;&\mathcal{L}_{I_L} \left(\frac{g_k\;n \;\theta_{\text{DC}} \; (r_x^2+h^2)^{\frac{\alpha_k}{2}}}{PG_{dM}G_{uM}}\right)
\mathcal{L}_{I_N} \left(\frac{g_k\;n \;\theta_{\text{DC}} \; (r_x^2+h^2)^{\frac{\alpha_k}{2}}}{PG_{dM}G_{uM}}\right).
\end{aligned}
\end{equation}
\normalsize
$\mathcal{L}_{I_L}$, $\mathcal{L}_{I_N}$ are given by (\ref{LOS}), (\ref{NLOS}), $g_L=m_L(m_L!)^{-\;\frac{1}{m_L}}$ and $g_N=m_N(m_N!)^{-\;\frac{1}{m_N}}$.
\begin{IEEEproof}
See Appendix B.
\end{IEEEproof}
\end{theorem}

The success probability calculated above is the average. In order to have a detailed characterization of the performance experienced by the devices, the meta-distribution is considered. It is defined as
$\overline{F}_{S_p}(X)=\mathbb{P}\left(S_p(\theta_{DC})> X\right)$, where $ X \in [0,1]$.
 The meta distribution identifies the fraction of devices having success probability at SINR threshold $\theta_{DC}$ with
reliability  $X$.\\
According to \cite{haenggi2015meta} and utilizing the Gil-Pelaez theorem \cite{gil1951note},  the meta distribution is given by
\begin{equation}\label{meta_app}
\overline{F}_{S_p}(X)= \frac{1}{2}+\frac{1}{\pi} \int_0^\infty \frac{\Im \left( e^{-jt \log X} M_{jt}(\theta_{DC}) \right)}{t} dt,
\end{equation}
where $\Im(f)$ is the imaginary part of $f$, $j \triangleq \sqrt{-1}$ and $ M_{jt}(\theta_{DC})$ are the imaginary moments of  the packet success probability at SINR threshold $\theta_{DC}$.\\
Following similar steps as in \cite{ibrahim2020meta} and utilizing our system parameters, the $b^\text{th}$ moment of the packet success probability at SINR threshold $\theta_{DC}$ is given by
\begin{equation}\label{moments}
\begin{aligned}
M_{b}(\theta_{DC})&=\sum_{z=0}^{b} \left(-1\right)^{z} {b \choose z}  \int_{-\infty}^{\infty}\left[z_L \; p_{\rm{LOS}}^{b}(r_x) +\right.\\
&\left. z_N\; \left(1-p_{\rm{LOS}}\left(r_x\right)\right)^{b}\right] \;\frac{1}{\sqrt{2\pi \eta^2}}\exp\left(\frac{- r_x^2}{2 \eta^2}\right) dr_x,
\end{aligned}
\end{equation}
where 
\begin{equation}
\begin{aligned}
&z_k=\sum_{n=0}^{m_k \;z} \left(-1\right)^{n} {m_k \;z \choose n}  \exp\left(-\;\frac{g_k\;n \;\theta_{\text{DC}} \; (r_x^2+h^2)^{\frac{\alpha_k}{2}}\; \sigma^2}{PG_{dM}G_{uM}}\right)\\
&\mathcal{L}_{I_L} \left(\frac{g_k\;n \;\theta_{\text{DC}} \; (r_x^2+h^2)^{\frac{\alpha_k}{2}}}{PG_{dM}G_{uM}}\right)
\mathcal{L}_{I_N} \left(\frac{g_k\;n \;\theta_{\text{DC}} \; (r_x^2+h^2)^{\frac{\alpha_k}{2}}}{PG_{dM}G_{uM}}\right).
\end{aligned}
\end{equation}
Note that $M_{1}(\theta_{DC})$ gives a similar result to the average  packet success probability as Theorem 1.
Substituting $b = jt$ in (\ref{moments}) and  utilizing Newton’s generalized binomial theorem, the $ M_{jt}(\theta_{DC})$ is given by 
\begin{equation}\label{moments_im}
\begin{aligned}
M_{jt}(\theta_{DC})&=\sum_{z=0}^{\infty} \left(-1\right)^{z} \frac{(jt)_z}{z!}  \int_{-\infty}^{\infty}\left[z_L \; p_{\rm{LOS}}^{jt}(r_x)  \right.\\
&+\left. z_N\left(1-p_{\rm{LOS}}\left(r_x\right)\right)^{jt}\right] \frac{1}{\sqrt{2\pi \eta^2}}\exp\left(\frac{- r_x^2}{2 \eta^2}\right) dr_x,
\end{aligned}
\end{equation}
where  $(.)_z$ is the Pochhammer symbol of falling
factorial.

\subsection{SUC-UL}\label{sta_model_ana}

The UAV-BSs in the SUC-UL scheme utilizes the entire time for uplink transmissions. Hence, there is always an active device that is scheduled in each of the hexagonal cells. As such, for simplicity, we utilize Approximation 1 for the interfering devices' locations but with intensity  
\begin{equation}
    \lambda_{\text{UC}} =\frac{2}{3\sqrt{3}  R^2},
\end{equation}
and with interference protection region of $R$. Considering that the typical UAV is hovering at the origin, the SINR of the SUC-UL model is given as in (\ref{sinr_mob}), but here $r_x$ is deterministic, not a RV. The LOS and NLOS interference $I_L$ and $I_N$ are similar to the URDC-UL model 
where $\bold\Phi_L$ is the PPP of LOS interfering devices with intensity $p_{\rm{LOS}}\left(r\right)\times\lambda_{\text{UC}}$, and
$\bold\Phi_N$ is the PPP of NLOS interfering devices with intensity $\left(1-p_{\rm{LOS}}\left(r\right)\right)\times\lambda_{\text{UC}}$.

 To find the transmission success probability of the SUC-UL scheme,  we start by formulating the LT of the aggregate interference seen by the test UAV.

\begin{lemma} \label{lemm2}
For the SUC-UL model, the LT of the aggregate LOS and NLOS interference seen by the test UAV located at the origin is given by (\ref{LOS_s}) and (\ref{NLOS_s}) respectively.
\small
\begin{equation}\label{LOS_s}
\begin{aligned}
\mathcal{L}_{I_L} (s)&=\exp\left\{-2\:\pi\: \lambda_{\text{UC}} \sum_{\epsilon=1}^{4} \mathbb{P}_\epsilon \int_{R}^{\infty}p_{\rm{LOS}}\left(r\right)\right.\\
&\qquad \left. \left(1-\left( 1+\frac{sPG_\epsilon(r^2+h^2)^{\frac{-\alpha_L}{2}}}{m_L}\right)^{-m_L}    \right) rdr \right\}.
\end{aligned}
\end{equation}
\begin{equation}\label{NLOS_s}
\begin{aligned}
\mathcal{L}_{I_N} (s)&=\exp\left\{-2\:\pi\: \lambda_{\text{UC}} \sum_{\epsilon=1}^{4} \mathbb{P}_\epsilon \int_{R}^{\infty}\left(1-p_{\rm{LOS}}\left(r\right)\right)\right.\\
&\qquad \left. \left(1-\left( 1+\frac{sPG_\epsilon(r^2+h^2)^{\frac{-\alpha_N}{2}}}{m_N}\right)^{-m_N}    \right) rdr \right\}.
\end{aligned}
\end{equation}
\normalsize
\begin{IEEEproof}
Same proof as Lemma 1, except $\lambda_{\text{UC}}$ instead of $\lambda_{\text{DC}}$ and integrate from $R$ not $\frac{R}{2}$.
\end{IEEEproof}
\end{lemma}
Utilizing the results of Lemma 2, the packet success probability can be formulated as in the following theorem. 
\begin{theorem}\label{theorem2}
For the SUC-UL model, the packet success probability is given by
\begin{equation}\label{ps_sta}
S_p=\left(S_p|k=L\right) p_{\rm{LOS}}(r_x) + \left(S_p|k=N\right) \left(1-p_{\rm{LOS}}\left(r_x\right)\right),
\end{equation}
where $(S_p|k=L)$ and $(S_p|k=N)$ are given by (\ref{succ_given_L}) but $\theta_{\text{DC}}$ is replaced by $\theta_{\text{UC}}$, and $\mathcal{L}_{I_L}$, $\mathcal{L}_{I_N}$ are given by (\ref{LOS_s}) and (\ref{NLOS_s}).
\begin{IEEEproof}
  As in the proof of Theorem 1, except $r_x$, has a fixed value, not a RV.
\end{IEEEproof}
\end{theorem}
Similarly, the meta distribution can be calculated for the SUC-UL model as we have done for the URDC-UL scheme.
 
\section{Temporal Analysis} \label{temp_an}

In this section, the temporal analysis is carried out. This includes the average UAV travel time for the URDC-UL scheme, the transmission rate, delay, and energy efficiency.

\subsection{Average UAV Travel Time}

In the URDC-UL scheme, the average traveling time by UAVs during a single time slot is defined as $t_V=\mathbb{E}\left[\frac{1}{N_d}\sum_{i=0}^{N_d-1} t_{v, i}\right]$. To evaluate the travel time for each segment $t_{v, i}$, the segment length has to be calculated first. However, the exact characterization for the distribution of the segment length is quite complex. Hence, we resort to a simple yet accurate approximation, which is formally stated in the following proposition. 
\begin{proposition} \label{prop_dist}
The average segment length in the trajectory of a typical UAV can be approximated by 
\small
\begin{align}
\!\!\!\!d_{\rm{avg}}&\approx\mathbb{E}_{N_d}\left[\frac{1}{N_d}\sum_{i=0}^{{N_d-1}} \int_{0}^{\infty} \frac{4\pi\left(N_d-i\right)r^2\;\;e^{\left(-\frac{2\pi\left(N_d-i\right)r^2}{3\sqrt{3} R^2} \right)}}{3\sqrt{3}  R^2}  \;dr\right]\notag\\
&= \mathbb{E}_{N_d}\left[\frac{1}{N_d}\sum_{i=0}^{{N_d-1}} \sqrt{\frac{3\sqrt{3}R^2}{8(N_d-i)}}\right]\label{exp_dist} \\
&>  \frac{1}{N}\sum_{i=0}^{{N-1}} \sqrt{\frac{3\sqrt{3}R^2}{8(N-i)}}\;.
\label{exp_distll}
\end{align}
\normalsize
\begin{IEEEproof}
 The UAV trajectory that passes by the $N_d$ devices is divided into $N_d$ consecutive segments indexed as $i\in\{0,1,\cdots, N_d-1\}$. Since the device's locations are random, the length of the $i^{\text{th}}$ segment across different UAV trajectories is also random. To take the device's intensity into consideration and to ensure that no device is left behind in each scheduling cycle, the length of the $i^{\text{th}}$ segment for a randomly selected UAV trajectory is approximated by the contact distance in a PPP of intensity $\frac{2(N_d-i)}{3\sqrt{3}R^2}$. Averaging across all segments, the average trajectory length is given by \eqref{exp_dist}. Noting that the function $\left[\frac{1}{N_d}\sum_{i=0}^{{N_d-1}} \sqrt{\frac{3\sqrt{3}R^2}{8(N_d-i)}}\;\right]$ in $N_d$ is convex, then Jensen's inequality for convex functions can be applied. Moreover, since  $N_d$ has a Poisson distribution with mean $N$, the expression in \eqref{exp_distll} can be deduced, which is much simpler than  \eqref{exp_dist}.
\end{IEEEproof}
\end{proposition}

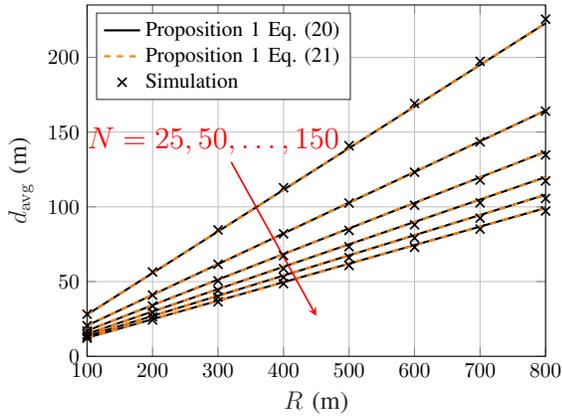
\begin{figure}[h]
\centering
    \begin{tikzpicture}[scale=0.8, transform shape,font=\Large]
%
%
\definecolor{mycolor1}{rgb}{0.00000,0.44700,0.74100}%
\definecolor{mycolor2}{rgb}{0.85000,0.32500,0.09800}%
\definecolor{mycolor3}{rgb}{0.92900,0.69400,0.12500}%
\definecolor{mycolor4}{rgb}{0.49400,0.18400,0.55600}%
\definecolor{mycolor5}{rgb}{0.46600,0.67400,0.18800}%
\definecolor{mycolor6}{rgb}{0.30100,0.74500,0.93300}%

\begin{axis}[%
width=3in,
height=2.3in,
at={(0.758in,0.481in)},
scale only axis,
xmin=100,
xmax=800,
xlabel style={font=\large \color{white!15!black}},
xlabel={$R$ (m)},
ymin=0,
ymax=235,
ticklabel style={font=\normalsize},
ylabel style={font=\large \color{white!15!black}},
ylabel={${d}_{\rm{avg}}\text{ (m)}$},
axis background/.style={fill=white},
legend style={at={(0.02,0.73)}, anchor=south west, legend cell align=left, align=left, draw=white!15!black},
  grid=both,
]
\addplot [color=black, line width =1pt]
  table[row sep=crcr]{%
100	27.8507628403101\\
200	55.7015256806656\\
300	83.5522885213381\\
400	111.403051361409\\
500	139.253814201424\\
600	167.104577042144\\
700	194.955339882638\\
800	222.806102722379\\
};
\addlegendentry{\normalsize{Proposition 1 Eq. (20)}}

\addplot [color=orange,line width =1pt, dashed]
  table[row sep=crcr]{%
100	27.8506353368049\\
200	55.7012706736099\\
300	83.5519060104149\\
400	111.40254134722\\
500	139.253176684025\\
600	167.10381202083\\
700	194.954447357635\\
800	222.80508269444\\
};
\addlegendentry{\normalsize{Proposition 1 Eq. (21)}}

\addplot [color=white, draw=none, mark=x, thick, mark size=3pt, mark options={ black}]
  table[row sep=crcr]{%
100	28.2019\\
200	56.4188\\
300	84.5835\\
400	112.7874\\
500	140.9123\\
600	169.1553\\
700	197.342\\
800	225.5463\\
};
\addlegendentry{\normalsize{Simulation}}

\addplot [color=black, line width =1pt]
  table[row sep=crcr]{%
100	20.5550705038231\\
200	41.1101410077712\\
300	61.6652115118192\\
400	82.2202820155721\\
500	102.775352519181\\
600	123.330423023327\\
700	143.885493527339\\
800	164.440564030902\\
};

\addplot [color=black, line width =1pt]
  table[row sep=crcr]{%
100	17.1048938359097\\
200	34.2097876719029\\
300	51.3146815080174\\
400	68.4195753439432\\
500	85.5244691795505\\
600	102.629363015719\\
700	119.734256851712\\
800	136.839150687453\\
};

\addplot [color=black, line width =1pt]
  table[row sep=crcr]{%
100	14.9819405747452\\
200	29.9638811495185\\
300	44.945821724374\\
400	59.927762299184\\
500	74.9097028735734\\
600	89.8916434484816\\
700	104.87358402326\\
800	119.855524597935\\
};

\addplot [color=black, line width =1pt]
  table[row sep=crcr]{%
100	13.5041929958849\\
200	27.0083859917922\\
300	40.5125789877627\\
400	54.016771983702\\
500	67.5209649793765\\
600	81.0251579752908\\
700	94.529350971229\\
800	108.033543967025\\
};

\addplot [color=black, line width =1pt]
  table[row sep=crcr]{%
100	12.3980927477888\\
200	24.7961854955262\\
300	37.1942782433514\\
400	49.5923709911449\\
500	61.9904637387498\\
600	74.3885564865665\\
700	86.7866492342847\\
800	99.1847419819858\\
};

\addplot [color=orange,line width =1pt, dashed]
  table[row sep=crcr]{%
100	20.5549764007678\\
200	41.1099528015356\\
300	61.6649292023034\\
400	82.2199056030711\\
500	102.774882003839\\
600	123.329858404607\\
700	143.884834805374\\
800	164.439811206142\\
};

\addplot [color=orange,line width =1pt,dashed]
  table[row sep=crcr]{%
100	17.104815528085\\
200	34.20963105617\\
300	51.3144465842551\\
400	68.4192621123401\\
500	85.5240776404251\\
600	102.62889316851\\
700	119.733708696595\\
800	136.83852422468\\
};

\addplot [color=orange,line width =1pt,dashed]
  table[row sep=crcr]{%
100	14.9818719859687\\
200	29.9637439719374\\
300	44.9456159579062\\
400	59.9274879438749\\
500	74.9093599298436\\
600	89.8912319158123\\
700	104.873103901781\\
800	119.85497588775\\
};

\addplot [color=orange,line width =1pt,dashed]
  table[row sep=crcr]{%
100	13.5041311723819\\
200	27.0082623447638\\
300	40.5123935171457\\
400	54.0165246895275\\
500	67.5206558619094\\
600	81.0247870342913\\
700	94.5289182066732\\
800	108.033049379055\\
};

\addplot [color=orange,line width =1pt, dashed]
  table[row sep=crcr]{%
100	12.398035988086\\
200	24.7960719761719\\
300	37.1941079642579\\
400	49.5921439523439\\
500	61.9901799404298\\
600	74.3882159285158\\
700	86.7862519166018\\
800	99.1842879046877\\
};

\addplot [color=white, draw=none, mark=x, thick, mark size=3pt, mark options={ black}]
  table[row sep=crcr]{%
100	20.5134998360431\\
200	40.9811981019515\\
300	61.5439089940629\\
400	81.9832725991519\\
500	102.549275621908\\
600	123.155992887674\\
700	143.480571695034\\
800	163.981290956046\\
};

\addplot [color=white, draw=none, mark=x, thick, mark size=3pt, mark options={ black}]
  table[row sep=crcr]{%
100	16.8477\\
200	33.6819\\
300	50.5278\\
400	67.3827\\
500	84.2149\\
600	101.0828\\
700	117.919\\
800	134.7231\\
};

\addplot [color=white, draw=none, mark=x, thick, mark size=3pt, mark options={ black}]
  table[row sep=crcr]{%
100	14.6684\\
200	29.3362\\
300	44.002\\
400	58.6636\\
500	73.3163\\
600	87.9867\\
700	102.6785\\
800	117.3449\\
};

\addplot [color=white, draw=none, mark=x, thick, mark size=3pt, mark options={ black}]
  table[row sep=crcr]{%
100	13.1987\\
200	26.4066\\
300	39.6011\\
400	52.7984\\
500	65.9931\\
600	79.1925\\
700	92.4234\\
800	105.6031\\
};

\addplot [color=white, draw=none, mark=x, thick, mark size=3pt, mark options={ black}]
  table[row sep=crcr]{%
100	12.1564\\
200	24.3158\\
300	36.4609\\
400	48.6145\\
500	60.7882\\
600	72.9434\\
700	85.0974\\
800	97.2355\\
};

\draw [stealth-,thick, red] (axis cs:450,27) -- (axis cs:320,130) node[above]{$\!\!\!\!\!\!\!N=25,50, \dots, 150$} ;

\end{axis}
\end{tikzpicture}
    \caption{Average traveling distance at different cell radii.}
  \label{avg_dist}
\end{figure}

The accuracy of Proposition~\ref{prop_dist} is validated in Fig. \ref{avg_dist} via Monte Carlo simulations. In each simulation run, $N_d$ devices following the Poisson distribution of mean $N=100$ are uniformly and randomly scattered over a hexagonal cell of radius $R$. Starting from an arbitrary location in the cell, the UAV moves towards the closest device among the $N_d$ devices, and the distance for the $i=0$ segment is recorded. The covered device is eliminated and the UAV moves towards the closest device among the remaining $N_d-1$ devices to record the length of the $i=1$ segment. This process is repeated until all $N_d$ devices are covered and the $N_d$ segments' lengths are recorded. The average segment length across the trajectory in each run is then computed. Towards this end,  Fig.~\ref{avg_dist} validates~\eqref{exp_dist} and~\eqref{exp_distll}  against the average segment length for trajectories over $5\times 10^4$ simulation runs, which illustrates the accuracy of Proposition~\ref{prop_dist}.

Applying the equations of motion, then the average traveling time is:
\begin{equation}
t_V =
\left\{
	\begin{array}{ll}
	\sqrt{\frac{d_{\rm{avg}}}{a_u + a_d}}  & \mbox{if } d_{\rm{avg}} \leq s_u + s_d \\
				t_u +t_d + \frac{d_{\rm{avg}}-s_u - s_d}{v}  & \mbox{if } d_{\rm{avg}} > s_u + s_d 
	\end{array}
\right.,
\end{equation}
where $a_u$ (resp. $a_d$) is the UAV acceleration (resp. deceleration), 
 $t_u=\frac{v}{a_u}$ (resp. $t_d=\frac{v}{a_d}$) are the times needed for the
UAV to change its speed from $0$ to $v$ (resp. $v$ to $0$), and $s_u=\frac{1}{2} a_u t_u^2$ (resp. $s_d=\frac{1}{2} a_d t_d^2$) are the distances needed by the
UAV to change its speed from $0$ to $v$ (resp. $v$ to $0$). From $t_V$, the transmission time for the URDC-UL scheme can be estimated as $t_{\text{DC}}=T_s-t_V$.

\subsection{Transmission Rate \& Delay}\label{qu_m}
 Here the transmission rate and delay analysis are presented for both models.
\subsubsection{URDC-UL}\label{qu_m}
The maximum achievable transmission rate is defined by the ergodic capacity $C = \zeta \,\, W\,\, \mathbb{E}[\log_2(1+\text{SINR})]$ bits/s. Operating at this rate requires instantaneous knowledge of the SINR \cite{george2017ergodic}. However, this is not applicable to large-scale IoT networks. Alternatively, a fixed transmission rate, lower than ergodic capacity, that experiences some outages are utilized.
In the URDC-UL model, the outage capacity is given by 

\begin{equation}\label{out_m}
C_{\text{DC}}=\mathbb{P} \left(\rm{SINR_{\text{DC}}}>\theta_{\text{DC}}\right) \times  \frac{t_{\text{DC}} } {T_s} \; \times  \zeta   W \; \log_2\left(1+\theta_{\text{DC}}\right),
\end{equation}
where the multiplicative term $\frac{t_{\text{DC}} } {T_s}$ emphasizes the shorter transmission during $t_{\text{DC}} $ when compared to the entire time slot $T_s$.
The matrix analytic method (MAM) \cite{alfa2016applied} is utilized to construct a Geo/PH/1 (see \cite[Sec. 5.8]{alfa2016applied}) queueing system at each device, where Geo stands for geometric inter-arrival process and PH stands for the Phase type departure process that accounts for the network dynamics \cite{alfa2016applied}. In particular, geometric inter-arrival times with parameter $\alpha$ are considered at each device. Moreover, since  each device attempts to transmit a packet once in each transmission cycle, a PH-type distribution is used to track the round-robin scheduling consisting of $N_d$ time slots, then to capture the probabilistic transmission which is SINR-dependant. Such traffic departure PH type distribution is defined by the initialization vector $\boldsymbol{\beta} = [1 \;\; 0 \;\; 0\; \cdots  \;0]$ of size $ 1 \times N_d$, transient transition matrix $\mathbf{S}$ of size $ N_d\times  N_d$ and an absorption vector $\mathbf{s}$ of size $ N_d \times 1$, given by
 \small
\begin{equation}\label{T_mat}
\mathbf{S}=\begin{bmatrix}
0 & 1 & 0 &\cdots & 0\\
0 & 0 & 1 &\cdots & 0\\
\vdots  & \vdots  & \ddots  & \ddots & \vdots\\
0  & 0  & \cdots  & 0 & 1\\
1-S_{p} & 0 & 0 & 0 & 0
\end{bmatrix}
 \qquad \text{and} \qquad \mathbf{s}=\begin{bmatrix}
0 \\
0 \\
\vdots \\
0  \\
S_{p}
\end{bmatrix},
\end{equation}
\normalsize
where $S_p$ is the packet transmission success probability calculated in Theorem 1 for the SINR threshold given by (\ref{rate_cond_DC}). 
 
The Geo/PH/1 queueing model has state-space $(q,d)$, where $q \in \mathbb{N}$ is the number of packets in the  buffer and $d \in \{0,\; 1,\;  \cdots,\;  N_d\}$ is the number of time slots that elapsed since the last transmission attempt. 

Owing to the fact that only one packet can arrive and/or depart during any time slot, the developed Geo/PH/1  model is a quasi-birth-death (QBD) process with a transition matrix 
\small
\begin{equation}\label{qbd}
\mathbf{P}=\begin{bmatrix}
\mathbf{B} & \mathbf{C} & \mathbf{0} & \mathbf{0} & \mathbf{0} & \ldots\\
\mathbf{E} & \mathbf{A_1} & \mathbf{A_0} & \mathbf{0} &\mathbf{ 0}& \ddots\\
\mathbf{0} & \mathbf{A_2} & \mathbf{A_1} & \mathbf{A_0} & \mathbf{0} & \ddots\\
\vdots & \ddots & \ddots & \ddots & \ddots& \ddots
\end{bmatrix}, 
\end{equation}
\normalsize
where the matrices $\mathbf{B}\!=\! 1-\alpha$, $\mathbf{E}\!=\! (1-\alpha) \mathbf{s}$, and $\mathbf{C} \!= \alpha \boldsymbol{\beta}$ are the sub-stochastic boundary matrices that track, respectively,
the transitions within, to, and from the idle state. The matrices $\mathbf{A_0} \!=\! \alpha  \mathbf{S}$, $\mathbf{\mathbf{A_1} \!=\! \alpha \mathbf{s} \boldsymbol{\beta} + (1-\alpha) \mathbf{S}}$, and $\mathbf{A_2} \!=\! (1-\alpha) \mathbf{s} \boldsymbol{\beta} $ are of size $N_d \times N_d$ that track, respectively, the upward transitions from $q$ to $q+1$, the transitions within the same level $q$, and the downward transitions from $q+1$ to $q$.

\begin{remark}
The queuing system is stable (packets have a finite delay)  if the  packet departure rate $\frac{S_{p}}{N_d}$ is higher
than the packet arrival rate $\alpha$.
\end{remark}
 
After ensuring the stability of the queue, the steady-state solution of the queueing model is obtained by solving (\ref{ss_queue}), where $\mathbf{1}$ is  a column vector of ones.
\begin{align} \label{ss_queue}
\boldsymbol{\pi} =\boldsymbol{\pi} \mathbf{P} \quad \text{and} \quad \boldsymbol{\pi} \mathbf{1} = 1.
\end{align}

The solution $\boldsymbol{\pi}=[\pi_0 \; \boldsymbol{\pi}_1  \cdots \boldsymbol{\pi}_q \cdots]$ is the joint distribution of the state space $(q,d)$, such that $\pi_0$ is the probability of having idle buffer. For $q\geq1$, the vector $\boldsymbol{\pi}_q$ of length $1 \times N_d$ tracks the probabilities of elapsed time slots, $d$, since the last attempt transmission of the HoB packet, when there are $q$ packets in the buffer. MAM\footnote{The MAM computation is carried out offline to  come up with long-term network design parameters (e.g., $\alpha$  and $L$). These design parameters remain fixed as long as the statistical network parameters (e.g, devices density, UAVs density, and fading distribution) remain constant.} is used to solve the system in \eqref{ss_queue} as follows
\begin{theorem}
 The steady-state solution  $\boldsymbol{\pi}$ is given by 
	 	\begin{equation}
  	 	   \label{SS_sol_theorem}
  	 	\!\!\!\!\!\!\!\!\!\!\!\!	\boldsymbol{\pi}_{q} \! =\!\left\{\begin{matrix}
  	 	(1+\alpha \boldsymbol{\beta} \mathbf{M}  (\mathbf{I}-\mathbf{R})^{-1}\mathbf{1})^{-1}, & \text{for} \; q=0 \\ 
  \pi_{0} \alpha \boldsymbol{\beta} \mathbf{M}, 	 & \text{for}\; q=1 \\
  	 	\boldsymbol{\pi}_{1} \mathbf{R}^{q-1}, &  \text{for}\; q\geq 2
  	 	\end{matrix}\right.	
  	 	\end{equation}
where $\mathbf{M}=(\mathbf{I}-\alpha \mathbf{s} \boldsymbol{\beta}-(1-\alpha )\mathbf{S}-\mathbf{R} (1-\alpha ) \mathbf{s} \boldsymbol{\beta})^{-1} $, and
$\mathbf{R}$ is the rate matrix given by $\mathbf{R}=\alpha \mathbf{S}(\mathbf{I}-\alpha \mathbf{s} \boldsymbol{\beta}-(1-\alpha )\mathbf{S}-\alpha\mathbf{S}\mathbf{1}\boldsymbol{\beta})^{-1}  $
\begin{IEEEproof}
	$\pi_{0}$ and $\boldsymbol{\pi}_{1}$ are found by solving the boundary equation $\boldsymbol{\pi}_{1}=\pi_{0} \mathbf{C}+ \boldsymbol{\pi}_{1}(\mathbf{A_1} + \mathbf{R} \mathbf{A_2})$ and the normalization condition $\pi_{0}+\boldsymbol{\pi}_{1}(\mathbf{I}-\mathbf{R})^{-1}\mathbf{1}=1$. After that, $\boldsymbol{\pi}_{q}$ follows from the definition of the rate matrix  $\mathbf{R}$  as the minimal non-negative solution of $\mathbf{R} =\mathbf{A_0}+\mathbf{R} \mathbf{A_1}+\mathbf{R}^2 \mathbf{A_2}$, and since $\mathbf{A_2}$ is of rank
one, an explicit expressions for   $\mathbf{R}$ is  obtained \cite{alfa2016applied}.
\end{IEEEproof}
\end{theorem}

The steady-state solution can be used to investigate certain performance indicators, such as the average queue length as specified by
\begin{equation} \label{av_size}
    Q_L=\sum_{i=1}^{\infty}i\boldsymbol{\pi}_i \mathbf{1} = \boldsymbol{\pi}_1 (\mathbf{I}-\mathbf{R})^{-2} \mathbf{1}.
\end{equation}
Utilizing Little’s Law, the average total delay\footnote{The  results of the steady-state solution and average total delay will remain the same if another queuing discipline such as last-come-first-served (LCFS) is utilized \cite{bose2013introduction,alfa2016applied}.} (queueing  and transmission delay) is calculated as
\begin{equation}
    Q_W =  \frac{\boldsymbol{\pi}_1 (\mathbf{I}-\mathbf{R})^{-2} \mathbf{1}}{\alpha}.
\end{equation}

\subsubsection{ SUC-UL}

In the SUC-UL model,  the transmission is during the whole time slot duration $T_s$, and the outage capacity is given by 
\begin{equation}\label{out_s}
C_{\text{UC}}=\mathbb{P} \left(\rm{SINR_{\text{UC}}}>\theta_{\text{UC}}\right) \times   \zeta W   \; \log_2\left(1+\theta_{\text{UC}}\right).
\end{equation}
Each device attempts to transfer a packet of size $L$ completely  during its assigned time slot that yields the SINR threshold in (\ref{s_th}). Similarly, a Geo/PH/1 queueing model is studied at each device, where all the equations are identical to the model described in \ref{qu_m}. The only variation is the term $S_p$; here, it refers to the packet transmission success probability derived in Theorem 2 for the SINR threshold specified in (\ref{s_th}).

 \subsection{Energy Efficiency}

To make an objective assessment of both models, the overall energy consumption is considered from both the UAV and devices point of view.

 \subsubsection{URDC-UL }
From the UAV viewpoint, the overall energy consumption is separated into propulsion energy and energy related to communication \cite{zeng2019energy}.
The power required for communication $P_c$ is considered constant. This power is employed in a variety of tasks, including communication and signal processing circuits, as well as signal transmission and reception. Propulsion energy, on the other hand, is the mechanical energy that is used during hovering or forward movement. Note that communication-related energy is negligible when compared to propulsion energy consumption.
According to \cite{zeng2019energy}, for a rotary-wing UAV that can hover or move forward with speed $v$, the propulsion power consumption
is given by
\small
\begin{equation}\label{pow_mob}
P_t(v)=P_0\left(1+\frac{3v^2}{U_{tip}^2}\right)+ P_i\left(\sqrt{1+\frac{v^4}{4 v_0^4}}-\frac{v^2}{2 v_0^2}\right)^{\frac{1}{2}}+\frac{1}{2}d_0 \rho s_r A v^3,
\end{equation}
\normalsize
where  $P_0$ and $P_i$ are two constants representing the blade profile power and induced power in hovering given by
\begin{equation}
P_0=\frac{\delta}{8} \rho s_r A  \Omega ^3 R_r^3 \quad \text{and} \quad P_i=(1+k_c)\frac{W_u^{\frac{3}{2}}}{\sqrt{2 \rho A}},
\end{equation} 
where $W_u$ is the UAV weight,
$\rho$ is the air density,
$R_r$ is the rotor radius,
$A$ is the rotor disc area,
$U_{tip}$ is the tip speed of the rotor blade,
$v_0$ is the mean rotor-induced velocity in hover,
$d_0$ is the fuselage drag ratio,
$s_r$ is the rotor solidity,
$\delta$ is the profile drag coefficient,
$\Omega$ is the blade angular velocity, and
$k_c$ is the incremental correction factor to induced power.
By substituting $v =0$ in (\ref{pow_mob}), we obtain the power consumption during hovering which is a constant given by
\begin{equation}\label{pow_hov}
P_h = P_0 + P_i.
\end{equation} 
For the  URDC-UL model, the total energy consumption by the UAV in a single time slot is the traveling energy of the UAV during $t_V$ plus
the hovering energy of the UAV during $t_{\text{DC}}$ plus communication-related energy during $t_{\text{DC}}$ given as
\begin{equation}\label{mob_po}
E_{\text{DC}_{\text{UAV}}}=P_t\;t_V+P_h\;t_{\text{DC}}+ P_c \;t_{\text{DC}},
\end{equation} 
where $P_t$ is the traveling power  at speed $v$ given by (\ref{pow_mob}).

The energy efficiency in bits/Joule is defined as the transmission throughput per unit of energy consumed during a fixed time interval.
Accordingly, the energy efficiency in the URDC-UL model from the UAV viewpoint $EE_{\text{DC}_{\text{UAV}}}$ is given by
\begin{equation}\label{EE_m}
EE_{\text{DC}_{\text{UAV}}}=\frac{C_{\text{DC}} \times T_s}{E_{\text{DC}_{\text{UAV}}}}.
\end{equation}
Considering only the uplink transmission power, the energy efficiency in the URDC-UL model from the device viewpoint  is given by
\begin{equation}\label{EE_m_iot}
EE_{\text{DC}_{\text{IoT}}}=\frac{C_{\text{DC}} \times T_s}{P \times t_{\text{DC}}}.
\end{equation} 
\subsubsection{SUC-UL}
From the UAV point of view, the total energy consumption in a single time slot  is the hovering energy  of the UAV during the whole  time slot duration $T_s$, plus the communication-related energy during $T_s$ given as
\begin{equation}\label{st_po}
E_{\text{UC}_{\text{UAV}}}=P_h\;T_s+ P_c \;T_s,
\end{equation} 
and the energy efficiency from the UAV viewpoint in the SUC-UL model $EE_{\text{UC}_{\text{UAV}}}$ is given by
\begin{equation}\label{EE_s}
EE_{\text{UC}_{\text{UAV}}}=\frac{C_{\text{UC}} \times T_s}{E_{\text{UC}_{\text{UAV}}}}=\frac{C_{\text{UC}}}{P_h + P_c},
\end{equation} 
while the energy efficiency from the device viewpoint in the SUC-UL model is given by
\begin{equation}\label{EE_s_iot}
EE_{\text{UC}_{\text{IoT}}}=\frac{C_{\text{UC}} \times T_s}{P \times T_s}=\frac{C_{\text{UC}}}{P}.
\end{equation}

\section{Numerical Results}\label{res}

This section validates the analytical results for success probability by performing Monte Carlo simulations to corroborate the derivations. Then, the two  models are compared in terms of outage capacity, delay, and energy efficiency. Furthermore, the effect of other parameters such as the UAV height, $t_{\text{DC}}$  duration, and transmission power is discussed.
Unless otherwise specified, the results in this section are based on the numerical parameters listed in Table   \ref{table1}, where the values of mechanical parameters are taken from \cite{zeng2019energy}. In addition, the environment parameters are given in Table \ref{table2} according to \cite{al2014optimal}. The $t_V$ value is calculated utilizing proposition 1, and the ratio between $t_{\text{DC}}$ and $t_V$ is selected to be unity which provides high gain while ensuring that the total energy consumption during one transmission cycle is still within the  practical limits as we will present in this section. Note that the values of $t_V$, $t_{\text{DC}}$, and $T_s$  remain fixed unless system parameters or devices' locations are changed. Moreover, the number of devices for the queueing analysis $N_d$ is configured to be equal to the mean number of devices in each cell $N$. As a result, the delay performance is calculated as the average of all cells.

\begin{table} 
  \begin{center}
     \caption{Parameters for numerical demonstration.}
     \begin{tabular}{|lcl|}
      \hline
      \textbf{Parameter Description} & \textbf{Symbol} & \textbf{Value} \\
      \hline
      Hexagonal cell radius & $R$  &    651.5 m \\
      UAV height &  $h$ & 30 m\\
  UAV intensity &$\lambda_u$   & $0.90689 / \text{km}^{2}$\\
  Device intensity   & $\lambda_d$   & $ 90.693 / \text{km}^{2}$\\
      Main lobes gain  & $G_{dM}$, $G_{uM}$  & 5 dBi  \\
    Side lobes gain  & $G_{dm}$, $G_{um}$  & 0 dBi    \\
    Main lobe beamwidths &  $\theta_d$, $\theta_u$  & 40$\degree$ \\ 
      Path loss exponents & $\alpha_L$, $\alpha_N$ & 2.5, 4 \\
      Gamma RVs  shape parameters   & $m_L$, $m_N$ &  3, 1 \\
      Theoretical-to-practical rate &$\zeta$ & 0.8\\
        Packet's generation probability   & $\alpha$           &   0.005   \\
      Average number of devices per cell &   $N$       &       100\\
    Average traveling  duration  &   $t_V$   & 6.4365 s\\
   Transmission duration & $t_{\text{DC}}$   & 6.4365 s\\
      Time slot duration & $T_s$  &  12.8729 s\\
      UAV speed   &   $v$   & 22 m/s\\
UAV acceleration and deceleration   &  $a_u$,$a_d$   & 11 m/s$^2$\\
     Bandwidth  &    $W$                &   125 kHz\\
   Noise power    & $\sigma^2$  &   $-90$ dBm\\
    Device transmission  power   & $P$ & 1 mW\\
    UAV communication related power   &   $P_c$            &        50 mW  \\
  SD of the horizontal distance   &    $\eta$   & 20 m \\
       Simulation area & -- & 20000 $\text{km}^2$ \\
        UAV weight      &   $W_u$                &  100 N       \\
 Air density    & $\rho$           &      1.225 kg/m$^3$  \\ 
   Rotor radius    &   $R_r$                &     0.5  m  \\
   Rotor disc area    &   $A$                  &   0.79 m$^2$    \\
   Tip speed of the rotor blade   &   $U_{tip}$           &      200 m/s  \\
   Mean rotor induced velocity in hover   &   $v_0$               &     7.2 m/s   \\
   Fuselage drag ratio   &   $d_0$                 &      0.3   \\
  Rotor solidity &   $s_r$                      &    0.05     \\
   Profile drag coefficient    &   $\delta$               &     0.012    \\ 
   Blade angular velocity   &   $\Omega$               &     400 radians/s    \\
  Incremental correction factor  &   $k_c$             &        0.1      \\
   \hline
    \end{tabular}
      \label{table1}
  \end{center}
\end{table}

\begin{table} 
  \begin{center}
  \caption{Environment  parameters \cite{al2014optimal}}
   \begin{tabular}{|lcl|}
      \hline
            \textbf{Environment} & $a$ & $b$ \\
      \hline
      Suburban & 4.88 & 0.429 \\
      Urban & 9.612 & 0.158 \\
      Dense urban & 12.081 & 0.114 \\
      High-rise urban & 27.23 & 0.078 \\
      \hline
      \end{tabular}
      \label{table2}
  \end{center}
\end{table}

 \begin{figure}
     \centering
   \begin{tikzpicture}[scale=0.7, transform shape,font=\Large]
  \input{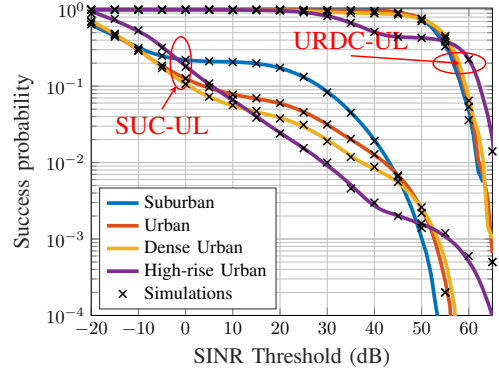}
\end{tikzpicture}
    \caption{Average success probability for both models.}
    \label{succ_mobility}
  \end{figure}
  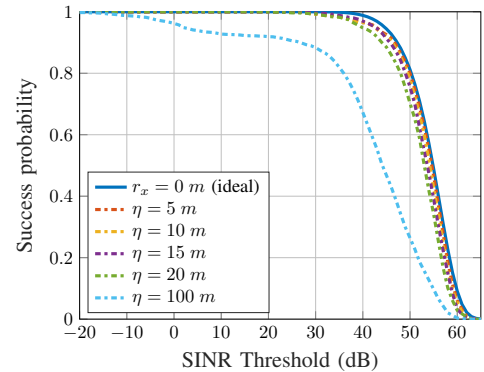
\begin{figure}
     \centering
%
%
\definecolor{mycolor1}{rgb}{0.00000,0.44700,0.74100}%
\definecolor{mycolor2}{rgb}{0.85000,0.32500,0.09800}%
\definecolor{mycolor3}{rgb}{0.92900,0.69400,0.12500}%
\definecolor{mycolor4}{rgb}{0.49400,0.18400,0.55600}%
\definecolor{mycolor5}{rgb}{0.46600,0.67400,0.18800}%
\definecolor{mycolor6}{rgb}{0.30100,0.74500,0.93300}%
  \begin{tikzpicture}[scale=0.7, transform shape,font=\Large]

\begin{axis}[%
width=3in,
height=2.3in,
at={(0.1158in,0.081in)},
scale only axis,
xmin=-20,
xmax=65,
xlabel style={font=\large \color{white!15!black}},
xlabel={SINR Threshold (dB)},
ymin=0,
ymax=1,
ticklabel style={font=\normalsize},
ylabel style={font=\large \color{white!15!black}},
ylabel={Success probability},
axis background/.style={fill=white},
legend style={font=\footnotesize,at={(0.02,0.02)}, anchor=south west, legend cell align=left, align=left, draw=white!15!black},
  grid=both,
]

\addplot [line width=0.6mm,color=mycolor1]
  table[row sep=crcr]{%
-20	1.00000000014376\\
-19	1.00000000017272\\
-18	1.00000000020786\\
-17	1.00000000024937\\
-16	1.00000000030019\\
-15	1.00000000036033\\
-14	1.00000000043389\\
-13	1.00000000052237\\
-12	1.00000000062671\\
-11	1.00000000075366\\
-10	1.00000000090621\\
-9	1.00000000108909\\
-8	1.00000000130871\\
-7	1.00000000157433\\
-6	1.0000000018937\\
-5	1.0000000022765\\
-4	1.00000000278237\\
-3	1.00000000329086\\
-2	1.00000000394708\\
-1	1.00000000476374\\
0	1.00000000571865\\
1	1.00000000687474\\
2	1.00000000821714\\
3	1.00000000993563\\
4	1.00000001193857\\
5	1.00000001439149\\
6	1.00000004499639\\
7	1.00000001927719\\
8	1.00000002491863\\
9	1.00000002993025\\
10	1.0000000025144\\
11	1.00000012039162\\
12	1.0000000508536\\
13	1.00000006025189\\
14	0.999999502178617\\
15	0.999999547751989\\
16	1.00000008994985\\
17	0.999999323571078\\
18	0.999999155663721\\
19	1.00000004190676\\
20	0.999999927091999\\
21	1.00000418908329\\
22	0.999997711967093\\
23	0.999998061374443\\
24	0.999998271469348\\
25	0.99999270244461\\
26	0.999987850171482\\
27	0.999970916908125\\
28	0.999967028229451\\
29	0.999949573855229\\
30	0.999860840728922\\
31	0.999765761065547\\
32	0.999617495618449\\
33	0.999429300090386\\
34	0.999016553703099\\
35	0.998438519561456\\
36	0.997542585441385\\
37	0.996265444529324\\
38	0.994426616541487\\
39	0.991795978083996\\
40	0.988234109320032\\
41	0.983588922387673\\
42	0.977110573962553\\
43	0.968816143495145\\
44	0.957906615253337\\
45	0.944383917074923\\
46	0.927991170674502\\
47	0.906770062587112\\
48	0.881484774746628\\
49	0.849595830319039\\
50	0.811723117067518\\
51	0.765367072491092\\
52	0.710211893388283\\
53	0.645270626408309\\
54	0.571248059899155\\
55	0.489172031760261\\
56	0.401315548305668\\
57	0.311845155898167\\
58	0.226660703702564\\
59	0.15131144859462\\
60	0.0910584297888256\\
61	0.0483571558388314\\
62	0.0220042285803634\\
63	0.00833019590101968\\
64	0.00252927179867436\\
65	0.00058632295695152\\
66	9.90870382555966e-05\\
67	1.11881936471075e-05\\
68	7.82600900704606e-07\\
69	3.02667171915485e-08\\
70	5.65529033703982e-10\\
71	4.37087429350815e-12\\
72	1.1166559522007e-14\\
73	7.27753966217466e-18\\
74	9.28887159597158e-22\\
75	1.50828717851807e-26\\
76	1.96863232153867e-32\\
77	1.14877090667334e-39\\
78	1.45111562121455e-48\\
79	1.61787257777467e-59\\
80	4.95458633867927e-73\\
};
\addlegendentry{\normalsize{$r_x=0 \;m$ (ideal)}}
\addplot [line width=0.6mm,color=mycolor2, dashdotted]
  table[row sep=crcr]{%
-20	1\\
-19	1\\
-18	1\\
-17	1\\
-16	1\\
-15	1\\
-14	1\\
-13	1\\
-12	1\\
-11	1\\
-10	1\\
-9	1\\
-8	1\\
-7	1\\
-6	1\\
-5	1\\
-4	1\\
-3	1\\
-2	1\\
-1	1\\
0	1\\
1	1\\
2	1\\
3	1\\
4	1\\
5	1\\
6	1\\
7	1\\
8	1\\
9	1\\
10	1\\
11	1\\
12	1\\
13	1\\
14	1\\
15	1\\
16	1\\
17	1\\
18	1\\
19	1\\
20	1\\
21	1\\
22	1\\
23	0.9995\\
24	0.999\\
25	0.9985\\
26	0.9985\\
27	0.9985\\
28	0.9965\\
29	0.9935\\
30	0.993\\
31	0.9905\\
32	0.9895\\
33	0.989\\
34	0.988\\
35	0.9845\\
36	0.981\\
37	0.9765\\
38	0.9735\\
39	0.9695\\
40	0.966\\
41	0.9615\\
42	0.958\\
43	0.948\\
44	0.9345\\
45	0.9255\\
46	0.914\\
47	0.8935\\
48	0.8655\\
49	0.8375\\
50	0.799\\
51	0.749\\
52	0.69\\
53	0.624\\
54	0.5515\\
55	0.466\\
56	0.3725\\
57	0.28\\
58	0.186\\
59	0.1135\\
60	0.0605\\
61	0.0325\\
62	0.01\\
63	0.002\\
64	0.0005\\
65	0\\
66	0\\
67	0\\
68	0\\
69	0\\
70	0\\
71	0\\
72	0\\
73	0\\
74	0\\
75	0\\
76	0\\
77	0\\
78	0\\
79	0\\
80	0\\
};
\addlegendentry{\normalsize{$\eta=5 \;m$}}

\addplot [line width=0.6mm,color=mycolor3, dashdotted]
  table[row sep=crcr]{%
-20	1\\
-19	1\\
-18	1\\
-17	1\\
-16	1\\
-15	1\\
-14	1\\
-13	1\\
-12	1\\
-11	1\\
-10	1\\
-9	1\\
-8	1\\
-7	1\\
-6	1\\
-5	1\\
-4	1\\
-3	1\\
-2	1\\
-1	1\\
0	1\\
1	1\\
2	1\\
3	1\\
4	1\\
5	1\\
6	1\\
7	1\\
8	1\\
9	1\\
10	1\\
11	1\\
12	1\\
13	1\\
14	1\\
15	1\\
16	1\\
17	1\\
18	1\\
19	1\\
20	1\\
21	1\\
22	1\\
23	0.9995\\
24	0.9985\\
25	0.9985\\
26	0.9985\\
27	0.998\\
28	0.9965\\
29	0.9935\\
30	0.992\\
31	0.9905\\
32	0.9895\\
33	0.9885\\
34	0.987\\
35	0.983\\
36	0.9795\\
37	0.975\\
38	0.9715\\
39	0.9665\\
40	0.965\\
41	0.96\\
42	0.9565\\
43	0.9425\\
44	0.93\\
45	0.921\\
46	0.9065\\
47	0.882\\
48	0.8545\\
49	0.8215\\
50	0.779\\
51	0.7265\\
52	0.666\\
53	0.5995\\
54	0.515\\
55	0.427\\
56	0.341\\
57	0.247\\
58	0.1565\\
59	0.0955\\
60	0.055\\
61	0.0245\\
62	0.006\\
63	0.0015\\
64	0\\
65	0\\
66	0\\
67	0\\
68	0\\
69	0\\
70	0\\
71	0\\
72	0\\
73	0\\
74	0\\
75	0\\
76	0\\
77	0\\
78	0\\
79	0\\
80	0\\
};
\addlegendentry{\normalsize{$\eta=10 \;m$}}

\addplot [line width=0.6mm,color=mycolor4, dashdotted]
  table[row sep=crcr]{%
-20	1\\
-19	1\\
-18	1\\
-17	1\\
-16	1\\
-15	1\\
-14	1\\
-13	1\\
-12	1\\
-11	1\\
-10	1\\
-9	1\\
-8	1\\
-7	1\\
-6	1\\
-5	1\\
-4	1\\
-3	1\\
-2	1\\
-1	1\\
0	1\\
1	1\\
2	1\\
3	1\\
4	1\\
5	1\\
6	1\\
7	1\\
8	1\\
9	1\\
10	1\\
11	1\\
12	1\\
13	1\\
14	1\\
15	1\\
16	1\\
17	1\\
18	1\\
19	1\\
20	1\\
21	1\\
22	0.9995\\
23	0.9995\\
24	0.9995\\
25	0.999\\
26	0.999\\
27	0.9985\\
28	0.9985\\
29	0.9985\\
30	0.996\\
31	0.9955\\
32	0.9945\\
33	0.993\\
34	0.991\\
35	0.988\\
36	0.986\\
37	0.982\\
38	0.98\\
39	0.976\\
40	0.9695\\
41	0.964\\
42	0.955\\
43	0.946\\
44	0.932\\
45	0.9125\\
46	0.8955\\
47	0.872\\
48	0.842\\
49	0.8005\\
50	0.7555\\
51	0.7005\\
52	0.624\\
53	0.547\\
54	0.4735\\
55	0.3905\\
56	0.3035\\
57	0.224\\
58	0.149\\
59	0.083\\
60	0.033\\
61	0.0155\\
62	0.006\\
63	0.0025\\
64	0\\
65	0\\
66	0\\
67	0\\
68	0\\
69	0\\
70	0\\
71	0\\
72	0\\
73	0\\
74	0\\
75	0\\
76	0\\
77	0\\
78	0\\
79	0\\
80	0\\
};
\addlegendentry{\normalsize{$\eta=15 \;m$}}

\addplot [line width=0.6mm,color=mycolor5, dashdotted]
  table[row sep=crcr]{%
-20	1\\
-19	1\\
-18	1\\
-17	1\\
-16	1\\
-15	1\\
-14	1\\
-13	1\\
-12	1\\
-11	1\\
-10	1\\
-9	1\\
-8	1\\
-7	1\\
-6	1\\
-5	1\\
-4	1\\
-3	1\\
-2	1\\
-1	1\\
0	1\\
1	1\\
2	1\\
3	1\\
4	1\\
5	1\\
6	1\\
7	1\\
8	1\\
9	1\\
10	1\\
11	1\\
12	1\\
13	1\\
14	1\\
15	1\\
16	1\\
17	1\\
18	1\\
19	0.9995\\
20	0.9995\\
21	0.999\\
22	0.9985\\
23	0.9985\\
24	0.9975\\
25	0.9965\\
26	0.9965\\
27	0.996\\
28	0.995\\
29	0.9935\\
30	0.9935\\
31	0.99\\
32	0.9875\\
33	0.9845\\
34	0.983\\
35	0.9825\\
36	0.978\\
37	0.974\\
38	0.964\\
39	0.9565\\
40	0.9475\\
41	0.9415\\
42	0.9325\\
43	0.9185\\
44	0.9005\\
45	0.885\\
46	0.8605\\
47	0.833\\
48	0.8\\
49	0.7535\\
50	0.702\\
51	0.6525\\
52	0.589\\
53	0.5015\\
54	0.418\\
55	0.342\\
56	0.254\\
57	0.1685\\
58	0.1015\\
59	0.0605\\
60	0.0285\\
61	0.0175\\
62	0.0075\\
63	0.0015\\
64	0\\
65	0\\
66	0\\
67	0\\
68	0\\
69	0\\
70	0\\
71	0\\
72	0\\
73	0\\
74	0\\
75	0\\
76	0\\
77	0\\
78	0\\
79	0\\
80	0\\
};
\addlegendentry{\normalsize{$\eta=20 \; m$}}

\addplot [line width=0.6mm,color=mycolor6, dashdotted]
  table[row sep=crcr]{%
-20	0.9975\\
-19	0.9975\\
-18	0.9965\\
-17	0.9965\\
-16	0.996\\
-15	0.995\\
-14	0.993\\
-13	0.992\\
-12	0.9905\\
-11	0.9885\\
-10	0.9875\\
-9	0.9865\\
-8	0.9855\\
-7	0.983\\
-6	0.982\\
-5	0.979\\
-4	0.9745\\
-3	0.9715\\
-2	0.969\\
-1	0.966\\
0	0.962\\
1	0.9575\\
2	0.9485\\
3	0.9455\\
4	0.941\\
5	0.939\\
6	0.936\\
7	0.934\\
8	0.933\\
9	0.9305\\
10	0.9275\\
11	0.9265\\
12	0.926\\
13	0.9255\\
14	0.9245\\
15	0.9225\\
16	0.922\\
17	0.922\\
18	0.921\\
19	0.9205\\
20	0.919\\
21	0.9185\\
22	0.9155\\
23	0.9135\\
24	0.9095\\
25	0.9055\\
26	0.9035\\
27	0.8985\\
28	0.893\\
29	0.8895\\
30	0.8835\\
31	0.874\\
32	0.866\\
33	0.8545\\
34	0.839\\
35	0.8195\\
36	0.799\\
37	0.7745\\
38	0.745\\
39	0.711\\
40	0.6725\\
41	0.633\\
42	0.593\\
43	0.553\\
44	0.5055\\
45	0.468\\
46	0.4305\\
47	0.388\\
48	0.3425\\
49	0.299\\
50	0.264\\
51	0.2285\\
52	0.187\\
53	0.157\\
54	0.127\\
55	0.097\\
56	0.066\\
57	0.043\\
58	0.0205\\
59	0.0125\\
60	0.0035\\
61	0.0015\\
62	0.0005\\
63	0\\
64	0\\
65	0\\
66	0\\
67	0\\
68	0\\
69	0\\
70	0\\
71	0\\
72	0\\
73	0\\
74	0\\
75	0\\
76	0\\
77	0\\
78	0\\
79	0\\
80	0\\
};
\addlegendentry{\normalsize{$\eta=100 \;m$}}

\end{axis}
\end{tikzpicture}%
    \caption{Success probability of URDC-UL model  for the suburban area at different $\eta$ values.}
    \label{suc_sd}
\end{figure}

Fig. \ref{succ_mobility} illustrates the average probability of success for the URDC-UL and SUC-UL schemes in various environments. The close correspondence between analysis and numerical simulation of the exact network  demonstrates the validation of proposed  analytical models and particularly, the accuracy of Approximation 1. Moreover, the figure demonstrates that the URDC-UL model has an incredibly high probability of success even at very high SINR thresholds. This result can be exploited to increase the size of the packets $L$ or even to lower the assigned bandwidth $W$, or more importantly to lower the transmission power of IoT devices  while keeping a reliable communication link.
Furthermore, the figure demonstrates that the environment affects both the intended and interference links. At very low and very high  SINR thresholds, the interference link's influence is prominent, resulting in the maximum success probability for high-rise urban locations. However, for  most of the SINR thresholds range, the intended link is the dominant one, resulting in the highest probability of success for the suburban region.

The success probability in a suburban area is plotted in Fig. \ref{suc_sd} for various values of the SD, $\eta$, of $r_x$. By treating $r_x$ as a Gaussian RV, the system covers a more realistic and practical scenario in which $P_{LOS}$ can take values other than 1 depending on the $\eta$ value. The figure confirms that the URDC-UL model still offers ultra-reliable links at reasonable values for localization error, not only at the ideal LOS probability of 1 (UAV hovering above the device). For the remaining results, we use $\eta$ = 20 m, which represents a large localization error. However, the URDC-UL model still preserves the ultra-reliable transmission probability at higher thresholds.

\begin{figure}
     \centering
  \input{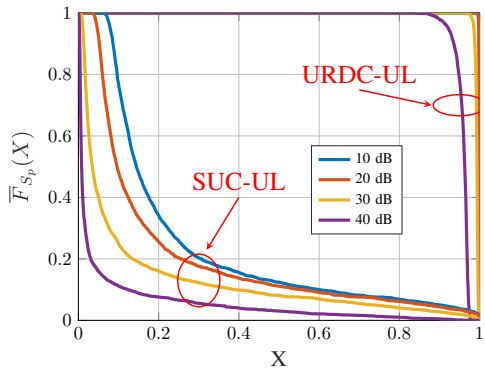}
    \caption{Meta distribution at $h=30$ m in a suburban area  for different SINR thresholds.}
    \label{meta_dist}
    \end{figure}
\begin{figure}
     \centering
     \begin{tikzpicture}[scale=0.7, transform shape,font=\Large]
  \input{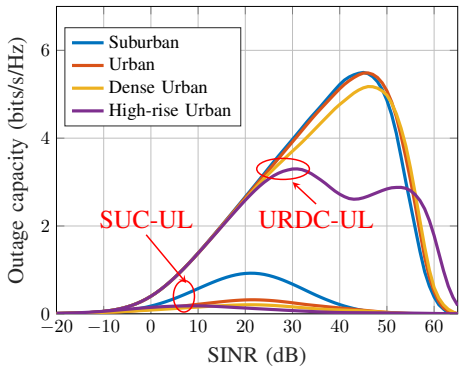}
\end{tikzpicture}
\caption{Average outage capacity for both models.}
\label{out cap_mo}
\end{figure}

Fig. \ref{meta_dist} plots the meta distribution for both models. The
results for the URDC-UL scheme show an ultra-reliable
and almost unified performance for all devices, while at a
high SINR threshold, i.e. 40 dB, there is a slight difference in
performance between devices. This result can be explained as
follow, the UAVs travel to the devices to offer high quality (i.e., short distance and high LOS probability) intended link for all devices. Moreover, the interference is relieved and randomized due to the unsynchronized movement of the UAVs, which results in low correlated interference at different transmission attempts. Based on the previous two reasons, the URDC-UL scheme guarantees high and unified success probability for all devices.  On
the other hand, the meta distribution for the SUC-UL model
confirms the location-dependent performance among devices,
where the devices near the cell center have better performance.
The results of the URDC-UL model highlight the device-centric advantage in terms of fairness. Consequently, all the following results for the URDC-UL scheme can be considered general for any device.

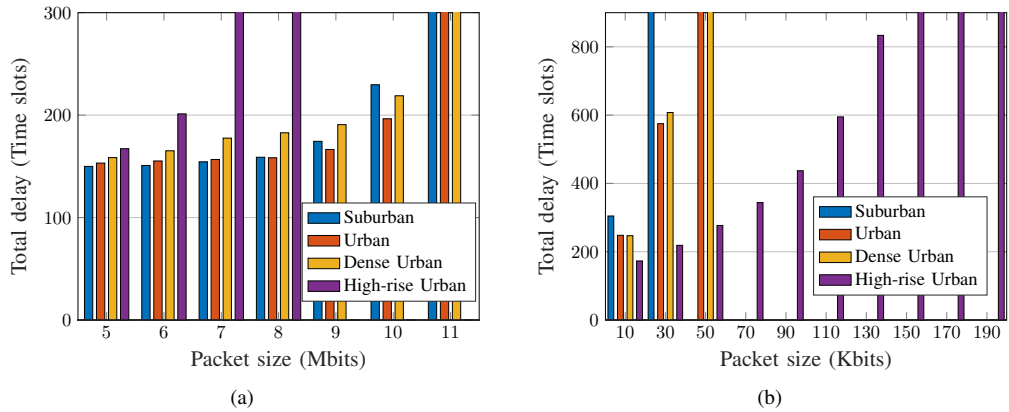
\begin{figure*}[ht!]
\centering
\subfloat[\label{delay_mob}]{%
%
%
\definecolor{mycolor1}{rgb}{0.00000,0.44700,0.74100}%
\definecolor{mycolor2}{rgb}{0.85000,0.32500,0.09800}%
\definecolor{mycolor3}{rgb}{0.92900,0.69400,0.12500}%
\definecolor{mycolor4}{rgb}{0.49400,0.18400,0.55600}%
\begin{tikzpicture}[scale=0.7, transform shape,font=\Large]

\begin{axis}[%
width=3in,
height=2.3in,
at={(0.758in,0.481in)},
scale only axis,
bar shift auto,
xmin=4.5,
xmax=11.5,
xtick={ 1,  2,  3,  4,  5,  6,  7,  8,  9, 10, 11, 12},
xlabel style={font=\large \color{white!15!black}},
xlabel={Packet size (Mbits)},
ymin=0,
ymax=300,
ticklabel style={font=\normalsize},
ylabel style={font=\large \color{white!15!black}},
ylabel={ Total delay (Time slots)},
axis background/.style={fill=white},
legend style={font=\footnotesize,at={(0.992,0.382)},legend cell align=left, align=left, draw=white!15!black},
ymajorgrids
]
\addplot[ybar, bar width=0.145, fill=mycolor1, draw=black, area legend] table[row sep=crcr] {%
1	149.499999678754\\
2	149.499999678754\\
3	149.499999678754\\
4	149.499999678754\\
5	149.949849197465\\
6	150.857719349258\\
7	154.442148507048\\
8	158.873538607634\\
9	174.445740687924\\
10	229.64669691901\\
11	633.474574869046\\
12	0\\
};
\addplot[forget plot, color=white!15!black] table[row sep=crcr] {%
0.509090909090909	0\\
12.4909090909091	0\\
};
\addlegendentry{\normalsize{Suburban}}

\addplot[ybar, bar width=0.145, fill=mycolor2, draw=black, area legend] table[row sep=crcr] {%
1	149.499999678754\\
2	149.499999678754\\
3	150.251255908838\\
4	151.010100880123\\
5	153.176229308912\\
6	155.244028782025\\
7	156.708595005241\\
8	158.368643902581\\
9	166.48106882175\\
10	196.452036484215\\
11	393.421051841646\\
12	0\\
};
\addplot[forget plot, color=white!15!black] table[row sep=crcr] {%
0.509090909090909	0\\
12.4909090909091	0\\
};
\addlegendentry{\normalsize{Urban}}

\addplot[ybar, bar width=0.145, fill=mycolor3, draw=black, area legend] table[row sep=crcr] {%
1	149.649649318702\\
2	150.100401244699\\
3	150.553877021539\\
4	153.019447094257\\
5	158.536585195564\\
6	165.193369987153\\
7	177.553443758722\\
8	182.762835856762\\
9	190.688775141724\\
10	218.887261745434\\
11	397.606381981568\\
12	0\\
};
\addplot[forget plot, color=white!15!black] table[row sep=crcr] {%
0.509090909090909	0\\
12.4909090909091	0\\
};
\addlegendentry{\normalsize{Dense Urban}}

\addplot[ybar, bar width=0.145, fill=mycolor4, draw=black, area legend] table[row sep=crcr] {%
1	149.949849197465\\
2	150.402414103206\\
3	152.551020231602\\
4	157.037814720512\\
5	167.225950534337\\
6	201.211305286956\\
7	328.571427830531\\
8	1494.99999668994\\
9	0\\
10	0\\
11	0\\
12	0\\
};
\addplot[forget plot, color=white!15!black] table[row sep=crcr] {%
0.509090909090909	0\\
12.4909090909091	0\\
};
\addlegendentry{\normalsize{High-rise Urban}}

\end{axis}
\end{tikzpicture}%
} 
 \qquad  \subfloat[\label{td_1_100}]{%
%
%
\definecolor{mycolor1}{rgb}{0.00000,0.44700,0.74100}%
\definecolor{mycolor2}{rgb}{0.85000,0.32500,0.09800}%
\definecolor{mycolor3}{rgb}{0.92900,0.69400,0.12500}%
\definecolor{mycolor4}{rgb}{0.49400,0.18400,0.55600}%
\begin{tikzpicture}[scale=0.7, transform shape,font=\Large]

\begin{axis}[%
width=3in,
height=2.3in,
at={(0.758in,0.481in)},
scale only axis,
bar shift auto,
xmin=0.18181818181818,
xmax=199.818181818182,
xtick={ 10,  30,  50,  70,  90, 110, 130, 150, 170, 190},
xlabel style={font=\large \color{white!15!black}},
xlabel={Packet size (Kbits)},
ticklabel style={font=\normalsize},
ymin=0,
ymax=900,
ylabel style={font=\large \color{white!15!black}},
ylabel={ Total delay (Time slots)},
axis background/.style={fill=white},
legend style={font=\footnotesize,at={(0.952,0.402)},legend cell align=left, align=left, draw=white!15!black},
ymajorgrids
]
\addplot[ybar, bar width=2.909, fill=mycolor1, draw=black, area legend] table[row sep=crcr] {%
10	304.65182853461\\
30	1011.39473494754\\
50	0\\
70	0\\
90	0\\
110	0\\
130	0\\
150	0\\
170	0\\
190	0\\
};
\addplot[forget plot, color=white!15!black] table[row sep=crcr] {%
0.18181818181818	0\\
199.818181818182	0\\
};
\addlegendentry{\normalsize{Suburban}}

\addplot[ybar, bar width=2.909, fill=mycolor2, draw=black, area legend] table[row sep=crcr] {%
10	248.064158657767\\
30	574.999998660108\\
50	1984.51327004945\\
70	0\\
90	0\\
110	0\\
130	0\\
150	0\\
170	0\\
190	0\\
};
\addplot[forget plot, color=white!15!black] table[row sep=crcr] {%
0.18181818181818	0\\
199.818181818182	0\\
};
\addlegendentry{\normalsize{Urban}}

\addplot[ybar, bar width=2.909, fill=mycolor3, draw=black, area legend] table[row sep=crcr] {%
10	246.971365064631\\
30	607.723575894044\\
50	3449.999991886\\
70	0\\
90	0\\
110	0\\
130	0\\
150	0\\
170	0\\
190	0\\
};
\addplot[forget plot, color=white!15!black] table[row sep=crcr] {%
0.18181818181818	0\\
199.818181818182	0\\
};
\addlegendentry{\normalsize{Dense Urban}}

\addplot[ybar, bar width=2.909, fill=mycolor4, draw=black, area legend] table[row sep=crcr] {%
10	173.032407189907\\
30	218.780487475072\\
50	276.851851324845\\
70	343.941717075223\\
90	437.134501970555\\
110	594.827585081338\\
130	833.643120698103\\
150	1218.74999730043\\
170	2177.18446472151\\
190	5901.3157753415\\
};
\addplot[forget plot, color=white!15!black] table[row sep=crcr] {%
0.18181818181818	0\\
199.818181818182	0\\
};
\addlegendentry{\normalsize{High-rise Urban}}

\end{axis}
\end{tikzpicture}%
}
\caption{Average packet delay for  (a) URDC-UL  model  and (b) SUC-UL model. Infinite delays are omitted from the figure.}
\end{figure*}

Fig. \ref{out cap_mo} illustrates the average outage capacity of both the URDC-UL   and  the SUC-UL models  in  various environments. The results are consistent with the success probability.\\
Fig.  \ref{delay_mob} illustrates the packet delay associated with the URDC-UL  model in various environments. The figure demonstrates how the URDC-UL  model enables the transmission of large packets (around 11 Mbits) with a finite delay. Additionally, it replicates the success probability results and demonstrates that the high-rise urban environment has the highest delay as higher success probability thresholds are used at these packet sizes in this model. In Fig. \ref{td_1_100}, the average packet delay for the SUC-UL model in various environments is illustrated. It replicates the success probability results and demonstrates that the high-rise urban region has the least delay provided that low success probability thresholds are used at these packet sizes.

Results in Fig.  \ref{H_suburban} illustrate the average outage probability at various elevations in the suburban region. The results clearly show the superiority of the URDC-UL  model at different heights. The findings  indicate that the outage probability improves with increasing height up to the ideal height with the minimum outage probability. However, as the height increases further, the probability of an outage rapidly increases. This can be explained by the fact that as the height increases, the likelihood of the intended device becoming LOS increases until the optimal height is reached. Any further increase results in a rise in the number of interferer devices with LOS links, which lowers the success probability. It is worth noting that these optimal heights will be lower if the cell size is decreased.

\begin{figure}
     \centering
    \input{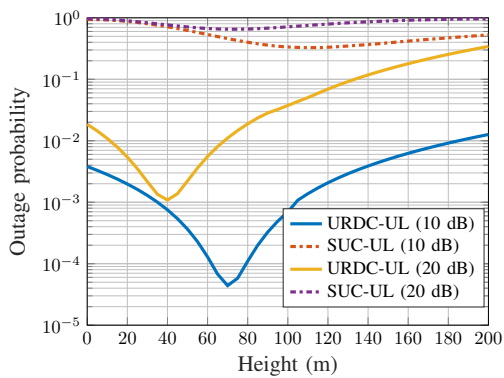}
    \caption{Effect of height on outage probability in a suburban area.}
    \label{H_suburban}
\end{figure}
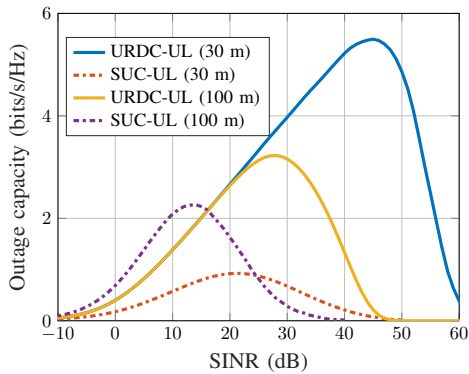
\begin{figure}
     \centering
%
%
\definecolor{mycolor1}{rgb}{0.00000,0.44700,0.74100}%
\definecolor{mycolor2}{rgb}{0.85000,0.32500,0.09800}%
\definecolor{mycolor3}{rgb}{0.92900,0.69400,0.12500}%
\definecolor{mycolor4}{rgb}{0.49400,0.18400,0.55600}%
\begin{tikzpicture}[scale=0.7, transform shape,font=\Large]

\begin{axis}[%
width=3in,
height=2.3in,
scale only axis,
xmin=-10,
xmax=60,
xlabel style={font=\large \color{white!15!black}},
xlabel={SINR (dB)},
ymin=0,
ymax=6,
ylabel style={font=\large \color{white!15!black}},
ylabel={Outage capacity (bits/s/Hz)},
ticklabel style={font=\normalsize},
axis background/.style={fill=white},
legend style={font=\footnotesize,at={(0.02,0.601)}, anchor=south west, legend cell align=left, align=left, draw=white!15!black},
grid=both
]
\addplot [line width=0.6mm,color=mycolor1]
  table[row sep=crcr]{%
-20	0.00574216179788647\\
-19	0.00721968774759149\\
-18	0.00907441977567453\\
-17	0.0114009447662056\\
-16	0.0143165921899733\\
-15	0.0179663519804947\\
-14	0.0225285469706735\\
-13	0.0282211968464762\\
-12	0.0353088827249508\\
-11	0.044109727753327\\
-10	0.0550018368153983\\
-9	0.0684281870446402\\
-8	0.0848985569250982\\
-7	0.104986698580606\\
-6	0.129320733913716\\
-5	0.158564896595157\\
-4	0.193391484179647\\
-3	0.234443392562029\\
-2	0.282289814319473\\
-1	0.337380152537483\\
0	0.400003109613457\\
1	0.470258310194114\\
2	0.548046129793046\\
3	0.633077866179741\\
4	0.724904116465877\\
5	0.822955694801287\\
6	0.926589696532435\\
7	1.03513381979683\\
8	1.14792383336406\\
9	1.26433161724538\\
10	1.38378345800025\\
11	1.50576954441483\\
12	1.62984687465144\\
13	1.75563704561773\\
14	1.88282211932899\\
15	2.01113806868972\\
16	2.14036647911\\
17	2.27032845039086\\
18	2.40087992558842\\
19	2.53190403253707\\
20	2.66292715770661\\
21	2.79401282523269\\
22	2.92548447798503\\
23	3.0575460529513\\
24	3.19067572222397\\
25	3.32542100706576\\
26	3.4599399754367\\
27	3.59082424934946\\
28	3.71815806181917\\
29	3.84533830865744\\
30	3.97202451143386\\
31	4.1005834498124\\
32	4.23179841547513\\
33	4.36350521162758\\
34	4.48797506899986\\
35	4.60690314634468\\
36	4.72779497194992\\
37	4.8534853195918\\
38	4.97894887827489\\
39	5.10246603127854\\
40	5.22177040856831\\
41	5.32251361263259\\
42	5.39328119905528\\
43	5.44323160499949\\
44	5.4777324871908\\
45	5.49674294909842\\
46	5.46875175517876\\
47	5.39963926522603\\
48	5.28729508196174\\
49	5.11127803583795\\
50	4.8621355266287\\
51	4.53870358065822\\
52	4.10906986474082\\
53	3.57457793017023\\
54	3.0119025848017\\
55	2.4854619307564\\
56	1.94860692373664\\
57	1.4232626071664\\
58	0.958744214797088\\
59	0.630633523012359\\
60	0.384613924756794\\
61	0.20604238817654\\
62	0.106747879356907\\
63	0.0153181134628144\\
64	-0.0659170756453951\\
65	-0.142034620441757\\
66	-0.216490768185799\\
67	-0.290833204779718\\
68	-0.365843913312727\\
69	-0.441957589803616\\
70	-0.519434182212467\\
71	-0.598438273184233\\
72	-0.679078948287929\\
73	-0.761431274112739\\
74	-0.845548550050167\\
75	-0.931469636944764\\
76	-1.0192235200801\\
77	-1.10883224925129\\
78	-1.20031289016463\\
79	-1.29367885363672\\
80	-1.38894082190285\\
};
\addlegendentry{\normalsize{URDC-UL (30 m)}}

\addplot [line width=0.6mm,color=mycolor2, dashdotted]
  table[row sep=crcr]{%
-10	0.0353317159116749\\
-9	0.0412307140875739\\
-8	0.0481215915528547\\
-7	0.0566123805427622\\
-6	0.0666892171986161\\
-5	0.078697521299028\\
-4	0.0929196146369301\\
-3	0.109533946784857\\
-2	0.128783550774752\\
-1	0.15092368342931\\
0	0.176136904347826\\
1	0.204424087083749\\
2	0.235797581783043\\
3	0.27012898251351\\
4	0.307326473289323\\
5	0.347184730530702\\
6	0.389549308666588\\
7	0.433839583817054\\
8	0.479839247827207\\
9	0.526480686783933\\
10	0.573909424168007\\
11	0.621045374514371\\
12	0.667564297697593\\
13	0.712326951642806\\
14	0.755317850440461\\
15	0.794986106733636\\
16	0.830852864222096\\
17	0.861693022129533\\
18	0.887469967458758\\
19	0.906904468731553\\
20	0.919347931914435\\
21	0.92466455537083\\
22	0.922927595244462\\
23	0.913345717592468\\
24	0.896756383889646\\
25	0.872940219736711\\
26	0.842691774249165\\
27	0.807445313148815\\
28	0.767085188384812\\
29	0.722210346899775\\
30	0.675322086607836\\
31	0.626418151263533\\
32	0.575800158986652\\
33	0.523658149321982\\
34	0.472830763237154\\
35	0.42293815871611\\
36	0.375886062001694\\
37	0.330323492856483\\
38	0.28819428165583\\
39	0.247303030655955\\
40	0.210474308387079\\
41	0.176260268523587\\
42	0.146392136836048\\
43	0.118770821929245\\
44	0.095040336968675\\
45	0.0750399849046922\\
46	0.0580848034565911\\
47	0.042905437349155\\
48	0.0308732621284078\\
49	0.0209609724199128\\
50	0.0136220353518773\\
};
\addlegendentry{\normalsize{SUC-UL (30 m)}}

\addplot [line width=0.6mm,color=mycolor3]
  table[row sep=crcr]{%
-20	0.00574216180956564\\
-19	0.00721968776518049\\
-18	0.00907441980242416\\
-17	0.011400944806006\\
-16	0.014316592250386\\
-15	0.0179663520716336\\
-14	0.022528547107585\\
-13	0.0282211970522702\\
-12	0.035308883036562\\
-11	0.0441097282187118\\
-10	0.0550018375113578\\
-9	0.0684281880743216\\
-8	0.0848985584633047\\
-7	0.104986699002321\\
-6	0.129320737200336\\
-5	0.158564901253057\\
-4	0.193391503506049\\
-3	0.234443401254529\\
-2	0.282289584693229\\
-1	0.337380162809308\\
0	0.400003112443223\\
1	0.470258286252972\\
2	0.548046039673411\\
3	0.633076825818821\\
4	0.724903453143145\\
5	0.8229541267765\\
6	0.926585927629568\\
7	1.03512641913407\\
8	1.14791020268748\\
9	1.26429742974\\
10	1.38371346783447\\
11	1.5056261008126\\
12	1.62958239227669\\
13	1.75513332875746\\
14	1.8818684049036\\
15	2.00940292530447\\
16	2.13726797955912\\
17	2.26493049446949\\
18	2.39164890316841\\
19	2.51660060537072\\
20	2.63853292746646\\
21	2.75585548372156\\
22	2.86654572452853\\
23	2.96810374855891\\
24	3.0575644101043\\
25	3.13143558902168\\
26	3.18661469155044\\
27	3.21924542339418\\
28	3.22659097144224\\
29	3.20669751752783\\
30	3.15806514735682\\
31	3.08301578219732\\
32	2.98050654302422\\
33	2.85249880063478\\
34	2.70011452805827\\
35	2.52408783222679\\
36	2.32496856133138\\
37	2.10341597421802\\
38	1.86116192410617\\
39	1.60188618414045\\
40	1.33169614679129\\
41	1.0600626216701\\
42	0.7989240718784\\
43	0.562425520034647\\
44	0.364214239410145\\
45	0.21279009682095\\
46	0.1099176124533\\
47	0.0487915517789884\\
48	0.0180479765883758\\
49	0.00540523201824084\\
50	0.00124102264756252\\
51	0.000209414394578049\\
52	2.38880426280323e-05\\
53	1.7133756707432e-06\\
54	6.93112615864282e-08\\
55	1.39027002457471e-09\\
56	1.1800950875674e-11\\
57	3.53098405095824e-14\\
58	2.86364587533776e-17\\
59	4.72922653431556e-21\\
60	1.10348648780015e-25\\
61	2.31690170876478e-31\\
62	2.50750811959608e-38\\
63	7.02640085928787e-47\\
64	2.17616971517249e-57\\
65	2.46131079806055e-70\\
66	3.24967996188208e-86\\
67	7.83197038607989e-106\\
68	6.22913683595723e-130\\
69	9.66841746869422e-160\\
70	2.05606632572306e-196\\
71	7.9694172985633e-242\\
72	9.03502251598939e-298\\
73	0\\
74	0\\
75	0\\
76	0\\
77	0\\
78	0\\
79	0\\
80	0\\
};
\addlegendentry{\normalsize{URDC-UL (100 m)}}

\addplot [line width=0.6mm,color=mycolor4, dashdotted]
  table[row sep=crcr]{%
-20	0.0101153136433626\\
-19	0.0126487946724696\\
-18	0.0158292948873312\\
-17	0.0198283691005133\\
-16	0.0248362311894083\\
-15	0.0311174798902928\\
-14	0.0389830948302595\\
-13	0.0487997144724728\\
-12	0.0610273987866238\\
-11	0.0761680861331865\\
-10	0.0949764339245551\\
-9	0.118078762168178\\
-8	0.146431893270212\\
-7	0.180995662044349\\
-6	0.222947212929674\\
-5	0.273300332072297\\
-4	0.333326971834646\\
-3	0.403989714420249\\
-2	0.486325112269044\\
-1	0.580559253764178\\
0	0.688\\
1	0.807333189775957\\
2	0.939124544834116\\
3	1.07850306393093\\
4	1.23029769435017\\
5	1.38222561647039\\
6	1.54072133419302\\
7	1.69180952485991\\
8	1.83344965858351\\
9	1.97133050310607\\
10	2.08396160706711\\
11	2.16768885231796\\
12	2.22961304054025\\
13	2.25562518450722\\
14	2.26765408177224\\
15	2.22912881005669\\
16	2.16860308779387\\
17	2.04600513900262\\
18	1.91876937201855\\
19	1.78548529055937\\
20	1.62779954330316\\
21	1.47239830007119\\
22	1.30892148642641\\
23	1.14286439196257\\
24	0.958679677255798\\
25	0.820301523812528\\
26	0.671895574569992\\
27	0.552680223068571\\
28	0.442111033894601\\
29	0.363832952397425\\
30	0.280677091448822\\
31	0.224108908051422\\
32	0.173499223896495\\
33	0.149097964055702\\
34	0.106626024892507\\
35	0.081855519540721\\
36	0.059318145284102\\
37	0.0353992947202783\\
38	0.0262569952199555\\
39	0.0228020340823861\\
40	0.0127563423761669\\
};
\addlegendentry{\normalsize{SUC-UL (100 m)}}

\end{axis}
\end{tikzpicture}%
    \caption{Outage capacity of the URDC-UL and SUC-UL schemes for a suburban area.}
     \label{comp_out_suburban}
\end{figure}

Following validation of the mathematical analysis, the two models will be compared in terms of success probability, outage capacity, packet delay, and energy efficiency. Results are displayed for suburban regions only for brevity; however, other environments will exhibit comparable trends. The performance characteristics are plotted at $h=30$ and $h=100$ meters.

Fig.  \ref{comp_out_suburban} compares the URDC-UL model's outage capacity with the SUC-UL model's average outage capacity.
Despite the fact that transmission happens exclusively during the $t_{\text{DC}}$ duration  in the URDC-UL model and throughout the time slot in the SUC-UL model, the results reveal that the URDC-UL model's outage capacity is always greater than the SUC-UL model's. This demonstrates that the URDC-UL model's higher probability of success is the decisive factor.
Additionally, the figure demonstrates that as the height approaches the SUC-UL model's optimal value, the outage capacity improves for the SUC-UL model while it drops for the URDC-UL model. As a result, the advantage of the URDC-UL model over the SUC-UL model decreases around the SUC-UL model's ideal height ($h=100$ m).

\begin{figure*}[ht!]
\centering
 \subfloat[\label{comp_succ_suburban_100}]{%
%
%
\definecolor{mycolor1}{rgb}{0.00000,0.44700,0.74100}%
\definecolor{mycolor2}{rgb}{0.96000,0.9200,0.9200}%
\definecolor{mycolor3}{rgb}{0.85000,0.32500,0.09800}%
\begin{tikzpicture}[scale=0.7, transform shape,font=\Large]

\begin{axis}[%
width=3in,
height=2.3in,
at={(1.133in,0.407in)},
scale only axis,
xmin=-20,
xmax=50,
xlabel style={font=\Large \color{white!15!black}},
xlabel={SINR Threshold (dB)},
ymin=0,
ymax=1,
ylabel style={font=\large \color{white!15!black}},
ylabel={Succcess probabilty},
axis background/.style={fill=white},
ticklabel style={font=\normalsize},
legend style={font=\footnotesize,at={(1.25in,0.507in)},legend cell align=left, align=left, draw=white!15!black},
grid=both
]

\addplot[area legend, draw=mycolor2, fill=mycolor2, forget plot]
table[row sep=crcr] {%
x	y\\
-20	1.00951275050032\\
-19	1.00707970374461\\
-18	1.0053432095142\\
-17	1.00452793408882\\
-16	1.00382114310986\\
-15	1.00362128597331\\
-14	1.00378418978947\\
-13	1.00400092406038\\
-12	1.00417489220608\\
-11	1.00369745373438\\
-10	1.00401995609393\\
-9	1.00357238323121\\
-8	1.00322518958306\\
-7	1.00285502523832\\
-6	1.00292560879879\\
-5	1.00278955108346\\
-4	1.00291668089446\\
-3	1.00290510548713\\
-2	1.00302917758068\\
-1	1.00249942019836\\
0	1.00311927003908\\
1	1.00282545008208\\
2	1.00329139476115\\
3	1.00085313746412\\
4	1.00177877637344\\
5	0.997844560635114\\
6	0.996545624342829\\
7	0.990707778590968\\
8	0.98185517494796\\
9	0.9745524726751\\
10	0.960363889140488\\
11	0.938023456116432\\
12	0.913586952392349\\
13	0.881362392910687\\
14	0.849806624054142\\
15	0.805666196810754\\
16	0.760104993221812\\
17	0.698307141770576\\
18	0.639980932393389\\
19	0.583004982074757\\
20	0.522366877383331\\
21	0.464901282359948\\
22	0.407909181855189\\
23	0.351913303569992\\
24	0.292055929863658\\
25	0.248610728295738\\
26	0.202593927598643\\
27	0.165359870218427\\
28	0.131653005597479\\
29	0.108441189102453\\
30	0.0836581673153358\\
31	0.0668173897813687\\
32	0.0516504555462955\\
33	0.0445287204844822\\
34	0.0315155683022842\\
35	0.0241919083876779\\
36	0.017586091277225\\
37	0.0104298782801184\\
38	0.00789327382883395\\
39	0.00691694472831294\\
40	0.00367454152587384\\
40	-0.00127454152587384\\
39	-0.00251694472831294\\
38	-0.00269327382883395\\
37	-0.00322987828011838\\
36	-0.005186091277225\\
35	-0.0065919083876779\\
34	-0.0079155683022842\\
33	-0.0105287204844822\\
32	-0.0108504555462955\\
31	-0.0124173897813687\\
30	-0.0132581673153358\\
29	-0.0140411891024529\\
28	-0.012853005597479\\
27	-0.0113598702184266\\
26	-0.008193927598643\\
25	-0.00181072829573801\\
24	0.008344070136342\\
23	0.021686696430008\\
22	0.039290818144811\\
21	0.061898717640052\\
20	0.088833122616669\\
19	0.122195017925243\\
18	0.159219067606611\\
17	0.202892858229424\\
16	0.253095006778188\\
15	0.302733803189246\\
14	0.354593375945858\\
13	0.403437607089313\\
12	0.454413047607651\\
11	0.501576543883568\\
10	0.545636110859512\\
9	0.5846475273249\\
8	0.61534482505204\\
7	0.643692221409032\\
6	0.666254375657171\\
5	0.681755439364886\\
4	0.695421223626563\\
3	0.702746862535883\\
2	0.710308605238846\\
1	0.713974549917916\\
0	0.716880729960918\\
-1	0.718300579801638\\
-2	0.719770822419322\\
-3	0.720294894512871\\
-4	0.720683319105542\\
-5	0.720810448916541\\
-6	0.721074391201212\\
-7	0.721144974761683\\
-8	0.721574810416945\\
-9	0.722027616768786\\
-10	0.722780043906066\\
-11	0.723102546265624\\
-12	0.724225107793915\\
-13	0.725199075939619\\
-14	0.726615810210533\\
-15	0.72837871402669\\
-16	0.730978856890135\\
-17	0.734672065911177\\
-18	0.739056790485797\\
-19	0.744920296255391\\
-20	0.752087249499679\\
}--cycle;
\addplot [color=mycolor3, line width=2.0pt, dashdotted]
  table[row sep=crcr]{%
-20	0.8808\\
-19	0.876\\
-18	0.8722\\
-17	0.8696\\
-16	0.8674\\
-15	0.866\\
-14	0.8652\\
-13	0.8646\\
-12	0.8642\\
-11	0.8634\\
-10	0.8634\\
-9	0.8628\\
-8	0.8624\\
-7	0.862\\
-6	0.862\\
-5	0.8618\\
-4	0.8618\\
-3	0.8616\\
-2	0.8614\\
-1	0.8604\\
0	0.86\\
1	0.8584\\
2	0.8568\\
3	0.8518\\
4	0.8486\\
5	0.8398\\
6	0.8314\\
7	0.8172\\
8	0.7986\\
9	0.7796\\
10	0.753\\
11	0.7198\\
12	0.684\\
13	0.6424\\
14	0.6022\\
15	0.5542\\
16	0.5066\\
17	0.4506\\
18	0.3996\\
19	0.3526\\
20	0.3056\\
21	0.2634\\
22	0.2236\\
23	0.1868\\
24	0.1502\\
25	0.1234\\
26	0.0972\\
27	0.077\\
28	0.0594\\
29	0.0472\\
30	0.0352\\
31	0.0272\\
32	0.0204\\
33	0.017\\
34	0.0118\\
35	0.0088\\
36	0.0062\\
37	0.0036\\
38	0.0026\\
39	0.0022\\
40	0.0012\\
};
\addlegendentry{\normalsize{SUC-UL}}

\addplot [color=mycolor1, line width=2.0pt]
  table[row sep=crcr]{%
-20	1.00000000217576\\
-19	1.00000000261165\\
-18	1.00000000315749\\
-17	1.00000000374301\\
-16	1.00000000452263\\
-15	1.00000000543693\\
-14	1.00000000651541\\
-13	1.00000000781909\\
-12	1.00000000945848\\
-11	1.00000001131161\\
-10	1.00000001356828\\
-9	1.00000001614723\\
-8	1.00000001944005\\
-7	1.00000000560656\\
-6	1.00000002732635\\
-5	1.00000003168709\\
-4	1.00000010271721\\
-3	1.00000004041034\\
-2	0.999999190551534\\
-1	1.00000003524917\\
0	1.00000001285061\\
1	0.999999956017464\\
2	0.999999843886771\\
3	0.999998366679046\\
4	0.999999097018469\\
5	0.999998117094177\\
6	0.999995961370872\\
7	0.99999287878145\\
8	0.999988150538534\\
9	0.999972980662411\\
10	0.99994946526845\\
11	0.999904804072893\\
12	0.999837799071993\\
13	0.99971298380076\\
14	0.999493140639317\\
15	0.999136923294265\\
16	0.998552066629047\\
17	0.997621885990937\\
18	0.996154609881319\\
19	0.993955520115098\\
20	0.99069864244249\\
21	0.985992303173001\\
22	0.979363012297815\\
23	0.970259423807458\\
24	0.958073103619937\\
25	0.942131981199371\\
26	0.921978216938739\\
27	0.897010465218632\\
28	0.867013035917933\\
29	0.832002412457709\\
30	0.792106182019545\\
31	0.748362736485552\\
32	0.700889142760226\\
33	0.650473039236677\\
34	0.597623403066055\\
35	0.542707455801182\\
36	0.4860129403632\\
37	0.427818618643176\\
38	0.368586227911557\\
39	0.309105952222295\\
40	0.250545636856152\\
41	0.194576513814832\\
42	0.143152781318013\\
43	0.098433011430622\\
44	0.0622943911369408\\
45	0.0355863937964234\\
46	0.0179826972250217\\
47	0.00781254225764762\\
48	0.00282965224948505\\
49	0.000830164548751839\\
50	0.000186790907822512\\
51	3.09017068351885e-05\\
52	3.45719121064131e-06\\
53	2.43289269931123e-07\\
54	9.65953704297922e-09\\
55	1.90231647216267e-10\\
56	1.58589817091925e-12\\
57	4.6619460302401e-15\\
58	3.71567459746755e-18\\
59	6.03232140713994e-22\\
60	1.38408303389339e-26\\
61	2.85840759802375e-32\\
62	3.04366637620979e-39\\
63	8.39341633399966e-48\\
64	2.5589345833547e-58\\
65	2.8497023366754e-71\\
66	3.70546791446184e-87\\
67	8.79716293327647e-107\\
68	6.89390606304672e-131\\
69	1.05451486826307e-160\\
70	2.21047437588495e-197\\
71	8.44723594017138e-243\\
72	9.44372100840134e-299\\
73	0\\
74	0\\
75	0\\
76	0\\
77	0\\
78	0\\
79	0\\
80	0\\
};

\addlegendentry{\normalsize{URDC-UL}}

\end{axis}
\end{tikzpicture}%
} 
 \qquad \subfloat[\label{comp_succ_suburban_30}]{%
%
%
\definecolor{mycolor1}{rgb}{0.00000,0.44700,0.74100}%
\definecolor{mycolor2}{rgb}{0.96000,0.9200,0.9200}%
\definecolor{mycolor3}{rgb}{0.85000,0.32500,0.09800}%
\begin{tikzpicture}[scale=0.7, transform shape,font=\Large]

\begin{axis}[%
width=3in,
height=2.3in,
at={(1.133in,0.413in)},
scale only axis,
xmin=-20,
xmax=63,
xlabel style={font=\large \color{white!15!black}},
xlabel={SINR Threshold (dB)},
ymin=0,
ymax=1,
ylabel style={font=\large \color{white!15!black}},
ylabel={Success probability},
axis background/.style={fill=white},
ticklabel style={font=\normalsize},
legend style={font=\footnotesize,at={(1.25in,2.45in)},legend cell align=left, align=left, draw=white!15!black},
grid=both
]
\addplot [color=mycolor1, line width=2.0pt]
  table[row sep=crcr]{%
-20	1.00000000021299\\
-19	1.00000000024226\\
-18	1.00000000029105\\
-17	1.00000000034977\\
-16	1.00000000042037\\
-15	1.00000000050626\\
-14	1.00000000060746\\
-13	1.00000000073082\\
-12	1.00000000087859\\
-11	1.00000000105707\\
-10	1.00000000128146\\
-9	1.0000000015253\\
-8	1.00000000183512\\
-7	1.00000000220515\\
-6	1.00000000264998\\
-5	1.00000000324018\\
-4	1.00000000383838\\
-3	1.00000000461329\\
-2	1.00000000546183\\
-1	1.00000000669474\\
0	1.0000000080039\\
1	1.00000000965598\\
2	1.00000001159594\\
3	1.00000001360123\\
4	1.00000001589201\\
5	1.00000000518357\\
6	1.00000002426979\\
7	1.00000002897538\\
8	1.00000003457799\\
9	1.00000004156821\\
10	1.00000004948937\\
11	0.999999545365303\\
12	1.00000006868962\\
13	1.00000007906218\\
14	1.00000008854965\\
15	1.00000009275921\\
16	0.999998937717529\\
17	0.999998331932017\\
18	0.999998301732323\\
19	0.999999573971716\\
20	0.999999355246087\\
21	0.999998522034183\\
22	0.999994493211755\\
23	0.999994878871364\\
24	0.999980460208659\\
25	0.999974952723959\\
26	0.999960784863582\\
27	0.999916906704934\\
28	0.99987616513883\\
29	0.999791990160108\\
30	0.999647658766066\\
31	0.99943863714122\\
32	0.999097829528976\\
33	0.998500815866842\\
34	0.997697767356397\\
35	0.99655575933568\\
36	0.994806907447847\\
37	0.992372638394558\\
38	0.98894135769786\\
39	0.984462748363689\\
40	0.978361700040204\\
41	0.970281538223165\\
42	0.95999407671599\\
43	0.947106939083549\\
44	0.930912854414167\\
45	0.910747729399855\\
46	0.885749821012839\\
47	0.855283971520109\\
48	0.818743293707885\\
49	0.773798580090494\\
50	0.72018400629417\\
51	0.656831479995576\\
52	0.584385897078134\\
53	0.503562137814559\\
54	0.416407954211362\\
55	0.3268209522594\\
56	0.240459036200795\\
57	0.16309364306815\\
58	0.100194729890616\\
59	0.0543737091531918\\
60	0.0254392565850384\\
61	0.00993506715680106\\
62	0.00314814090490899\\
63	0.000766265422268544\\
64	0.000136811537640352\\
65	1.67644169848042e-05\\
66	1.27506705641274e-06\\
67	5.49866978194225e-08\\
68	1.18475040549778e-09\\
69	1.06946124345788e-11\\
70	3.34366629735354e-14\\
71	2.83791136819799e-17\\
72	4.84150239335332e-21\\
73	1.12079310087905e-25\\
74	2.37855533572626e-31\\
75	2.45977727980583e-38\\
76	6.29465558778307e-47\\
77	1.67746989025874e-57\\
78	1.59527191468083e-70\\
79	1.43775065970029e-86\\
80	2.19390756609044e-106\\
};
\addlegendentry{\normalsize{URDC-UL}}

\addplot[area legend, draw=mycolor2, fill=mycolor2, forget plot]
table[row sep=crcr] {%
x	y\\
-20	0.794609445493415\\
-19	0.764110292883174\\
-18	0.735063051603354\\
-17	0.708021869271142\\
-16	0.682950269650176\\
-15	0.658864064080882\\
-14	0.636651569433461\\
-13	0.614815927649458\\
-12	0.594270394146801\\
-11	0.575419686789486\\
-10	0.556396598514203\\
-9	0.538410880316577\\
-8	0.521551984804311\\
-7	0.510171074827283\\
-6	0.500512743061555\\
-5	0.49212508485956\\
-4	0.485795713679115\\
-3	0.480049693234653\\
-2	0.475022111833665\\
-1	0.470595454022284\\
0	0.467000587130703\\
1	0.463857984517022\\
2	0.460666872818301\\
3	0.458011445862087\\
4	0.455662291753513\\
5	0.453828468237675\\
6	0.452369573214598\\
7	0.450793559167986\\
8	0.449982439313245\\
9	0.448097548861467\\
10	0.446106811073183\\
11	0.444507888304296\\
12	0.442458232749\\
13	0.439886579046292\\
14	0.437282502885523\\
15	0.433696105705719\\
16	0.429123655024265\\
17	0.423420993239678\\
18	0.417082417334308\\
19	0.410046272037785\\
20	0.401672760944724\\
21	0.391384112412562\\
22	0.380477810870697\\
23	0.368454170344689\\
24	0.355615255776284\\
25	0.341062140191832\\
26	0.326527729109084\\
27	0.311186246466275\\
28	0.294882779027775\\
29	0.27760349719594\\
30	0.261065983569247\\
31	0.242572002625156\\
32	0.224643279797701\\
33	0.206150777264352\\
34	0.188398287674033\\
35	0.171134031024665\\
36	0.154039364492514\\
37	0.137183248761475\\
38	0.121557245722371\\
39	0.105814262302451\\
40	0.0911481148145769\\
41	0.0799095958177032\\
42	0.0652903542356924\\
43	0.0530742813905317\\
44	0.0426286817906609\\
45	0.0333335666295503\\
46	0.0255992604841758\\
47	0.0188413223229551\\
48	0.0130936350783247\\
49	0.00883078873016138\\
50	0.00555575920705136\\
50	-0.00350544036647166\\
49	-0.0056114650586638\\
48	-0.00825313266286573\\
47	-0.0119712160427619\\
46	-0.0160964199044656\\
45	-0.020784001412159\\
44	-0.0263730682640909\\
43	-0.0322874214871501\\
42	-0.0390593204192672\\
41	-0.0475564363974134\\
40	-0.0515491196454949\\
39	-0.0580933734135625\\
38	-0.0644825404083616\\
37	-0.0699974709836971\\
36	-0.075463287197828\\
35	-0.0801968329570326\\
34	-0.0837446065146129\\
33	-0.0867368255735304\\
32	-0.0892383715851408\\
31	-0.0905158093884415\\
30	-0.0916803217334986\\
29	-0.0902189947804816\\
28	-0.088758721056761\\
27	-0.086197956611203\\
26	-0.08271114939894\\
25	-0.078425000095214\\
24	-0.074618772684496\\
23	-0.069883349088651\\
22	-0.065154680435915\\
21	-0.06055427666377\\
20	-0.056479485582406\\
19	-0.051852880733437\\
18	-0.0474361178174\\
17	-0.043872703384606\\
16	-0.040938225072575\\
15	-0.038401497010067\\
14	-0.036116899020789\\
13	-0.034146443780592\\
12	-0.032867894584748\\
11	-0.032060815840528\\
10	-0.031364241024873\\
9	-0.031684022291419\\
8	-0.031973009361555\\
7	-0.03167580071388\\
6	-0.031954423456144\\
5	-0.031949820894679\\
4	-0.031704340062691\\
3	-0.031316566635033\\
2	-0.030412186827963\\
1	-0.029148651183688\\
0	-0.026658326261137\\
-1	-0.023251840495714\\
-2	-0.018808160142843\\
-3	-0.012837557968953\\
-4	-0.005317800635637\\
-5	0.00418989098585001\\
-6	0.015179778677575\\
-7	0.029066954158225\\
-8	0.045265251910665\\
-9	0.064133717230803\\
-10	0.085981773838083\\
-11	0.110898874340206\\
-12	0.140094769732797\\
-13	0.171702252952548\\
-14	0.206126041863455\\
-15	0.244279391883444\\
-16	0.28466544495406\\
-17	0.328172517945506\\
-18	0.373816422198206\\
-19	0.4215606814758\\
-20	0.471959443395473\\
}--cycle;
\addplot [color=mycolor3, line width=2.0pt, dashdotted]
  table[row sep=crcr]{%
-20	0.633284444444444\\
-19	0.592835487179487\\
-18	0.55443973690078\\
-17	0.518097193608324\\
-16	0.483807857302118\\
-15	0.451571727982163\\
-14	0.421388805648458\\
-13	0.393259090301003\\
-12	0.367182581939799\\
-11	0.343159280564846\\
-10	0.321189186176143\\
-9	0.30127229877369\\
-8	0.283408618357488\\
-7	0.269619014492754\\
-6	0.257846260869565\\
-5	0.248157487922705\\
-4	0.240238956521739\\
-3	0.23360606763285\\
-2	0.228106975845411\\
-1	0.223671806763285\\
0	0.220171130434783\\
1	0.217354666666667\\
2	0.215127342995169\\
3	0.213347439613527\\
4	0.211978975845411\\
5	0.210939323671498\\
6	0.210207574879227\\
7	0.209558879227053\\
8	0.209004714975845\\
9	0.208206763285024\\
10	0.207371285024155\\
11	0.206223536231884\\
12	0.204795169082126\\
13	0.20287006763285\\
14	0.200582801932367\\
15	0.197647304347826\\
16	0.194092714975845\\
17	0.189774144927536\\
18	0.184823149758454\\
19	0.179096695652174\\
20	0.172596637681159\\
21	0.165414917874396\\
22	0.157661565217391\\
23	0.149285410628019\\
24	0.140498241545894\\
25	0.131318570048309\\
26	0.121908289855072\\
27	0.112494144927536\\
28	0.103062028985507\\
29	0.0936922512077294\\
30	0.0846928309178744\\
31	0.0760280966183575\\
32	0.0677024541062802\\
33	0.0597069758454106\\
34	0.0523268405797101\\
35	0.0454685990338164\\
36	0.039288038647343\\
37	0.0335928888888889\\
38	0.0285373526570048\\
39	0.0238604444444444\\
40	0.019799497584541\\
41	0.0161765797101449\\
42	0.0131155169082126\\
43	0.0103934299516908\\
44	0.00812780676328502\\
45	0.00627478260869565\\
46	0.00475142028985507\\
47	0.00343505314009662\\
48	0.00242025120772947\\
49	0.00160966183574879\\
50	0.00102515942028985\\
};
\addlegendentry{\normalsize{SUC-UL}}

\end{axis}
\end{tikzpicture}%
}
\caption{Success probability for suburban area at (a) $h= 100$ m  and (b) $h= 30$ m. The dashed  line represents the mean while  the shaded area represents the  SD of the SUC-UL.}
\end{figure*}
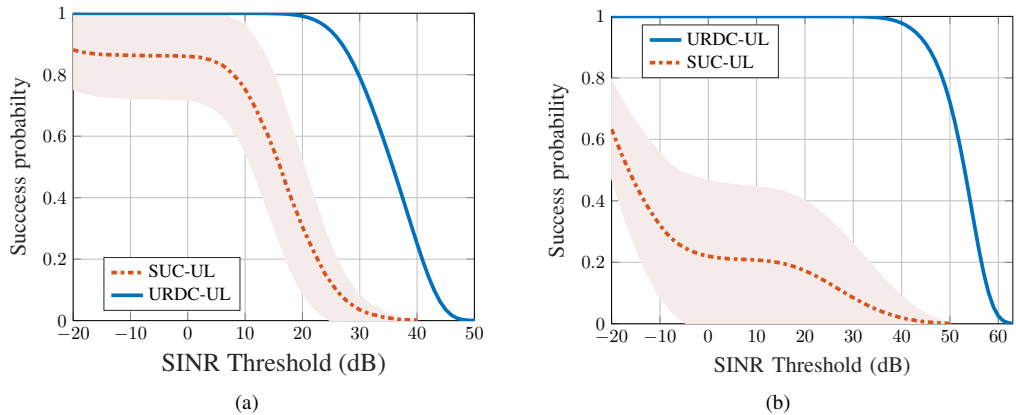

Fig.  \ref{comp_succ_suburban_100} compares the almost unified success probability of the URDC-UL model with the average success probability of the SUC-UL model  at its optimal height ($h= 100$ m), where the shaded region shows the SUC-UL model's SD.
The results indicate that the URDC-UL model always has a greater probability of success, whereas the SUC-UL model's performance is location-dependent. Even in the best-case scenario for the SUC-UL model, if a device is positioned in the cell's center right beneath the UAV, the URDC-UL model has a better success probability.
This can be explained by the fact that in the SUC-UL model, all cells always interfere throughout each time slot. However, due to the asynchronous system in the URDC-UL model, some UAVs travel while their selected devices wait and do not contribute to interference.
Fig. \ref{comp_succ_suburban_30} demonstrates that as the height is reduced to 30 m, the success probability for the SUC-UL  model declines while it improves for the URDC-UL model. Moreover, the shaded region increases, which indicates that around the optimal height ($h= 100$ m) the performance gap between devices is reduced.

\begin{figure*}[h]
\centering
%
%
\definecolor{mycolor1}{rgb}{0.00000,0.44700,0.74100}%
\definecolor{mycolor2}{rgb}{0.85000,0.32500,0.09800}%
\definecolor{mycolor3}{rgb}{0.92900,0.69400,0.12500}%
\definecolor{mycolor4}{rgb}{0.49400,0.18400,0.55600}%
\begin{tikzpicture}[scale=0.7, transform shape,font=\Large]

\begin{axis}[%
width=6in,
height=2in,
at={(0.758in,0.481in)},
scale only axis,
bar shift auto,
xmin=-0.45090909090909,
xmax=11.5309090909091,
xtick={0.04,    1,    2,    3,    4,    5,    6,    7,    8,    9,   10,   11},
xticklabel style={/pgf/number format/fixed},
xlabel style={font=\large \color{white!15!black}},
xlabel={Packet size (Mbits)},
ymin=0,
ymax=400,
ylabel style={font=\large \color{white!15!black}},
ylabel={ Total delay (Time slots)},
ticklabel style={font=\normalsize},
axis background/.style={fill=white},
legend style={font=\footnotesize,at={(0.05,0.577)}, anchor=south west, legend cell align=left, align=left, draw=white!15!black},
ymajorgrids
]
\addplot[ybar, bar width=0.145, fill=mycolor1, draw=black, area legend] table[row sep=crcr] {%
1	149.499999678754\\
2	149.499999678754\\
3	149.499999678754\\
4	149.499999678754\\
5	149.949849197465\\
6	150.857719349258\\
7	154.442148507048\\
8	158.873538607634\\
9	174.445740687924\\
10	229.64669691901\\
11	633.474574869046\\
12	0\\
};
\addplot[forget plot, color=white!15!black] table[row sep=crcr] {%
0.509090909090909	0\\
12.4909090909091	0\\
};
\addlegendentry{\normalsize{URDC-UL (30 m)}}

\addplot[ybar, bar width=0.145, fill=mycolor2, draw=black, area legend] table[row sep=crcr] {%
0.04	2629.62630941854\\
1.04	0\\
2.04	0\\
3.04	0\\
4.04	0\\
5.04	0\\
6.04	0\\
7.04	0\\
8.04	0\\
9.04	0\\
10.04	0\\
11.04	0\\
};
\addplot[forget plot, color=white!15!black] table[row sep=crcr] {%
-0.45090909090909	0\\
11.5309090909091	0\\
};
\addlegendentry{\normalsize{SUC-UL (30 m)}}

\addplot[ybar, bar width=0.145, fill=mycolor3, draw=black, area legend] table[row sep=crcr] {%
1	149.499999678754\\
2	149.649649318702\\
3	150.553877021539\\
4	157.53424643341\\
5	178.400954476291\\
6	256.432246327935\\
7	638.888887614834\\
8	0\\
9	0\\
10	0\\
11	0\\
12	0\\
};
\addplot[forget plot, color=white!15!black] table[row sep=crcr] {%
-0.2	0\\
13.2	0\\
};
\addlegendentry{\normalsize{URDC-UL (100 m)}}

\addplot[ybar, bar width=0.145, fill=mycolor4, draw=black, area legend] table[row sep=crcr] {%
0.04	200.940859683053\\
1.04	204.794520193517\\
2.04	210.365853367009\\
3.04	225.150602112092\\
4.04	264.75796893986\\
5.04	372.508304699705\\
6.04	732.843135790895\\
7.04	0\\
8.04	0\\
9.04	0\\
10.04	0\\
11.04	0\\
};
\addplot[forget plot, color=white!15!black] table[row sep=crcr] {%
-0.45090909090909	0\\
11.5309090909091	0\\
};
\addlegendentry{\normalsize{SUC-UL (100 m)}}

\end{axis}
\end{tikzpicture}%
    \caption{ Packet delay for a suburban area. Infinite delays are omitted from the figure.}
   \label{comp_delay_suburban}
\end{figure*}
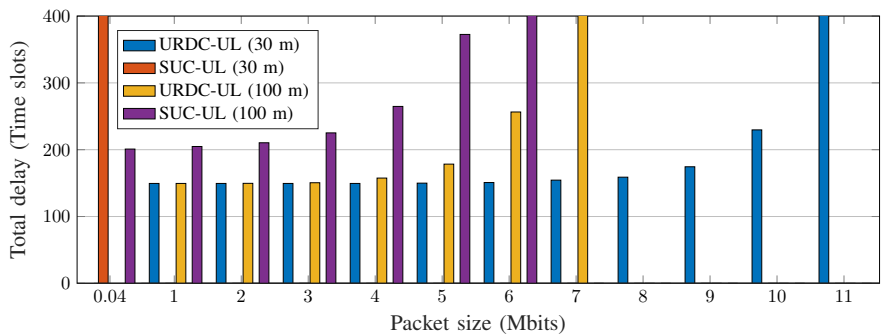

Fig.  \ref{comp_delay_suburban} compares the URDC-UL model's delay with the SUC-UL model's average delay.
The results indicate that the URDC-UL model has a shorter delay; these findings corroborate the success probability and outage capacity findings.
Additionally, the figure demonstrates that large packet sizes are permitted to be transmitted with finite delay in the SUC-UL situation when $h= 100$ m, but significantly smaller packet sizes are permitted when $h= 30$ m. However, in the URDC-UL model, because the delay at $h= 30$ m is less than that at $h= 100$ m, larger packet sizes with a finite delay can be delivered at $h= 30$ m.

Fig.  \ref{pow_pro} illustrates the propulsion power consumption at various speeds specified in (\ref{pow_mob}).
As shown in the figure, there is a speed $v$ at which the Propulsion Power usage is the least. This speed, based on the numerical values used, is $v=22$ m/s, which was used in this section.
According to (\ref{pow_mob}) and (\ref{pow_hov}), the hovering power  $P_h=1.371$ KW, while at $v=22$ m/s the traveling power $P_t=0.9363$ KW.
From (\ref{mob_po}) and (\ref{st_po}), the total energy consumption by UAV during a single time slot is $ 17.649$ kJ for the SUC-UL model and $ 15.568$ kJ for the URDC-UL model, taking into account additional energy consumed during acceleration before reaching $v$.

Fig.  \ref{comp_EE_suburban}  compares the URDC-UL model's energy efficiency to the SUC-UL model's average energy efficiency versus bandwidth at 20 dB SINR threshold from the UAV viewpoint.
Given that the URDC-UL model has a greater outage capacity and consumes less energy on average than the  SUC-UL model, it is predicted that the URDC-UL model is more energy efficient.
Additionally, the figure demonstrates that as the height approaches the optimal value for the  SUC-UL model, the energy efficiency of the  SUC-UL model improves. Furthermore, the figure confirms that increasing the bandwidth can greatly enhance energy efficiency. This is predictable as increasing the bandwidth increases  the outage capacity while the mechanical energy consumption is unaffected.

\begin{figure}
     \centering
%
%
\definecolor{mycolor1}{rgb}{0.00000,0.44700,0.74100}%
\definecolor{mycolor2}{rgb}{0.85000,0.32500,0.09800}%
\begin{tikzpicture}[scale=0.7, transform shape,font=\Large]

\begin{axis}[%
width=3in,
height=2.3in,
at={(0.758in,0.481in)},
scale only axis,
xmin=0,
xmax=80,
xlabel style={font=\large \color{white!15!black}},
xlabel={UAV speed (m/s)},
ymin=0,
ymax=6,
ylabel style={font=\large \color{white!15!black}},
ylabel={Power (kw)},
axis background/.style={fill=white},
legend style={font=\normalsize,at={(0.05,0.78)}, anchor=south west, legend cell align=left, align=left, draw=white!15!black},
grid=both
]
\addplot [line width=0.6mm,color=mycolor1]
  table[row sep=crcr]{%
0	1.37132152281327\\
1	1.36756853018281\\
2	1.35645150697737\\
3	1.33836729516248\\
4	1.31399039221448\\
5	1.28431239366354\\
6	1.25064763593369\\
7	1.21456415870572\\
8	1.17772320059898\\
9	1.14166439874458\\
10	1.10761841297427\\
11	1.07641696530332\\
12	1.04851079595271\\
13	1.02405473109379\\
14	1.00300928366668\\
15	0.985226929933693\\
16	0.970512807310602\\
17	0.958662126246532\\
18	0.949480674067965\\
19	0.942794518349733\\
20	0.93845340586517\\
21	0.93633076804423\\
22	0.936322074697304\\
23	0.938342518507138\\
24	0.942324553958394\\
25	0.948215550101419\\
26	0.955975670180856\\
27	0.965576013360067\\
28	0.976997014667265\\
29	0.990227081439443\\
30	1.00526143800852\\
31	1.02210114960845\\
32	1.04075229843865\\
33	1.06122528787033\\
34	1.08353425408871\\
35	1.10769656761878\\
36	1.13373241000969\\
37	1.16166441340189\\
38	1.19151735277665\\
39	1.22331788242546\\
40	1.25709430961979\\
41	1.29287639965436\\
42	1.33069520742049\\
43	1.37058293147608\\
44	1.4125727872458\\
45	1.45669889653539\\
46	1.50299619099813\\
47	1.55150032756696\\
48	1.60224761417747\\
49	1.65527494436527\\
50	1.71061973953716\\
51	1.76831989789509\\
52	1.82841374914257\\
53	1.8909400142297\\
54	1.95593776949923\\
55	2.02344641468601\\
56	2.09350564429792\\
57	2.16615542197091\\
58	2.24143595744552\\
59	2.319387685859\\
60	2.40005124908674\\
61	2.48346747890174\\
62	2.56967738174881\\
63	2.65872212495689\\
64	2.75064302423394\\
65	2.84548153230741\\
66	2.94327922859047\\
67	3.04407780976753\\
68	3.1479190812055\\
69	3.25484494910753\\
70	3.36489741333574\\
71	3.47811856083713\\
72	3.59455055961538\\
73	3.71423565319482\\
74	3.83721615553238\\
75	3.96353444633421\\
76	4.09323296674111\\
77	4.2263542153485\\
78	4.36294074453184\\
79	4.50303515705006\\
80	4.64668010290303\\
81	4.79391827642097\\
82	4.94479241356637\\
83	5.09934528943032\\
84	5.2576197159076\\
85	5.41965853953467\\
86	5.58550463947895\\
87	5.75520092566638\\
88	5.92879033703567\\
89	6.10631583991082\\
90	6.28782042648169\\
91	6.47334711338423\\
92	6.66293894037363\\
93	6.85663896908249\\
94	7.05449028185883\\
95	7.25653598067602\\
96	7.46281918611263\\
97	7.67338303639348\\
98	7.88827068649054\\
99	8.10752530727765\\
100	8.33119008473641\\
};
\addlegendentry{\normalsize{Travelling}}

\addplot [line width=0.6mm,color=mycolor2]
  table[row sep=crcr]{%
0	1.37132152281327\\
1	1.37132152281327\\
2	1.37132152281327\\
3	1.37132152281327\\
4	1.37132152281327\\
5	1.37132152281327\\
6	1.37132152281327\\
7	1.37132152281327\\
8	1.37132152281327\\
9	1.37132152281327\\
10	1.37132152281327\\
11	1.37132152281327\\
12	1.37132152281327\\
13	1.37132152281327\\
14	1.37132152281327\\
15	1.37132152281327\\
16	1.37132152281327\\
17	1.37132152281327\\
18	1.37132152281327\\
19	1.37132152281327\\
20	1.37132152281327\\
21	1.37132152281327\\
22	1.37132152281327\\
23	1.37132152281327\\
24	1.37132152281327\\
25	1.37132152281327\\
26	1.37132152281327\\
27	1.37132152281327\\
28	1.37132152281327\\
29	1.37132152281327\\
30	1.37132152281327\\
31	1.37132152281327\\
32	1.37132152281327\\
33	1.37132152281327\\
34	1.37132152281327\\
35	1.37132152281327\\
36	1.37132152281327\\
37	1.37132152281327\\
38	1.37132152281327\\
39	1.37132152281327\\
40	1.37132152281327\\
41	1.37132152281327\\
42	1.37132152281327\\
43	1.37132152281327\\
44	1.37132152281327\\
45	1.37132152281327\\
46	1.37132152281327\\
47	1.37132152281327\\
48	1.37132152281327\\
49	1.37132152281327\\
50	1.37132152281327\\
51	1.37132152281327\\
52	1.37132152281327\\
53	1.37132152281327\\
54	1.37132152281327\\
55	1.37132152281327\\
56	1.37132152281327\\
57	1.37132152281327\\
58	1.37132152281327\\
59	1.37132152281327\\
60	1.37132152281327\\
61	1.37132152281327\\
62	1.37132152281327\\
63	1.37132152281327\\
64	1.37132152281327\\
65	1.37132152281327\\
66	1.37132152281327\\
67	1.37132152281327\\
68	1.37132152281327\\
69	1.37132152281327\\
70	1.37132152281327\\
71	1.37132152281327\\
72	1.37132152281327\\
73	1.37132152281327\\
74	1.37132152281327\\
75	1.37132152281327\\
76	1.37132152281327\\
77	1.37132152281327\\
78	1.37132152281327\\
79	1.37132152281327\\
80	1.37132152281327\\
81	1.37132152281327\\
82	1.37132152281327\\
83	1.37132152281327\\
84	1.37132152281327\\
85	1.37132152281327\\
86	1.37132152281327\\
87	1.37132152281327\\
88	1.37132152281327\\
89	1.37132152281327\\
90	1.37132152281327\\
91	1.37132152281327\\
92	1.37132152281327\\
93	1.37132152281327\\
94	1.37132152281327\\
95	1.37132152281327\\
96	1.37132152281327\\
97	1.37132152281327\\
98	1.37132152281327\\
99	1.37132152281327\\
100	1.37132152281327\\
};
\addlegendentry{\normalsize{Hovering ($v=0$)}}

\end{axis}
\end{tikzpicture}%
    \caption{Propulsion Power consumption  at different speeds $v$ versus the constant hovering power.}
   \label{pow_pro}
   \end{figure}
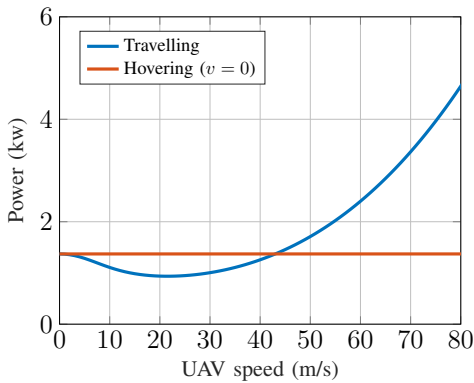
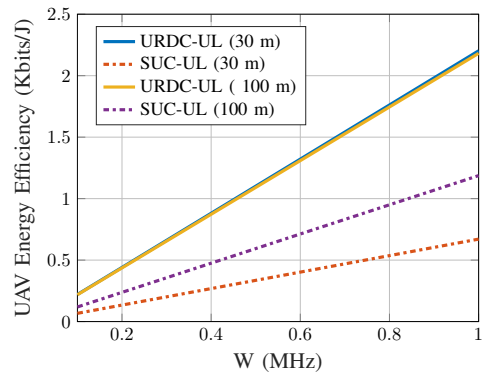
\begin{figure}
     \centering
%
%
\definecolor{mycolor1}{rgb}{0.00000,0.44700,0.74100}%
\definecolor{mycolor2}{rgb}{0.85000,0.32500,0.09800}%
\definecolor{mycolor3}{rgb}{0.92900,0.69400,0.12500}%
\definecolor{mycolor4}{rgb}{0.49400,0.18400,0.55600}%
\begin{tikzpicture}[scale=0.7, transform shape,font=\Large]

\begin{axis}[%
width=3in,
height=2.3in,
at={(0.758in,0.481in)},
scale only axis,
xmin=0.1,
xmax=1,
xlabel style={font=\large\color{white!15!black}},
xlabel={W (MHz)},
ymin=0,
ymax=2.5,
ylabel style={font=\large\color{white!15!black}},
ylabel={UAV Energy Efficiency (Kbits/J)},
ticklabel style={font=\normalsize,/pgf/number format/fixed},
axis background/.style={fill=white},
legend style={font=\footnotesize,at={(0.57,0.961)},legend cell align=left, align=left, draw=white!15!black},
grid=both
]
\addplot [line width=0.6mm,color=mycolor1]
  table[row sep=crcr]{%
0.1	0.220225291746978\\
0.2	0.440450583493956\\
0.3	0.660675875240934\\
0.4	0.880901166987912\\
0.5	1.10112645873489\\
0.6	1.32135175048187\\
0.7	1.54157704222885\\
0.8	1.76180233397582\\
0.9	1.9820276257228\\
1	2.20225291746978\\
};
\addlegendentry{\normalsize{URDC-UL (30 m)}}

\addplot [line width=0.6mm,color=mycolor2, dashdotted]
  table[row sep=crcr]{%
0.1	0.0670542964818522\\
0.2	0.134108592963704\\
0.3	0.201162889445557\\
0.4	0.268217185927409\\
0.5	0.335271482409261\\
0.6	0.402325778891113\\
0.7	0.469380075372966\\
0.8	0.536434371854818\\
0.9	0.60348866833667\\
1	0.670542964818522\\
};
\addlegendentry{\normalsize{SUC-UL (30 m)}}

\addplot [line width=0.6mm,color=mycolor3]
  table[row sep=crcr]{%
0.1	0.218177038235732\\
0.2	0.436354076471463\\
0.3	0.654531114707194\\
0.4	0.872708152942926\\
0.5	1.09088519117866\\
0.6	1.30906222941439\\
0.7	1.52723926765012\\
0.8	1.74541630588585\\
0.9	1.96359334412158\\
1	2.18177038235732\\
};
\addlegendentry{\normalsize{URDC-UL ( 100 m)}}

\addplot [line width=0.6mm,color=mycolor4, dashdotted]
  table[row sep=crcr]{%
0.1	0.118726490157409\\
0.2	0.237452980314818\\
0.3	0.356179470472228\\
0.4	0.474905960629637\\
0.5	0.593632450787046\\
0.6	0.712358940944455\\
0.7	0.831085431101864\\
0.8	0.949811921259273\\
0.9	1.06853841141668\\
1	1.18726490157409\\
};
\addlegendentry{\normalsize{SUC-UL (100 m)}}

\end{axis}
\end{tikzpicture}%
    \caption{Energy efficiency from UAV viewpoint versus bandwidth for a suburban area  at 20 dB SINR threshold.}
     \label{comp_EE_suburban}
\end{figure}

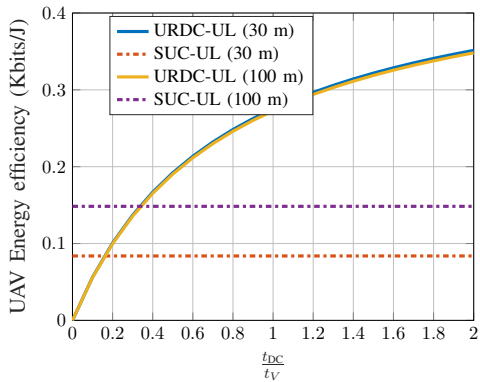
\begin{figure}
     \centering
%
%
\definecolor{mycolor1}{rgb}{0.00000,0.44700,0.74100}%
\definecolor{mycolor2}{rgb}{0.85000,0.32500,0.09800}%
\definecolor{mycolor3}{rgb}{0.92900,0.69400,0.12500}%
\definecolor{mycolor4}{rgb}{0.49400,0.18400,0.55600}%
\begin{tikzpicture}[scale=0.7, transform shape,font=\Large]

\begin{axis}[%
width=3in,
height=2.3in,
at={(0.758in,0.509in)},
scale only axis,
xmin=0,
xmax=2,
xlabel style={font=\large \color{white!15!black}},
xlabel={$\frac{t_{\text{DC}}}{t_V}$},
ymin=0,
ymax=0.4,
ticklabel style={font=\normalsize},
ylabel style={font=\large \color{white!15!black}},
ylabel={UAV Energy efficiency (Kbits/J)},
axis background/.style={fill=white},
legend style={font=\footnotesize,at={(0.6,0.99)},legend cell align=left, align=left, draw=white!15!black}, 
grid=both
]
\addplot [line width=0.6mm,color=mycolor1]
  table[row sep=crcr]{%
0	0\\
0.1	0.0561995252395273\\
0.2	0.100740784412536\\
0.3	0.136910528469208\\
0.4	0.166866151911952\\
0.5	0.192082391381009\\
0.6	0.213601545775162\\
0.7	0.232181155172572\\
0.8	0.248385029586748\\
0.9	0.26264145711502\\
1	0.275281614683722\\
1.1	0.286565615579997\\
1.2	0.29670059995363\\
1.3	0.305853566744173\\
1.4	0.314160647960912\\
1.5	0.321733925239254\\
1.6	0.328666516390555\\
1.7	0.335036423401662\\
1.8	0.340909480006802\\
1.9	0.346341635420504\\
2	0.351380742345868\\
};
\addlegendentry{\normalsize{URDC-UL (30 m)}}

\addplot [line width=0.6mm,color=mycolor2, dashdotted]
  table[row sep=crcr]{%
0	0.0838178706023153\\
0.1	0.0838178706023153\\
0.2	0.0838178706023153\\
0.3	0.0838178706023153\\
0.4	0.0838178706023153\\
0.5	0.0838178706023153\\
0.6	0.0838178706023153\\
0.7	0.0838178706023153\\
0.8	0.0838178706023153\\
0.9	0.0838178706023153\\
1	0.0838178706023153\\
1.1	0.0838178706023153\\
1.2	0.0838178706023153\\
1.3	0.0838178706023153\\
1.4	0.0838178706023153\\
1.5	0.0838178706023153\\
1.6	0.0838178706023153\\
1.7	0.0838178706023153\\
1.8	0.0838178706023153\\
1.9	0.0838178706023153\\
2	0.0838178706023153\\
};
\addlegendentry{\normalsize{SUC-UL (30 m)}}

\addplot [line width=0.6mm,color=mycolor3]
  table[row sep=crcr]{%
0	0\\
0.1	0.0556768292585657\\
0.2	0.0998038227049961\\
0.3	0.13563716214312\\
0.4	0.165314176755735\\
0.5	0.190295887072291\\
0.6	0.211614897862604\\
0.7	0.23002170353759\\
0.8	0.246074870272364\\
0.9	0.260198702777099\\
1	0.272721297794664\\
1.1	0.28390034937166\\
1.2	0.29394107110559\\
1.3	0.303008908725825\\
1.4	0.311238728116116\\
1.5	0.318741568472096\\
1.6	0.325609681542541\\
1.7	0.331920343839754\\
1.8	0.337738776796912\\
1.9	0.343120409260632\\
2	0.348112648869478\\
};
\addlegendentry{\normalsize{URDC-UL (100 m)}}

\addplot [line width=0.6mm,color=mycolor4, dashdotted]
  table[row sep=crcr]{%
0	0.148408112696761\\
0.1	0.148408112696761\\
0.2	0.148408112696762\\
0.3	0.148408112696761\\
0.4	0.148408112696762\\
0.5	0.148408112696761\\
0.6	0.148408112696761\\
0.7	0.148408112696761\\
0.8	0.148408112696761\\
0.9	0.148408112696762\\
1	0.148408112696761\\
1.1	0.148408112696761\\
1.2	0.148408112696761\\
1.3	0.148408112696761\\
1.4	0.148408112696762\\
1.5	0.148408112696761\\
1.6	0.148408112696761\\
1.7	0.148408112696761\\
1.8	0.148408112696762\\
1.9	0.148408112696761\\
2	0.148408112696761\\
};
\addlegendentry{\normalsize{SUC-UL (100 m)}}

\end{axis}
\end{tikzpicture}%
    \caption{ Transmission to traveling duration ratio vs energy efficiency from UAV viewpoint at 20 dB SINR threshold.}
    \label{trn_subslot_time}
 \end{figure}

Fig.   \ref{trn_subslot_time} illustrates the ratio of transmission  duration $t_{\text{DC}}$ to the average traveling time duration  $t_V$   versus the average energy efficiency of the URDC-UL and SUC-UL models from the UAV viewpoint at an SINR threshold of 20 dB. The figure demonstrates that as this ratio increases, the URDC-UL model's energy efficiency increases. On the other hand, the SUC-UL model's energy efficiency is always constant for any time slot duration $T_s$. This implies that increasing the transmission  duration   $t_{\text{DC}}$ increases the gain of the URDC-UL model (i.e., higher rate, higher energy efficiency, and lower delay). However, increasing $t_{\text{DC}}$ affects the transmission cycle's overall energy consumption. By increasing  $t_{\text{DC}}$, the duration of the total time slot $T_s$ is extended, resulting in an increase in the total energy consumed during the time slot, $E_{\text{DC}}$. The transmission cycle's total energy usage $E_{cycle}=E_{\text{DC}} \times N_d $ must be less than the capacity of the batteries carried by the UAVs. To obtain the numerical findings, in this section, we select a ratio of unity between $t_{DC}$ and $t_V$. This results in a remarkable gain while maintaining an acceptable energy consumption that can be handled by the available types of batteries.

Fig.  \ref{pow_diff} plots the outage probability versus transmission power $P$ at SINR threshold = 10 dB for the URDC-UL model and the average performance of the SUC-UL model.
The results demonstrate that the URDC-UL model can reliably provide very low outage probabilities at extremely low transmission power levels. These results illustrate why the URDC-UL model is well-suited for data aggregation in IoT devices, as it overcomes their power limitation issue. Moreover, these results demonstrate that the URDC-UL model offers ultra-reliable service. Note that  the outage probability saturates even with increasing  transmission power. This can be explained by the transformation of the system from noise-limited to interference-limited, where increasing $P$ only can not enhance the outage probability.

Fig. \ref{iotee} plots the energy efficiency from the IoT device point of view  against the transmission power at $10$ dB SINR threshold for the URDC-UL model and the average performance of the SUC-UL model.
The results confirm  that the URDC-UL model will always offer higher energy efficiency for IoT devices. Moreover, the figure shows that the peak of the energy efficiency for the URDC-UL scheme is at lower transmission powers. This can be explained by the results of Fig. \ref{pow_diff} where we can see very low outage probability at extremely low  transmission power, owing to the ultra-reliable transmission.

 \begin{figure}
     \centering
%
%
\definecolor{mycolor1}{rgb}{0.00000,0.44700,0.74100}%
\definecolor{mycolor2}{rgb}{0.85000,0.32500,0.09800}%
\definecolor{mycolor3}{rgb}{0.92900,0.69400,0.12500}%
\definecolor{mycolor4}{rgb}{0.49400,0.18400,0.55600}%
\begin{tikzpicture}[scale=0.7, transform shape,font=\Large]

\begin{axis}[%
width=3in,
height=2.3in,
at={(0.758in,0.509in)},
scale only axis,
xmin=-60,
xmax=20,
xlabel style={font=\large\color{white!15!black}},
xlabel={Transmission power (dBm)},
ymode=log,
ymin=1e-7,
ymax=1,
yminorticks=true,
ticklabel style={font=\normalsize},
ylabel style={font=\large \color{white!15!black}},
ylabel={Outage probability},
axis background/.style={fill=white},
legend style={font=\footnotesize,at={(0.41,0.48)}, anchor=south west, legend cell align=left, align=left, draw=white!15!black},
grid=both
]
\addplot [line width=0.6mm,color=mycolor1]
  table[row sep=crcr]{%
-60	0.999987746746246\\
-59	0.999842763222895\\
-58	0.998806625199389\\
-57	0.994038032749735\\
-56	0.978688144212824\\
-55	0.941876342171474\\
-54	0.872922415194364\\
-53	0.768358823197784\\
-52	0.636245285379109\\
-51	0.493671176480258\\
-50	0.3592715069676\\
-49	0.246315861226287\\
-48	0.16009972498593\\
-47	0.0993517098416707\\
-46	0.0592757006636571\\
-45	0.034221686764376\\
-44	0.0192283163004605\\
-43	0.0105668902176672\\
-42	0.00570351000501934\\
-41	0.00303427668126333\\
-40	0.00159575480546659\\
-39	0.000831686698346545\\
-38	0.000430516322856289\\
-37	0.000221801778525932\\
-36	0.000113994199816725\\
-35	5.86203022296949e-05\\
-34	3.03010541848892e-05\\
-33	1.58651310009361e-05\\
-32	8.5235894333735e-06\\
-31	4.79578541823233e-06\\
-30	2.90451175755102e-06\\
-29	1.94512987194972e-06\\
-28	1.45819342078912e-06\\
-27	1.21071714231924e-06\\
-26	1.08466014936415e-06\\
-25	1.02023580683674e-06\\
-24	9.8715426288809e-07\\
-23	9.7005606003453e-07\\
-22	9.6114044889628e-07\\
-21	9.56436293608753e-07\\
-20	9.53915306078912e-07\\
-19	9.52536877485421e-07\\
-18	9.51763928225269e-07\\
-17	9.5131707855689e-07\\
-16	9.51049508923774e-07\\
-15	9.50883029093852e-07\\
-14	9.50775299157769e-07\\
-13	9.50702919388924e-07\\
-12	9.50652599418511e-07\\
-11	9.5061658944573e-07\\
-10	9.50590189452427e-07\\
-9	9.50570469449019e-07\\
-8	9.50555519407814e-07\\
-7	9.50544059463709e-07\\
-6	9.50535199550906e-07\\
-5	9.50528309506815e-07\\
-4	9.50522929588082e-07\\
-3	9.50518699638359e-07\\
-2	9.50515369635418e-07\\
-1	9.5051274950908e-07\\
0	9.50510679609273e-07\\
1	9.50509039587821e-07\\
2	9.50507749508667e-07\\
3	9.50506719443744e-07\\
4	9.5050590964707e-07\\
5	9.50505259500467e-07\\
6	9.5050474946401e-07\\
7	9.50504349561676e-07\\
8	9.50504019603393e-07\\
9	9.50503769581168e-07\\
10	9.50503559526972e-07\\
11	9.50503399654856e-07\\
12	9.50503279528725e-07\\
13	9.50503169616645e-07\\
14	9.50503089458543e-07\\
15	9.505030295065e-07\\
16	9.50502979546464e-07\\
17	9.50502939578435e-07\\
18	9.50502909602413e-07\\
19	9.50502879626391e-07\\
20	9.50502859642377e-07\\
};
\addlegendentry{\normalsize{URDC-UL (30 m)}}

\addplot [line width=0.6mm,color=mycolor2, dashdotted]
  table[row sep=crcr]{%
-60	1\\
-50	0.99798128\\
-40	0.9775188\\
-30	0.90062392\\
-20	0.81152496\\
-10	0.79702688\\
0	0.79344824\\
10	0.78179472\\
20	0.77739024\\
};
\addlegendentry{\normalsize{SUC-UL (30 m)}}

\addplot [line width=0.6mm,color=mycolor3]
  table[row sep=crcr]{%
-60	1\\
-59	1\\
-58	1\\
-57	1\\
-56	1\\
-55	1\\
-54	1\\
-53	0.999999999999997\\
-52	0.999999999996082\\
-51	0.999999998911914\\
-50	0.999999905013908\\
-49	0.999996692893103\\
-48	0.999944521212789\\
-47	0.999478970364034\\
-46	0.996915281713826\\
-45	0.987363665194258\\
-44	0.961509128845363\\
-43	0.907859635405085\\
-42	0.819009766271382\\
-41	0.697820829602739\\
-40	0.558047418253161\\
-39	0.418499231685104\\
-38	0.295286231997403\\
-37	0.197184416513589\\
-36	0.125512145402615\\
-35	0.0767234539432253\\
-34	0.0453680504563272\\
-33	0.0261275178082091\\
-32	0.014746789317987\\
-31	0.00820539005394771\\
-30	0.00452665758985693\\
-29	0.00249041948129802\\
-28	0.00137523468872092\\
-27	0.00076800210162109\\
-26	0.000437728904943757\\
-25	0.000257456093865738\\
-24	0.000158228289096618\\
-23	0.000102867436090559\\
-22	7.13940919069289e-05\\
-21	5.30654720034551e-05\\
-20	4.20803080696253e-05\\
-19	3.52801279548398e-05\\
-18	3.09242067387805e-05\\
-17	2.8037313691609e-05\\
-16	2.60616778432476e-05\\
-15	2.4670309922703e-05\\
-14	2.36660380269482e-05\\
-13	2.2926271179835e-05\\
-12	2.23723311466584e-05\\
-11	2.1952114749868e-05\\
-10	2.16300817339476e-05\\
-9	2.13813316226608e-05\\
-8	2.11880055139835e-05\\
-7	2.10370380789859e-05\\
-6	2.09187123122456e-05\\
-5	2.08257038523119e-05\\
-4	2.07524318586172e-05\\
-3	2.06946071054093e-05\\
-2	2.06489103550256e-05\\
-1	2.06127589812422e-05\\
0	2.05841348359659e-05\\
1	2.05614554875666e-05\\
2	2.0543476810575e-05\\
3	2.05292185677663e-05\\
4	2.05179071350869e-05\\
5	2.05089311410278e-05\\
6	2.05018069181495e-05\\
7	2.04961515148927e-05\\
8	2.04916615164086e-05\\
9	2.04880963999843e-05\\
10	2.04852654207555e-05\\
11	2.04830172579884e-05\\
12	2.04812318337355e-05\\
13	2.04798138451201e-05\\
14	2.04786876373264e-05\\
15	2.0477793148177e-05\\
16	2.04770826869316e-05\\
17	2.04765183825506e-05\\
18	2.04760701619788e-05\\
19	2.04757141413214e-05\\
20	2.04754313536393e-05\\
};
\addlegendentry{\normalsize{URDC-UL (100 m)}}

\addplot [line width=0.6mm,color=mycolor4, dashdotted]
  table[row sep=crcr]{%
-60	1\\
-50	1\\
-40	0.9954\\
-30	0.8643\\
-20	0.4156\\
-10	0.2656\\
0	0.2593\\
10	0.2548\\
20	0.254\\
};
\addlegendentry{\normalsize{SUC-UL (100 m)}}

\end{axis}
\end{tikzpicture}%
    \caption{Outage probability versus transmission power at $10$ dB SINR threshold. }
    \label{pow_diff}
\end{figure}
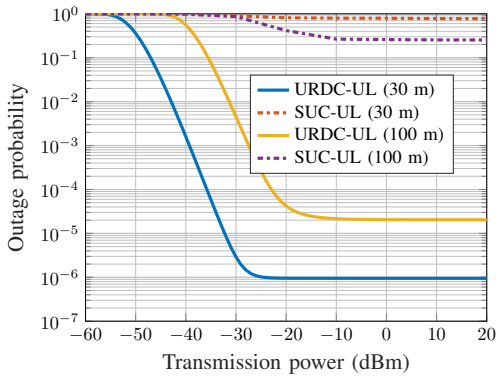
 \begin{figure}
     \centering
%
%
\definecolor{mycolor1}{rgb}{0.00000,0.44700,0.74100}%
\definecolor{mycolor2}{rgb}{0.85000,0.32500,0.09800}%
\definecolor{mycolor3}{rgb}{0.92900,0.69400,0.12500}%
\definecolor{mycolor4}{rgb}{0.49400,0.18400,0.55600}%
\begin{tikzpicture}[scale=0.7, transform shape,font=\Large]

\begin{axis}[%
width=3in,
height=2.3in,
at={(0.758in,0.481in)},
scale only axis,
xmin=-44,
xmax=10,
xlabel style={font=\large\color{white!15!black}},
xlabel={Transmission power (dBm)},
ymode=log,
ymin=0.01,
ymax=10000,
yminorticks=true,
ylabel style={font=\large\color{white!15!black}},
ylabel={Device Energy Efficiency (Gbits/J)},
axis background/.style={fill=white},
legend style={font=\footnotesize,at={(0.9,0.961)},legend cell align=left, align=left, draw=white!15!black},
grid=both
]
\addplot [line width=0.6mm,color=mycolor1]
  table[row sep=crcr]{%
-60	4.23892934686611\\
-59	43.2074746632283\\
-58	260.484336594707\\
-57	1033.69957223735\\
-56	2935.12106427867\\
-55	6358.54410305492\\
-54	11042.6600527057\\
-53	15988.9709362368\\
-52	19944.0512452486\\
-51	22051.4626692027\\
-50	22165.5640775811\\
-49	20710.6689221831\\
-48	18332.9550641446\\
-47	15615.646869439\\
-46	12955.8858422941\\
-45	10565.3089084645\\
-44	8522.6110560558\\
-43	6829.53585975254\\
-42	5451.55824855716\\
-41	4341.95159529183\\
-40	3453.91121400767\\
-39	2745.63879876736\\
-38	2181.81407637191\\
-37	1733.43839785073\\
-36	1377.06753781034\\
-35	1093.90420369541\\
-34	868.94360365396\\
-33	690.236403123185\\
-32	548.278288877435\\
-31	435.514548862332\\
-30	345.942157067749\\
-29	274.791886571206\\
-28	218.275060462899\\
-27	173.382086369836\\
-26	137.722303959843\\
-25	109.396721634402\\
-24	86.8969076551651\\
-23	69.024668440925\\
-22	54.8282435239366\\
-21	43.5516220962753\\
-20	34.5942831863253\\
-19	27.4792159328238\\
-18	21.8275171003989\\
-17	17.3382131345215\\
-16	13.7722322361028\\
-15	10.9396729221376\\
-14	8.68969108163876\\
-13	6.90246697767704\\
-12	5.48282440989675\\
-11	4.35516223497331\\
-10	3.45942833013554\\
-9	2.74792159868592\\
-8	2.18275171267754\\
-7	1.73382131479243\\
-6	1.3772232243186\\
-5	1.09396729260182\\
-4	0.868969108383177\\
-3	0.690246697894861\\
-2	0.548282441064916\\
-1	0.435516223542556\\
0	0.34594283304106\\
1	0.274792159885473\\
2	0.218275171278181\\
3	0.173382131485717\\
4	0.137722322435894\\
5	0.109396729262703\\
6	0.0868969108398975\\
7	0.0690246697904767\\
8	0.0548282441071139\\
9	0.0435516223546467\\
10	0.0345942833043523\\
11	0.0274792159887022\\
12	0.0218275171279157\\
13	0.0173382131486332\\
14	0.0137722322436283\\
15	0.0109396729262947\\
16	0.00868969108400513\\
17	0.0069024669790574\\
18	0.00548282441071747\\
19	0.00435516223546854\\
20	0.00345942833043765\\
};
\addlegendentry{\normalsize{URDC-UL (30 m)}}

\addplot [line width=0.6mm,color=mycolor2, dashdotted]
  table[row sep=crcr]{%
-60	0\\
-59	249.390136946173\\
-58	330.60214206354\\
-57	325.098529374246\\
-56	281.329277301891\\
-55	226.621614396245\\
-54	175.132992286997\\
-53	133.091403530527\\
-52	102.184589444962\\
-51	81.700416921984\\
-50	69.8362379717548\\
-49	64.4645186159425\\
-48	63.550231538528\\
-47	65.3514059127659\\
-46	68.4898234937585\\
-45	71.9483770302877\\
-44	75.0309248442128\\
-43	77.3066242349985\\
-42	78.5516012310531\\
-41	78.6949267654785\\
-40	77.7721741049088\\
-39	75.8439248758111\\
-38	72.907181206147\\
-37	69.0803527202247\\
-36	64.5683372404167\\
-35	59.5998614135041\\
-34	54.3921030976849\\
-33	49.1328327364509\\
-32	43.9734442104302\\
-31	39.0284417746689\\
-30	34.378475328823\\
-29	30.0699426140377\\
-28	26.1133169231302\\
-27	22.5125501476954\\
-26	19.2674501386558\\
-25	16.371549978467\\
-24	13.8120147376799\\
-23	11.5705874365728\\
-22	9.62496164512974\\
-21	7.95022282240896\\
-20	6.5201651269993\\
-19	5.30970965381219\\
-18	4.2975477291386\\
-17	3.46053573766825\\
-16	2.77469943786549\\
-15	2.21705653262564\\
-14	1.76660123925513\\
-13	1.40474801155085\\
-12	1.11543785211815\\
-11	0.885044518817325\\
-10	0.702171629061462\\
-9	0.55733801045746\\
-8	0.442627893895197\\
-7	0.35172469167735\\
-6	0.279646712783612\\
-5	0.222462994439549\\
-4	0.177070029349234\\
-3	0.14101626171652\\
-2	0.112364069404986\\
-1	0.0895811748627009\\
0	0.0714551689429183\\
1	0.0570253760867278\\
2	0.0455287044076304\\
3	0.0363617114332907\\
4	0.0290472604540253\\
5	0.0232075136870045\\
6	0.0185428381565018\\
7	0.0148152987413129\\
8	0.0118357165152304\\
9	0.00945350334346936\\
10	0.00754866244985605\\
11	0.0060254820897651\\
12	0.00480755526631905\\
13	0.00383384001614269\\
14	0.00305553781326568\\
15	0.00243361640666423\\
16	0.00193684121259154\\
17	0.00154020874707781\\
18	0.00122369843653905\\
19	0.000971276965101936\\
20	0.00077010324236126\\
};
\addlegendentry{\normalsize{SUC-UL (30 m)}}

\addplot [line width=0.6mm,color=mycolor3]
  table[row sep=crcr]{%
-60	1.12063668067947e-69\\
-59	2.32997694982262e-54\\
-58	3.26749116865018e-42\\
-57	1.38753302783588e-32\\
-56	5.88154411294095e-25\\
-55	6.41893007844302e-19\\
-54	3.8284814003123e-14\\
-53	2.26894832395268e-10\\
-52	2.14831249224761e-07\\
-51	4.7387940598441e-05\\
-50	0.00328597889329762\\
-49	0.0908767911042738\\
-48	1.21096532910882\\
-47	9.03373147173176\\
-46	42.4834970240505\\
-45	138.237501157135\\
-44	334.474097806669\\
-43	635.996428571188\\
-42	992.33861480947\\
-41	1316.04056215971\\
-40	1528.9047352334\\
-39	1597.92003984056\\
-38	1538.21664621544\\
-37	1391.94009359665\\
-36	1204.36612752846\\
-35	1010.03530347964\\
-34	829.546462531994\\
-33	672.212903954491\\
-32	540.197548885853\\
-31	431.943053620946\\
-30	344.37719562442\\
-29	274.108072679387\\
-28	217.975198878069\\
-27	173.249138318822\\
-26	137.662168243212\\
-25	109.3686683636\\
-24	86.8832438733036\\
-23	69.0176350012306\\
-22	54.8243818052103\\
-21	43.5493526610737\\
-20	34.5928604468904\\
-19	27.4782726366376\\
-18	21.8268628757682\\
-17	17.3377435112925\\
-16	13.7718864063689\\
-15	10.9394134391188\\
-14	8.68949369283515\\
-13	6.9023152918986\\
-12	5.48270695848306\\
-11	4.3550707699547\\
-10	3.45935679084863\\
-9	2.7478654565027\\
-8	2.18270753930692\\
-7	1.73378648836785\\
-6	1.3771957236552\\
-5	1.09394554978917\\
-4	0.868951901119675\\
-3	0.690233069590082\\
-2	0.548271640770563\\
-1	0.435507660307191\\
0	0.34593604092304\\
1	0.274786770946018\\
2	0.218270894615439\\
3	0.173378736883698\\
4	0.137719627565392\\
5	0.109394589632849\\
6	0.0868952118905213\\
7	0.0690233206573529\\
8	0.0548271726987421\\
9	0.0435507714599819\\
10	0.0345936075126238\\
11	0.0274786792500249\\
12	0.0218270908201985\\
13	0.0173378745449602\\
14	0.0137719632966797\\
15	0.0109394593039246\\
16	0.00868952140389827\\
17	0.00690233220125247\\
18	0.00548271735535895\\
19	0.00435507719992499\\
20	0.00345936078528267\\
};
\addlegendentry{\normalsize{URDC-UL (100 m)}}

\addplot [line width=0.6mm,color=mycolor4, dashdotted]
  table[row sep=crcr]{%
-60	0\\
-59	183.353816157595\\
-58	241.066014758385\\
-57	232.709861431413\\
-56	194.356186216861\\
-55	146.83281527536\\
-54	101.305762699275\\
-53	63.0005440962457\\
-52	33.6430924347916\\
-51	13.0281200722541\\
-50	0\\
-49	-6.95614290739906\\
-48	-9.37728936887327\\
-47	-8.64699412303642\\
-46	-5.91950879995242\\
-45	-2.10264570436458\\
-44	2.12709579911658\\
-43	6.29382890945115\\
-42	10.0861458919511\\
-41	13.3213877630992\\
-40	15.9133854457316\\
-39	18.0956912842393\\
-38	20.7388250602906\\
-37	24.1452513679302\\
-36	28.1472206257234\\
-35	32.3968991648172\\
-34	36.5258054014474\\
-33	40.2232581615253\\
-32	43.2665662344423\\
-31	45.5240913821451\\
-30	46.9444870649081\\
-29	47.4589465352729\\
-28	46.853579721387\\
-27	45.1382340611733\\
-26	42.4969692601648\\
-25	39.1802856094354\\
-24	35.4434597660189\\
-23	31.5143287275278\\
-22	27.5793045476988\\
-21	23.7801837389987\\
-20	20.2169183793164\\
-19	16.9610968247994\\
-18	14.0732237345693\\
-17	11.5723751326957\\
-16	9.44567460217016\\
-15	7.66275965234237\\
-14	6.18513424366598\\
-13	4.97202074518321\\
-12	3.98384903890719\\
-11	3.18417878240485\\
-10	2.54060658072723\\
-9	2.02484474254974\\
-8	1.61235083662719\\
-7	1.28291231234942\\
-6	1.02013791883599\\
-5	0.810773144688171\\
-4	0.644127591236071\\
-3	0.511599122946282\\
-2	0.406280683529301\\
-1	0.322637027911973\\
0	0.256240099992465\\
1	0.203548167603589\\
2	0.161723595601716\\
3	0.128514945611615\\
4	0.102140007943785\\
5	0.0811872968556246\\
6	0.0645383969357197\\
7	0.051306706987017\\
8	0.0407890879808055\\
9	0.0324276728956456\\
10	0.0257796844220851\\
11	0.020493567327621\\
12	0.0162901035678909\\
13	0.0129474616954989\\
14	0.0102893546812614\\
15	0.0081756551341925\\
16	0.00649495440442159\\
17	0.0051586602421799\\
18	0.00409631287385749\\
19	0.00325186648004475\\
20	0.00258073598750342\\
};
\addlegendentry{\normalsize{SUC-UL (100 m)}}

\end{axis}
\end{tikzpicture}%
    \caption{Energy efficiency from IoT device perspective versus transmission power at $10$ dB SINR threshold.}
    \label{iotee}
 \end{figure}
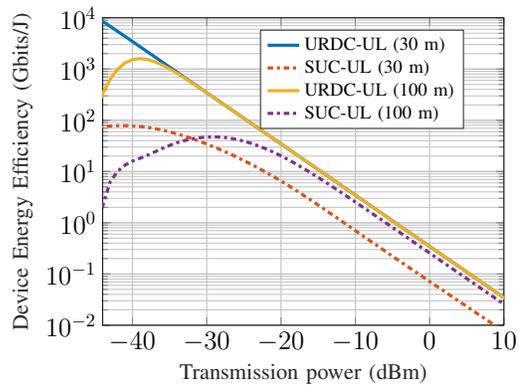

\section{Summary \& Conclusion}\label{conc}

This study examines whether it is worthwhile to travel toward IoT devices for data aggregation. A novel spatiotemporal URDC-UL model for data aggregation in large-scale IoT networks employing UAVs, that accounts for interference from other devices, is presented to answer this topic. The UAV in particular
navigates to the nearest IoT device and attempts to communicate from the closest distance possible.
However, time is lost due to mechanical movement. A second stationary SUC-UL model with comparable parameters is presented for comparison with the proposed model, where the entire duration is allocated for  transmission, but the probability of retransmission is increased.

The results indicate that the transmission-to-traveling duration ratio is a crucial design parameter that is constrained by the total amount of energy consumed.
In conclusion, by increasing the transmission duration to a suitable value that can be supported by UAV batteries, the proposed URDC-UL model outperforms the SUC-UL model. It delivers a higher probability
of success, an improved rate, enhanced energy efficiency, and a shorter delay. In addition, the URDC-UL paradigm assures that performance is almost identical across all  devices, regardless of location. However, the SUC-UL model's performance varies by location. Moreover, the proposed URDC-UL model enables IoT devices to greatly reduce their transmission power while maintaining link quality due to the ultra-reliable transmission introduced by the model. This helps preserve the device's battery and addresses a crucial issue in the IoT network design.

\appendices
\section{Proof of Lemma 1 }
We begin with the LOS interference, 
\small
\begin{equation}
\begin{aligned}[b]
\mathcal{L}_{I_L} (s)&\stackrel{a}{=}\mathbb{E}_{\bold\Phi_L}\left[\exp\left\{-s \sum\limits_{\substack{x_i\in \bold\Phi_L}} PH_{L,i}G_{\epsilon,i}(r_i^2+h^2)^{\frac{-\alpha_L}{2}} \right\}\right]\\
&\stackrel{b}{=}\mathbb{E}_{r}\left[\prod\limits_{\substack{x_i\in \bold\Phi_L}}\mathbb{E}_{H_L}\exp\left\{- s PH_{L,i}G_{\epsilon,i}(r_i^2+h^2)^{\frac{-\alpha_L}{2}}\right\}\right]\\
&\stackrel{c}{=}\mathbb{E}_{r}\left[\prod\limits_{\substack{x_i\in \bold\Phi_L}}\left(1+\frac{ sPG_{\epsilon,i}(r_i^2+h^2)^{\frac{-\alpha_L}{2}}}{m_L}\right)^{-m_L}\right]\\
&\stackrel{d}{=}\exp\left\{-2\:\pi\: \lambda_{\text{DC}} \int_{\frac{R}{2}}^{\infty}p_{\rm{LOS}}\left(r\right) \right.\\
 &\left. \left(1-\left( 1+\frac{sPG_\epsilon(r^2+h^2)^{\frac{-\alpha_L}{2}}}{m_L}\right)^{-m_L}    \right) rdr \right\}.
\end{aligned}
\end{equation}
\normalsize
 (a) follows from the definition of Laplace Transform. (b) follows from the independence between the distributions of channel fading gains and interferers distance. (c) follows 
from the Gamma distribution's moment-generating function. (d) follows from
the definition of probability generating functional (PGFL) in PPP.
When a sectored antenna is employed, four discrete probabilities of the antenna gain are available, and the final equation is given by (\ref{LOS}).
For the NLOS interference, we replace $p_{\rm{LOS}}\left(r\right)$ by $\left(1-p_{\rm{LOS}}\left(r\right)\right)$, $m_L$ by $m_N$, and $\alpha_L$ by $\alpha_N$ to get (\ref{NLOS}). 

\section{Proof of theorem 1 }
The packet's success probability given the intended link is LOS is 
\small
\begin{align}
(S_p|k=L)&=\mathbb{P}\left(\frac{PH_{L,o}G_{dM}G_{uM}(r_x^2+h^2)^{\frac{-\alpha_L}{2}}}{I_L+I_N+\sigma^2}>\theta_{\text{DC}}\right)\notag \\
&=\mathbb{P}\left(H_{L,o}>\frac{\theta_{\text{DC}} \; (r_x^2+h^2)^{\frac{\alpha_L}{2}}\left(I_L+I_N+\sigma^2\right)}{PG_{dM}G_{uM} } \right).
\end{align}
\normalsize
Since $H_{L,o}$ is a Gamma RV, then using Alzer’s inequality \cite{alzer1997some},
\begin{equation}
\begin{aligned}
&(S_p|k=L)\approx \sum_{n=1}^{m_L} \left(-1\right)^{n+1} {m_L \choose n}\\
&\mathbb{E}\left[\exp\left(-\;\frac{g_L\;n \;\theta_{\text{DC}} \; (r_x^2+h^2)^{\frac{\alpha_L}{2}}\; (I_L+I_N+\sigma^2)}{PG_{dM}G_{uM}}\right)\right],
\end{aligned}
\end{equation}
where $g_L=m_L(m_L!)^{-\;\frac{1}{m_L}}$. Since $\bold\Phi_L$ and $\bold\Phi_N$ are independent and utilizing the definition of Laplace Transform, then
\small
\begin{equation}
\begin{aligned}
&(S_p|k=L)=\sum_{n=1}^{m_L} \left(-1\right)^{n+1} {m_L \choose n} \mathcal{L}_{I_L} \left(\frac{g_L\;n \;\theta_{\text{DC}} \; (r_x^2+h^2)^{\frac{\alpha_L}{2}}}{PG_{dM}G_{uM}}\right)\\
&\!\!\!\!\!\mathcal{L}_{I_N} \left(\frac{g_L\;n \;\theta_{\text{DC}} \; (r_x^2+h^2)^{\frac{\alpha_L}{2}}}{PG_{dM}G_{uM}}\right) \exp\left(-\;\frac{g_L\;n \;\theta_{\text{DC}} \; (r_x^2+h^2)^{\frac{\alpha_L}{2}}\; \sigma^2}{PG_{dM}G_{uM}}\right).
\end{aligned}
\end{equation}
\normalsize
Similarly, $(S_p|k=N)$ is formulated by replacing $g_L$ by $g_N=m_N(m_N!)^{-\;\frac{1}{m_N}}$, $m_L$ by $m_N$, and $\alpha_L$ by $\alpha_N$. Finally, by using the law of total probability and since $r_x$ is a Gaussian RV with zero mean and variance   $\eta^2$, then 
\small
\begin{equation}
\begin{aligned}
S_p&=\int_{-\infty}^{\infty}\left[\left(S_p|k=L\right) p_{\rm{LOS}} (r_x) + \left(S_p|k=N\right) \left(1-p_{\rm{LOS}}\left(r_x\right)\right)\right]\\
&\;\frac{1}{\sqrt{2\pi \eta^2}}\exp\left(\frac{- r_x^2}{2 \eta^2}\right) dr_x.
\end{aligned}
\end{equation}

\bibliographystyle{ieeetr}
\bibliography{bibliography.bib}

\begin{thebibliography}{10}

\bibitem{cao2018airborne}
X.~Cao, P.~Yang, M.~Alzenad, X.~Xi, D.~Wu, and H.~Yanikomeroglu, ``Airborne
  communication networks: A survey,'' {\em IEEE Journal on Selected Areas in
  Communications}, vol.~36, no.~9, pp.~1907--1926, 2018.

\bibitem{bor20195g}
I.~Bor-Yaliniz, M.~Salem, G.~Senerath, and H.~Yanikomeroglu, ``Is {5G} ready
  for drones: A look into contemporary and prospective wireless networks from a
  standardization perspective,'' {\em IEEE Wireless Communications}, vol.~26,
  no.~1, pp.~18--27, 2019.

\bibitem{zeng2019accessing}
Y.~Zeng, Q.~Wu, and R.~Zhang, ``Accessing from the sky: A tutorial on {UAV}
  communications for {5G} and beyond,'' {\em Proceedings of the IEEE},
  vol.~107, no.~12, pp.~2327--2375, 2019.

\bibitem{li2018uav}
B.~Li, Z.~Fei, and Y.~Zhang, ``{UAV} communications for {5G} and beyond: Recent
  advances and future trends,'' {\em IEEE Internet of Things Journal}, vol.~6,
  no.~2, pp.~2241--2263, 2018.

\bibitem{darwish2021leo}
T.~Darwish, G.~K. Kurt, H.~Yanikomeroglu, M.~Bellemare, and G.~Lamontagne,
  ``{LEO} satellites in {5G} and beyond networks: A review from a
  standardization perspective,'' {\em arXiv preprint arXiv:2110.08654}, 2021.

\bibitem{kurt2021vision}
G.~K. Kurt, M.~G. Khoshkholgh, S.~Alfattani, A.~Ibrahim, T.~S. Darwish, M.~S.
  Alam, H.~Yanikomeroglu, and A.~Yongacoglu, ``A vision and framework for the
  high altitude platform station {(HAPS)} networks of the future,'' {\em IEEE
  Communications Surveys \& Tutorials}, vol.~23, no.~2, pp.~729--779, 2021.

\bibitem{alam2021high}
M.~S. Alam, G.~K. Kurt, H.~Yanikomeroglu, P.~Zhu, and N.~D. {\DJ}{\`a}o, ``High
  altitude platform station based super macro base station constellations,''
  {\em IEEE Communications Magazine}, vol.~59, no.~1, pp.~103--109, 2021.

\bibitem{mozaffari2019tutorial}
M.~Mozaffari, W.~Saad, M.~Bennis, Y.-H. Nam, and M.~Debbah, ``A tutorial on
  {UAVs} for wireless networks: Applications, challenges, and open problems,''
  {\em IEEE communications surveys \& tutorials}, vol.~21, no.~3,
  pp.~2334--2360, 2019.

\bibitem{lin2018sky}
X.~Lin, V.~Yajnanarayana, S.~D. Muruganathan, S.~Gao, H.~Asplund, H.-L.
  Maattanen, M.~Bergstrom, S.~Euler, and Y.-P.~E. Wang, ``The sky is not the
  limit: {LTE} for unmanned aerial vehicles,'' {\em IEEE Communications
  Magazine}, vol.~56, no.~4, pp.~204--210, 2018.

\bibitem{azizi2019profit}
A.~Azizi, S.~Parsaeefard, M.~R. Javan, N.~Mokari, and H.~Yanikomeroglu,
  ``Profit maximization in {5G+} networks with heterogeneous aerial and ground
  base stations,'' {\em IEEE Transactions on Mobile Computing}, vol.~19,
  no.~10, pp.~2445--2460, 2019.

\bibitem{cicek2019uav}
C.~T. Cicek, H.~Gultekin, B.~Tavli, and H.~Yanikomeroglu, ``{UAV} base station
  location optimization for next generation wireless networks: Overview and
  future research directions,'' in {\em 2019 1st International Conference on
  Unmanned Vehicle Systems-Oman (UVS)}, pp.~1--6, IEEE, 2019.

\bibitem{zeng2017energy}
Y.~Zeng and R.~Zhang, ``Energy-efficient {UAV} communication with trajectory
  optimization,'' {\em IEEE Transactions on Wireless Communications}, vol.~16,
  no.~6, pp.~3747--3760, 2017.

\bibitem{ahmad2020device}
M.~Ahmad, M.~Ali, M.~Naeem, A.~Ahmed, M.~Iqbal, W.~Ejaz, and A.~Anpalagan,
  ``{Device-centric communication in IoT: an energy efficiency perspective},''
  {\em Transactions on Emerging Telecommunications Technologies}, vol.~31,
  no.~2, p.~e3750, 2020.

\bibitem{jameel2018survey}
F.~Jameel, Z.~Hamid, F.~Jabeen, S.~Zeadally, and M.~A. Javed, ``{A survey of
  device-to-device communications: Research issues and challenges},'' {\em IEEE
  Communications Surveys \& Tutorials}, vol.~20, no.~3, pp.~2133--2168, 2018.

\bibitem{masaracchia2021uav}
A.~Masaracchia, Y.~Li, K.~K. Nguyen, C.~Yin, S.~R. Khosravirad, D.~B. Da~Costa,
  and T.~Q. Duong, ``{{UAV-enabled Ultra-Reliable Low-Latency Communications
  for 6G}} : {A Comprehensive Survey},'' {\em IEEE Access}, 2021.

\bibitem{gharbieh2017spatiotemporal}
M.~Gharbieh, H.~ElSawy, A.~Bader, and M.-S. Alouini, ``Spatiotemporal
  stochastic modeling of {IoT} enabled cellular networks: Scalability and
  stability analysis,'' {\em IEEE Transactions on Communications}, vol.~65,
  no.~8, pp.~3585--3600, 2017.

\bibitem{mankar2021spatial}
P.~D. Mankar, M.~A. Abd-Elmagid, and H.~S. Dhillon, ``Spatial distribution of
  the mean peak age of information in wireless networks,'' {\em IEEE
  Transactions on Wireless Communications}, 2021.

\bibitem{nabil2022data}
Y.~Nabil, H.~ElSawy, S.~Al-Dharrab, H.~Mostafa, and H.~Attia, ``Data
  aggregation in regular large-scale iot networks: Granularity, reliability,
  and delay tradeoffs,'' {\em IEEE Internet of Things Journal}, vol.~9, no.~18,
  pp.~17767--17784, 2022.

\bibitem{zeng2016wireless}
Y.~Zeng, R.~Zhang, and T.~J. Lim, ``Wireless communications with unmanned
  aerial vehicles: Opportunities and challenges,'' {\em IEEE Communications
  Magazine}, vol.~54, no.~5, pp.~36--42, 2016.

\bibitem{fotouhi2019survey}
A.~Fotouhi, H.~Qiang, M.~Ding, M.~Hassan, L.~G. Giordano, A.~Garcia-Rodriguez,
  and J.~Yuan, ``Survey on {UAV} cellular communications: Practical aspects,
  standardization advancements, regulation, and security challenges,'' {\em
  IEEE Communications Surveys \& Tutorials}, vol.~21, no.~4, pp.~3417--3442,
  2019.

\bibitem{matracia2021coverage}
M.~Matracia, M.~Kishk, and M.-S. Alouini, ``Coverage analysis for
  {UAV}-assisted cellular networks in rural areas,'' {\em IEEE Open Journal of
  Vehicular Technology}, 2021.

\bibitem{mei2019cellular}
W.~Mei, Q.~Wu, and R.~Zhang, ``Cellular-connected {UAV}: Uplink association,
  power control and interference coordination,'' {\em IEEE Transactions on
  wireless communications}, vol.~18, no.~11, pp.~5380--5393, 2019.

\bibitem{arshad2018integrating}
R.~Arshad, L.~Lampe, H.~ElSawy, and M.~J. Hossain, ``Integrating {UAVs} into
  existing wireless networks: A stochastic geometry approach,'' in {\em 2018
  IEEE Globecom Workshops (GC Wkshps)}, pp.~1--6, IEEE, 2018.

\bibitem{mozaffari2017mobile}
M.~Mozaffari, W.~Saad, M.~Bennis, and M.~Debbah, ``Mobile unmanned aerial
  vehicles {(UAVs)} for energy-efficient {Internet of Things} communications,''
  {\em IEEE Transactions on Wireless Communications}, vol.~16, no.~11,
  pp.~7574--7589, 2017.

\bibitem{mozaffari2016unmanned}
M.~Mozaffari, W.~Saad, M.~Bennis, and M.~Debbah, ``Unmanned aerial vehicle with
  underlaid device-to-device communications: Performance and tradeoffs,'' {\em
  IEEE Transactions on Wireless Communications}, vol.~15, no.~6,
  pp.~3949--3963, 2016.

\bibitem{bushnaq2019aeronautical}
O.~M. Bushnaq, A.~Celik, H.~ElSawy, M.-S. Alouini, and T.~Y. Al-Naffouri,
  ``Aeronautical data aggregation and field estimation in {IoT} networks:
  Hovering and traveling time dilemma of {UAVs},'' {\em IEEE Transactions on
  Wireless Communications}, vol.~18, no.~10, pp.~4620--4635, 2019.

\bibitem{choi2019modeling}
C.-S. Choi, F.~Baccelli, and G.~de~Veciana, ``Modeling and analysis of data
  harvesting architecture based on unmanned aerial vehicles,'' {\em IEEE
  Transactions on Wireless Communications}, vol.~19, no.~3, pp.~1825--1838,
  2019.

\bibitem{zhang2019stochastic}
S.~Zhang, J.~Liu, and W.~Sun, ``Stochastic geometric analysis of multiple
  unmanned aerial vehicle-assisted communications over {Internet of Things},''
  {\em IEEE Internet of Things Journal}, vol.~6, no.~3, pp.~5446--5460, 2019.

\bibitem{xiong2020uav}
Z.~Xiong, Y.~Zhang, W.~Y.~B. Lim, J.~Kang, D.~Niyato, C.~Leung, and C.~Miao,
  ``{UAV-assisted wireless energy and data transfer with deep reinforcement
  learning},'' {\em IEEE Transactions on Cognitive Communications and
  Networking}, vol.~7, no.~1, pp.~85--99, 2020.

\bibitem{ng2022uav}
J.~S. Ng, W.~C. Ng, W.~Y.~B. Lim, Z.~Xiong, D.~Niyato, C.~Leung, and C.~Miao,
  ``{UAV-assisted Wireless Power Charging for Efficient Hybrid Coded Edge
  Computing Network},'' in {\em ICC 2022-IEEE International Conference on
  Communications}, pp.~4037--4042, IEEE, 2022.

\bibitem{yang2021joint}
Y.~Yang, X.~Wei, R.~Xu, and L.~Peng, ``{Joint Optimization of AoI, SINR,
  Completeness, and Energy in UAV-Aided SDCNs: Coalition Formation Game and
  Cooperative Order},'' {\em IEEE Transactions on Green Communications and
  Networking}, vol.~6, no.~1, pp.~265--280, 2021.

\bibitem{han2022age}
Z.~Han, Y.~Yang, W.~Wang, L.~Zhou, T.~N. Nguyen, and C.~Su, ``{Age Efficient
  Optimization in UAV-Aided VEC Network: A Game Theory Viewpoint},'' {\em IEEE
  Transactions on Intelligent Transportation Systems}, vol.~23, no.~12,
  pp.~25287--25296, 2022.

\bibitem{yang2022aoi}
Y.~Yang, W.~Wang, R.~Xu, G.~Srivastava, M.~Alazab, T.~R. Gadekallu, and C.~Su,
  ``{AoI Optimization for UAV-aided MEC Networks under Channel Access Attacks:
  A Game Theoretic Viewpoint},'' in {\em ICC 2022-IEEE International Conference
  on Communications}, pp.~1--6, IEEE, 2022.

\bibitem{wang2022data}
W.~Wang, G.~Srivastava, J.~C.-W. Lin, Y.~Yang, M.~Alazab, and T.~R. Gadekallu,
  ``{Data freshness optimization under CAA in the UAV-aided MECN: a potential
  game perspective},'' {\em IEEE Transactions on Intelligent Transportation
  Systems}, 2022.

\bibitem{9470921}
N.~Kouzayha, H.~Elsawy, H.~Dahrouj, K.~Alshaikh, T.~Y. Al-Naffouri, and M.-S.
  Alouini, ``Analysis of large scale aerial terrestrial networks with mmwave
  backhauling,'' {\em IEEE Transactions on Wireless Communications}, vol.~20,
  no.~12, pp.~8362--8380, 2021.

\bibitem{hattab2020energy}
G.~Hattab and D.~Cabric, ``Energy-efficient massive {IoT} shared spectrum
  access over {UAV}-enabled cellular networks,'' {\em IEEE Transactions on
  Communications}, vol.~68, no.~9, pp.~5633--5648, 2020.

\bibitem{UAV_Dis}
A.~M. Hayajneh, S.~A.~R. Zaidi, D.~C. McLernon, M.~Di~Renzo, and M.~Ghogho,
  ``Performance analysis of {UAV} enabled disaster recovery networks: A
  stochastic geometric framework based on cluster processes,'' {\em IEEE
  Access}, vol.~6, pp.~26215--26230, 2018.

\bibitem{qin2020performance}
Y.~Qin, M.~A. Kishk, and M.-S. Alouini, ``Performance evaluation of
  {UAV}-enabled cellular networks with battery-limited drones,'' {\em IEEE
  Communications Letters}, vol.~24, no.~12, pp.~2664--2668, 2020.

\bibitem{lahmeri2019stochastic}
M.-A. Lahmeri, M.~A. Kishk, and M.-S. Alouini, ``Stochastic geometry-based
  analysis of airborne base stations with laser-powered {UAVs},'' {\em IEEE
  Communications Letters}, vol.~24, no.~1, pp.~173--177, 2019.

\bibitem{Tethered_kisk}
M.~A. Kishk, A.~Bader, and M.-S. Alouini, ``On the 3-d placement of airborne
  base stations using tethered {UAVs},'' {\em IEEE Transactions on
  Communications}, vol.~68, no.~8, pp.~5202--5215, 2020.

\bibitem{zhu2022aerial}
K.~Zhu, J.~Yang, Y.~Zhang, J.~Nie, W.~Y.~B. Lim, H.~Zhang, and Z.~Xiong,
  ``{Aerial Refueling: Scheduling Wireless Energy Charging for UAV Enabled Data
  Collection},'' {\em IEEE Transactions on Green Communications and
  Networking}, 2022.

\bibitem{popovski2014ultra}
P.~Popovski, ``Ultra-reliable communication in {5G} wireless systems,'' in {\em
  1st International Conference on 5G for Ubiquitous Connectivity},
  pp.~146--151, IEEE, 2014.

\bibitem{sutton2019enabling}
G.~J. Sutton, J.~Zeng, R.~P. Liu, W.~Ni, D.~N. Nguyen, B.~A. Jayawickrama,
  X.~Huang, M.~Abolhasan, Z.~Zhang, E.~Dutkiewicz, {\em et~al.}, ``Enabling
  technologies for ultra-reliable and low latency communications: {From {PHY}
  and {MAC}} layer perspectives,'' {\em IEEE Communications Surveys \&
  Tutorials}, vol.~21, no.~3, pp.~2488--2524, 2019.

\bibitem{wang2020packet}
K.~Wang, C.~Pan, H.~Ren, W.~Xu, L.~Zhang, and A.~Nallanathan, ``Packet error
  probability and effective throughput for ultra-reliable and low-latency {UAV}
  communications,'' {\em IEEE Transactions on Communications}, vol.~69, no.~1,
  pp.~73--84, 2020.

\bibitem{she2018uav}
C.~She, C.~Liu, T.~Q. Quek, C.~Yang, and Y.~Li, ``{UAV-assisted uplink
  transmission for ultra-reliable and low-latency communications},'' in {\em
  2018 IEEE International Conference on Communications Workshops (ICC
  Workshops)}, pp.~1--6, IEEE, 2018.

\bibitem{chen2020power}
K.~Chen, Y.~Wang, Z.~Fei, and X.~Wang, ``Power limited ultra-reliable and
  low-latency communication in {UAV}-enabled {IoT} networks,'' in {\em 2020
  IEEE Wireless Communications and Networking Conference (WCNC)}, pp.~1--6,
  IEEE, 2020.

\bibitem{li2021aerial}
Y.~Li, C.~Yin, T.~Do-Duy, A.~Masaracchia, and T.~Q. Duong, ``{Aerial
  Reconfigurable Intelligent Surface-Enabled URLLC UAV Systems},'' {\em IEEE
  Access}, vol.~9, pp.~140248--140257, 2021.

\bibitem{azari2017ultra}
M.~M. Azari, F.~Rosas, K.-C. Chen, and S.~Pollin, ``{Ultra reliable UAV
  communication using altitude and cooperation diversity},'' {\em IEEE
  Transactions on Communications}, vol.~66, no.~1, pp.~330--344, 2017.

\bibitem{ranjha2019quasi}
A.~Ranjha and G.~Kaddoum, ``{Quasi-optimization of distance and blocklength in
  URLLC aided multi-hop UAV relay links},'' {\em IEEE Wireless Communications
  Letters}, vol.~9, no.~3, pp.~306--310, 2019.

\bibitem{han2019uav}
A.~Han, T.~Lv, and X.~Zhang, ``{UAV beamwidth design for ultra-reliable and
  low-latency communications with NOMA},'' in {\em 2019 IEEE International
  Conference on Communications Workshops (ICC Workshops)}, pp.~1--6, IEEE,
  2019.

\bibitem{ranjha2021facilitating}
A.~Ranjha, G.~Kaddoum, and K.~Dev, ``{Facilitating URLLC in UAV-assisted relay
  systems with multiple-mobile robots for 6G Networks: A prospective of
  agriculture 4.0},'' {\em IEEE Transactions on Industrial Informatics},
  vol.~18, no.~7, pp.~4954--4965, 2021.

\bibitem{she2019ultra}
C.~She, C.~Liu, T.~Q. Quek, C.~Yang, and Y.~Li, ``{Ultra-reliable and
  low-latency communications in unmanned aerial vehicle communication
  systems},'' {\em IEEE Transactions on communications}, vol.~67, no.~5,
  pp.~3768--3781, 2019.

\bibitem{el2021uav}
E.~El~Haber, H.~A. Alameddine, C.~Assi, and S.~Sharafeddine, ``{UAV-aided
  ultra-reliable low-latency computation offloading in future IoT networks},''
  {\em IEEE Transactions on Communications}, vol.~69, no.~10, pp.~6838--6851,
  2021.

\bibitem{xi2020network}
X.~Xi, X.~Cao, P.~Yang, J.~Chen, T.~Q. Quek, and D.~Wu, ``{Network resource
  allocation for eMBB payload and URLLC control information communication
  multiplexing in a multi-UAV relay network},'' {\em IEEE Transactions on
  Communications}, vol.~69, no.~3, pp.~1802--1817, 2020.

\bibitem{elsawy2016modeling}
H.~ElSawy, A.~Sultan-Salem, M.-S. Alouini, and M.~Z. Win, ``Modeling and
  analysis of cellular networks using stochastic geometry: A tutorial,'' {\em
  IEEE Communications Surveys \& Tutorials}, vol.~19, no.~1, pp.~167--203,
  2016.

\bibitem{al2014optimal}
A.~Al-Hourani, S.~Kandeepan, and S.~Lardner, ``Optimal {LAP} altitude for
  maximum coverage,'' {\em IEEE Wireless Communications Letters}, vol.~3,
  no.~6, pp.~569--572, 2014.

\bibitem{andrews2016modeling}
J.~G. Andrews, T.~Bai, M.~N. Kulkarni, A.~Alkhateeb, A.~K. Gupta, and R.~W.
  Heath, ``Modeling and analyzing millimeter wave cellular systems,'' {\em IEEE
  Transactions on Communications}, vol.~65, no.~1, pp.~403--430, 2016.

\bibitem{ratnarajah2016performance}
T.~Ratnarajah, ``On the performance of relay aided millimeter wave networks,''
  {\em IEEE Journal of Selected Topics in Signal Processing}, 2016.

\bibitem{emara_Iot}
M.~Emara, H.~Elsawy, and G.~Bauch, ``A spatiotemporal model for peak {AoI} in
  uplink {IoT} networks: Time versus event-triggered traffic,'' {\em IEEE
  Internet of Things Journal}, vol.~7, no.~8, pp.~6762--6777, 2020.

\bibitem{gharbieh2018spatiotemporal}
M.~Gharbieh, H.~ElSawy, H.-C. Yang, A.~Bader, and M.-S. Alouini,
  ``{Spatiotemporal model for uplink IoT traffic: Scheduling and random access
  paradox},'' {\em IEEE Transactions on Wireless Communications}, vol.~17,
  no.~12, pp.~8357--8372, 2018.

\bibitem{kouzayha2017joint}
N.~Kouzayha, Z.~Dawy, J.~G. Andrews, and H.~ElSawy, ``{Joint downlink/uplink RF
  wake-up solution for IoT over cellular networks},'' {\em IEEE Transactions on
  Wireless Communications}, vol.~17, no.~3, pp.~1574--1588, 2017.

\bibitem{haenggi2015meta}
M.~Haenggi, ``{The meta distribution of the SIR in Poisson bipolar and cellular
  networks},'' {\em IEEE Transactions on Wireless Communications}, vol.~15,
  no.~4, pp.~2577--2589, 2015.

\bibitem{gil1951note}
J.~Gil-Pelaez, ``Note on the inversion theorem,'' {\em Biometrika}, vol.~38,
  no.~3-4, pp.~481--482, 1951.

\bibitem{ibrahim2020meta}
H.~Ibrahim, H.~Tabassum, and U.~T. Nguyen, ``{The meta distributions of the
  SIR/SNR and data rate in coexisting sub-6GHz and millimeter-wave cellular
  networks},'' {\em IEEE Open Journal of the Communications Society}, vol.~1,
  pp.~1213--1229, 2020.

\bibitem{george2017ergodic}
G.~George, R.~K. Mungara, A.~Lozano, and M.~Haenggi, ``Ergodic spectral
  efficiency in {MIMO} cellular networks,'' {\em IEEE Transactions on Wireless
  Communications}, vol.~16, no.~5, pp.~2835--2849, 2017.

\bibitem{alfa2016applied}
A.~S. Alfa, {\em {Applied Discrete-Time Queues}}.
\newblock Springer, 2016.

\bibitem{bose2013introduction}
S.~K. Bose, {\em An introduction to queueing systems}.
\newblock Springer Science \& Business Media, 2013.

\bibitem{zeng2019energy}
Y.~Zeng, J.~Xu, and R.~Zhang, ``Energy minimization for wireless communication
  with rotary-wing {UAV},'' {\em IEEE Transactions on Wireless Communications},
  vol.~18, no.~4, pp.~2329--2345, 2019.

\bibitem{alzer1997some}
H.~Alzer, ``On some inequalities for the incomplete gamma function,'' {\em
  Mathematics of Computation}, vol.~66, no.~218, pp.~771--778, 1997.

\end{thebibliography}

%








\end{document}